\documentclass[useAMS,usenatbib]{mn2e}
\usepackage{amsmath,amssymb}
\usepackage{times}
\usepackage[bf,big]{titlesec}
\usepackage[small,it]{caption}
\usepackage{stfloats}
\usepackage{graphicx}
\usepackage{color}
\usepackage{rotating}
\usepackage{booktabs,fixltx2e}
\usepackage{threeparttable}
\usepackage[section]{placeins}
\usepackage{multicol}
\numberwithin{equation}{section} 
\title[Surface Brightness and Luminosity of Ellipticals]{Surface Brightness and Intrinsic Luminosity of Ellipticals}
\author[Barun Kumar Dhar and Liliya L.R. Williams]{Barun Kumar Dhar$^{1}$\thanks{E-mail:dhar@physics.umn.edu}, Liliya L.R. Williams$^{1}$\\
  $^{1}$School of Physics and Astronomy, University of Minnesota,
  116 Church Street SE, Minneapolis, MN 55455 USA}
\date{Accepted 12th December, 2011; in original form 2nd June, 2011.}
\newcommand{\mgasc}{\ensuremath{mag~arcsec^{-2}}}
\begin{document}
  \maketitle
  \label{firstpage}  
  \begin{abstract}
    We show that the surface brightness profiles of elliptical galaxies can be parametrized using a linear superposition of two or three components, each of which is described by functions developed in Dhar \& Williams as the 2D projections of a 3D Einasto density profile. For a sample of 23 ellipticals in and around the Virgo Cluster with total absolute {\it V}-magnitude $-24$$<$$M_{VT}$$<$$-15$, our multi-component models span a dynamic range up to $10^6$ in surface brightness and up to $10^5$ in radius down to the resolution limit of the HST, have a median rms of 0.032 $\mgasc$ consistent with the rms of 0.03 from random errors of the data, and are statistically justified at $>$3$\sigma$. Our models indicate that i) the central component is more concentrated than the outer component; and ii) the central component of massive shallow-cusp ('core') galaxies is much more luminous, extended and concentrated than that of steep-cusp ('cuspy') galaxies, with their near exponential central profiles indicating disk-like systems, whose existence, must be verified spectroscopically.

Galaxy structure can thus be modelled extremely well with a central mass excess for all galaxies. This is not necessarily contrary to the notion of a mass deficit in 'core' galaxies, since mass ejection due to core-scouring by a supermassive black hole (SMBH) binary could have affected the shape of the central components. However, we show that the existence, amount, radial extent and sign of such deficits disagree substantially in the literature, both for a given galaxy and on an average over a sample. We discuss possible implications and suggest that SMBH binaries are unlikely to be the sole mechanism for producing the large 'cores' of massive galaxies.

Using results from the SAURON survey we deduce that under certain conditions of symmetry, inclination angles and degree of triaxiality, the intrinsic (3D) density of light can be well described with a multi-component Einasto model for both steep- as well as shallow- cusp galaxies. This indicates an universality in the functional form describing the 3D density distribution of light in galaxies and dark matter in $\Lambda$CDM N-body haloes. Finally, Planetary Nebulae and strong lensing observations, and the Einasto index $n$ of $\Lambda$CDM dark matter haloes indicate that our result -- the outer component of the surface brightness profiles of massive galaxies has 5 $\lesssim$ $n$ $\lesssim$ 8 -- could imply i) a common feature of collisionless systems; and ii) that galaxies with such $n$ for their outer component are dark matter dominated.
  \end{abstract}     
  \section{Introduction}\label{intro}
  \cite{deVauc48} showed that a remarkably simple parametrization of the radial surface brightness profile of galaxies exists for a wide range of ellipticals. He proposed a two parameter fitting function of the form of equation \eqref{sersic} with $m$$=$4 defining the shape of the distribution and $q$$=$7.67 ensuring that the {\it effective} or {\it half-light radius}, $R_{E}$, contains half the total projected light.
  \begin{align}\label{sersic}
    \Sigma(R)= \Sigma_{R_E}\exp\left\{-q\left[\left(\frac{R}{R_E}\right)^{\frac{1}{m}} -1\right]\right\}
  \end{align}
 where, $\Sigma(R)$ is the 2D surface brightness at a plane of sky projected radial distance $R$, $\Sigma_{R_E}$$=$$\Sigma(R_E)$, $q$$=$$q(m)$ and $\Sigma_S(0)$$=$$\Sigma_{R_E}$$e^q$.  

  Observing that our Galaxy is made up of multiple subsystems \cite{Einasto65} proposed a modification, equation \eqref{einasto}, of the 2D de Vaucouleurs law to model the intrinsic (3D) baryonic mass density $\rho(r)$ of each subsystem by allowing the shape parameter $n$ to be a free parameter, and $b$$=$$b(n)$. 
  \begin{align}\label{einasto}
    \rho(r) = \rho_{s}\exp\left\{-b\left[\left(\frac{r}{r_s}\right)^{\frac{1}{n}} -1\right]\right\}
  \end{align}
  where, $\rho_s$ is the density(3D) at a scale radius $r_s$ and $\rho(0)$$=$$\rho_s$$e^b$. 
  
  Around the same time \cite{Sersic68} observed that $m$ in equation \eqref{sersic}, characterized by $m=4$ in de Vaucouleurs law, was not the same for all galaxies and rendering it a free-parameter provided much better fits to the surface brightness profiles (hereafter, SB). Equation~\eqref{sersic} is the current standard paradigm for describing the global structural properties of galaxies.
  
  Over the past forty years, the pioneering works of Caon, Capaccioli, Einasto, Ferrarese, Graham, Kormendy, Lauer and their collaborators have shown that no single commonly used three parameter fitting function could model the SB over the entire dynamic radial range. They showed that the SB profiles of all galaxies reveal an inherently multi-component structure such that the outer regions can be described with a Sersic or a Sersic+exponential models while the central regions can be described with power-laws. 

  Additionally, for some nearby spirals and the giant elliptical M87 (NGC4486), Einasto and collaborators (Rummel, Haud and Tenjes) have shown that if spectroscopic and kinematic constraints are used in addition to the SB data, then the intrinsic 3D mass density including their central regions can be described with a multi-component Einasto model.
  \subsection{Motivation for this work}
  While the power-law+Sersic models are widely accepted as providing an accurate description of the SB profiles, the fit residuals of these models are often larger than measurement errors. Further, these functions do not have the flexibility to model deviations from power-laws within the central regions. Since these fits are often used to draw inferences on galaxy structure, formation and evolution, it is important to address two crucial issues when modelling galaxy structure:
  
  (i)~Models of the SB profiles must be consistent with measurement errors over the entire available radial range; and  
  
  (ii)~Model fitting functions must allow one to easily infer the intrinsic 3D luminosity density from the 2D SB profiles.

  Our goal in this paper is to address the above issues with a new function derived in \cite{DW10} (DW10) to model the projected surface mass density profile of Einasto-like systems. We show that a multi-component parametrization with this function (hereafter, the DW-function) provides excellent fits, consistent with measurement errors, to the SB of ellipticals over a large dynamic radial range down to the HST resolution limit.

  This suggests that the 3D density distribution of light in galaxies can be described with a multi-component Einasto model.

  Such an interpretation is similar to that of Einasto and collaborators, except that a) our work extends to a much larger sample of shallow and steep cusp ellipticals; and b) we deduce the intrinsic 3D Einasto model properties by fitting the 2D SB using a multi-component DW-function.

  For the rest of this paper, we draw a distinction between components deduced from fitting and physically distinct kinematic systems or stellar populations. We shall refer to {\it a component} as a single spherically symmetric closed form fitting function described by at most three parameters -- a scale length, $r_{\alpha}$, a shape parameter, $\alpha$, and some normalization, $\Sigma_{0}$ or $\Sigma_{\alpha}$. Our multi-component fits consist of the minimum number of such functions (i.e. DW-functions) required to model the entire dynamic radial range down to the resolution limit of the HST. The minimum number depends on the quality of data, the available degrees of freedom, and the amplitude of systematic patterns in the residuals.
  
  One should hence use caution in interpreting these {\it fit components} as kinematically {\it distinct systems or stellar populations} since the physical properties of the components, and even their presence will depend on the choice of parametrization. It is however possible that some of our fitted components do coincide with kinematically identified systems or stellar populations, or combinations thereof, which shall then facilitate interesting interpretations.
    
  Also note that neither fit components nor physically distinct systems have to correspond to structure in the total gravitational potential. Only where $\rho_{baryons}$$\gg$$\rho_{DM}$, or $\rho_{baryons}$$\propto$$\rho_{DM}$ features observed in the SB profiles will trace the total mass density, and hence will provide information about the gravitational potential.
 
  Further, since no galaxy has a truly flat density core, we refrain from using the terms 'core-' and 'cusp-' galaxies and instead refer to them as shallow-cusp and steep-cusp galaxies, respectively.
  \subsection{Brief outline and summary of results}
  In section \ref{MCstruc} we provide a brief history and basis for believing that galaxies have a multi-component structure. This is followed by an overview, in section \ref{fittingfuncs}, of the most flexible fitting functions tried to date, and the motivation for this work. The data we use are described in section \ref{data} and important features of our 2D parametrization and the fitting procedure are discussed in section \ref{DWfunction}. Section \ref{fitresults} discusses results of our 2D fits to a sample of 23 ellipticals in and around the Virgo cluster spanning absolute V-magnitudes in the range -24$<$$M_{VT}$$<$-15. Our models produce consistently low residuals, over a large dynamic radial range ($\sim$$10^5$) and in congruence with the measurement errors. 

  Section \ref{comparison} contains a comparison of fits with other parametrizations from the literature. It highlights the main result of this work: {\it of all the functions tried to date to model the 2D structure of galaxies, the best fitting model is the multi-component DW-function}. This suggests that the intrinsic 3D density structure of galaxies is best represented with a multi-component Einasto model.

  In section \ref{compprops} we discuss the properties of the components deduced from the multi-component fits with the DW-function, and in section \ref{compsys} we present three cases where the central DW-component of our multi-component fits coincide with spectroscopically identified systems.
  
 We next explore the conditions under which our 2D models can be used to infer that the intrinsic 3D luminosity density distribution can be described with a multi-component Einasto model. In section \ref{uniqueness} we therefore first discuss the key issues pertaining to non-uniqueness of deprojection, and then in section \ref{lumdensity} we present our models of the 3D Einasto luminosity density profiles for $14$ galaxies whose deprojections are less likely to suffer from non-uniqueness. Since these $14$ galaxies span a wide luminosity range and belong to both the steep- and shallow-cusp classes, we conclude that the intrinsic structure of all galaxies could be described with multi-component Einasto models as well.

 This is followed by section \ref{massdeficit}, where we discuss the notion of 'mass deficit' in massive ellipticals. Following a brief review of the phenomena of mass ejection by binary SMBHs and results from N-body simulations, we show that {\it estimates} of the amount, radial extent and sign of the 'observed' deficit in real galaxies, have large disagreements in the literature -- both for a given galaxy as well as on an average for a given sample.

 In this paper we have shown that the structure of galaxies can be modelled extremely well with a central mass excess for all galaxies. We suggest that, while mass ejection due to core-scouring by binary SMBHs could have shaped the central and the intermediate DW-components, this phonemena by itself is unlikely to be the sole cause for the observed shallow-cusps; other formation processes are likely to have also contributed to the existence of this feature.

  In section \ref{BaryonsDM} we show that an Einasto shape parameter of 5 $\lesssim$ $n$ $\lesssim$ 8 is found in two different systems - (i) the outer baryonic DW-component of the massive ellipticals in our sample, and (ii) in fits with an Einasto profile to the collisionless $\Lambda$CDM N-body haloes. This range of $n$ may thus be a common feature of collisionless systems. Finally, by comparing results from Planetary nebulae and strong lensing studies we show that it is possible that galaxies with 5 $\lesssim$ $n$ $\lesssim$ 8 for the outer component of their SB profiles are likely to be more dark matter dominated.
  
Section \ref{conclusion} contains a general discussion and a summary of conclusions drawn from the work presented in this paper.
  \section{Multi-component structure of galaxies}\label{MCstruc} 
  \subsection{Structure of baryonic (stellar) distribution}\label{MCbaryons}
  Hubble's (1926) classification of galaxies as ellipticals and spirals based on the identification of bulges and disks was probably the first systematic characterization of the multi-component structure of galaxies. \cite{deVauc59a} showed that the outer regions of spirals and lenticulars can be modelled as an exponential (also see \cite{Patt40}, de Vaucouleurs (1955, 1959b) with the central bulge following an $\exp(-r^{1/4})$ profile, similar to ellipticals (de Vaucouleurs 1948,1958). \cite{BertCap70} modelled the giant elliptical M87 as a linear superposition of a de Vaucouleurs and exponential profiles. The multi-component structure of lenticulars as a superposition of an exponential and a $\exp(-r^{1/4})$ component was also investigated in \cite{Free70} and \cite{Korm77}.
  
  However, it was also observed that not all ellipticals and bulges of lenticulars followed a pure de Vaucouleurs profile. Even then Sersic's (1968) generalized form, equation \eqref{sersic}, did not find much application until the comprehensive works of \cite{Caon93} and \cite{Graham96}, which revealed how an incorrect parametrization -- a de Vaucouleurs profile instead of a Sersic profile -- could lead to misleading conclusions (also see \cite{Davies88} and \cite{YC94}). Since then the Sersic profile has become the norm for describing the SB of galaxies over large dynamic radial ranges. Additionally, \cite{F94}, \cite{L95} and \cite{Graham03} also observed that a single Sersic profile cannot fit the entire dynamic radial range, especially the central regions down to the HST resolution. The failure of a single Sersic profile to fit any of the galaxies in their sample confirmed that multiple fit components are necessary to adequately model the structure of galaxies.

  In the absence of a sound theoretical model of galaxy formation and evolution it is important to understand under what conditions can analysis through fitting functions identify the true intrinsic components of a galaxy. This is especially important since a correct (or incorrect) identification of such components can have a serious bearing on our understanding of galaxy structure, formation and evolution. It is therefore important to ask, how many components in a galaxy can be realistically identified and how?
  
  In a series of papers on nearby spirals and the giant shallow-cusp ('core') elliptical M87 (NGC4486), Einasto and collaborators (following \cite{Einasto65}) demonstrated that the 3D baryonic mass density can be described as a multi-component system of nested Einasto profiles by parametrizing each kinematically and photometrically identified system or stellar population using equation \eqref{einasto} (see \cite{ER70}, \cite{HE89}, \cite{TEH91}, \cite{THE94}, \cite{THE98}). They observed that, depending on the galaxy and quality of data, the 3D mass density can be described as a certain combination of superposed systems -- a nucleus, a core, a bulge, a disk, a halo, a flat system -- all well fit with equation \eqref{einasto} and a massive corona of baryonic and dark matter described using an isothermal density profile. It is to be noted that in their construction each Einasto-component had an unique shape parameter $n$ and scale-length $r_s$ identifying physically distinct systems as opposed to apparent components generated by the use of \eqref{einasto} merely as a fitting function. To our knowledge, this was the first systematic modelling of the intrinsic 3D structure of the components in galaxies through a prior identification of systems.
  
  In the pre-HST era, while the detection of bulges and disks revealed a multi-component structure at larger radial scales, the seeing effects of the ground based observations did not allow one to adequately resolve the central regions. Some galaxies possibly showed central flattening of the density profiles, but because of limited resolution these claims remained inconclusive.

  The dawn of the HST and high resolution, 0.02 $arcsec$ $pixel^{-1}$, imaging improved the situation dramatically. \cite{Crane93}, using the HST Faint Object Camera (FOC) showed that the central regions of galaxies do not have flat density cores, but instead exhibit a wide range of inner power-law slopes. The low S/N of their observations however prevented them from obtaining psf-deconvolved images that would allow for detailed modelling. 
  
  Imaging with the Wide-Field Planetary Camera I (WFPC-I) on the HST, \cite{Jaffe94}, \cite{vB94}, \cite{F94} and \cite{L95} generated psf- deconvolved images of the central regions of ellipticals. They found that the SB of the central regions convincingly revealed a wide range of inner slopes inside a characteristic break, or transition radius, with shallow slopes and fainter central magnitudes for the largest and most luminous galaxies, and steeper slopes and brighter central magnitudes for the smaller and less luminous galaxies thus implying an inherently diverse multi-component structure. This led to the development of a number of power-law based parametrizations (\cite{F94}, \cite{L95}, \cite{Graham03}) to model the central and outer regions of galaxies.
  
  Thus analysis of the SB profiles through parametric fitting functions also revealed a multi-component structure for galaxies, indicating the presence of two or three regions; a double power-law domain around a transition radius, and a main body for the galaxy usually well described with a Sersic profile. This is similar to the work done in modelling the 3D mass density by Einasto and co-workers except that there is no reason to assume that components generated through the use of fitting functions should correspond to kinematically identified components or distinct stellar populations. Further, there is no convincing reason as to why the form of the fitting functions should be different in different regions.

  \subsection{Structure of $\Lambda$CDM haloes}
  \cite{Nav04} (N+04) showed that $3$-parameter fitting functions, especially those with a power-law logarithmic slope like the Einasto profile, are able to describe the 3D mass density distribution of spherically averaged $\Lambda$CDM N-body haloes better than any of the $2$-parameter fitting functions tried to date. Subsequent simulations by \cite{M06} (M+06), \cite{Pr06}, \cite{G08}, \cite{HW08}, \cite{SM09}, \cite{Nav10} have confirmed the need for a $3$-parameter function and for over 30 such dwarf, galaxy and cluster size N-body haloes the Einasto profile seems to be the best performing function in comparison to other $3$-parameter fitting functions. The Einasto index deduced from these simulations are typically in the range 5$\lesssim$$n$$\lesssim$8. \cite{SM09} proposed a $2$-parameter function that provides fractionally better fit in terms of $rms$ than the Einasto profile for two haloes they simulated. However, even for these cases the Einasto profile has comparable residuals. 
  
  Note that current resolution of N-body simulations does not allow one to probe the very central regions of galaxies where baryons reveal a multi-component structure. One may thus conclude that within the resolved and converged domain of N-body simulations, the dark matter distribution can be described as a single component system.
  \subsection{Intrinsic and Projected structure}
  Noting the similarity between the functional form of the Einasto and the Sersic profiles, \cite{M05} obtained spherically averaged non-parametric estimates of the 3D intrinsic and 2D projected mass densities of the N-body haloes described in N+04. They found that the same fitting function, the Sersic profile, which is used to describe the SB of ellipticals also describes the surface densities of $\Lambda$CDM N-body haloes, whose intrinsic 3D density is best described by a function of similar form - the Einasto profile.
  
  \cite{DW10} show that while it is possible to find limited radial ranges over which a Sersic profile can approximate a projected Einasto profile, over a large radial range a Sersic profile is not a good representation of a projected Einasto profile, and using such fits can lead to a misinterpretation of the best fit parameters. DW10 point out that the fitted Sersic profile parameters depend strongly on the radial range of a projected Einasto profile. In other words, fits with a Sersic profile to a single projected Einasto profile implies the existence of a variable Sersic shape parameter. DW10 provide an accurate analytical approximation for the 2D projection of an Einasto profile in terms of the 3D Einasto profile parameters, thereby allowing one to explore the intrinsic properties of systems believed to be Einasto-like from 2D observations of those systems. 

  \section{An overview of fitting functions}\label{fittingfuncs}
  \subsection{Combination of power-law functions}\label{powerlaw}
  With the ability to resolve the central regions of galaxies with the HST, it was found that a single $3$-parameter fitting function, like the Sersic profile, is not able to model the SB profile over a large dynamic radial range down to the HST limit; less so with $2$-parameter functions, for example, the de Vaucouleurs and the Jaffe profiles. \cite{F94} thus proposed the $4$-parameter double power-law while \cite{L95} proposed the $5$-parameter Nuker profile - which is a modified double power-law with an additional parameter to control the sharpness of transition between the power-laws.

  \cite{L95} pointed out that the more flexible $5$-parameter Nuker profile is designed to model only the small central regions, $\sim$ 10-20 arcsec, and not the rest of the large galaxy structure, which could be well fit with a $3$-parameter Sersic profile. This is because power-laws assign a fixed logarithmic slope to the density distribution while the light of galaxies at large $R$ exhibit a variable slope. However, even with 5 parameters modelling a small radial range it can be seen that the central-most regions show residuals larger than measurement errors. This implies that a total of 8 parameters are required for a near complete description excluding the very central inner regions. Moreover, in the central regions many galaxies exhibit a sharp change in slope, i.e. a transition radius. Even though the presence of the transition is unambiguous, the domain over which one should fit the Nuker profile to obtain robust parameters is not always obvious.
    
  \cite{Graham03} showed that the best-fitting Nuker parameters are extremely sensitive to the chosen domain of fit. Hence, observing that the central regions can be modelled as power-laws while the outer regions require a Sersic profile, they proposed the $6$-parameter Core-Sersic profile which has an inner power-law coupled to an outer Sersic profile along with a parameter to control the sharpness of transition. \cite{Truj04} suggested a modification of the Core-Sersic profile by allowing an infinitely sharp transition along with a step-function that reduces the Core-Sersic profile to a $5$-parameter function, but as expected produces an unphysically sharp break in the profile quite unlike the much smoother transition observed in galaxies. Even then, the large fit residuals in the central region continue to exist.

  \cite{F06} found that the light excess in the central regions (often referred to as 'nuclei') of the steep-cusp galaxies in their sample could be best described with a $3$-parameter King model. Such 'nuclei' are not as rare as was originally believed; but may not always be very prominent. \cite{Cotenuclei} showed that in the ACS Virgo Cluster survey \citep{ACSVCS}, hereafter ACSVCS, 66-82 per cent of the galaxies have such a central feature. It thus appears that one needs a King model for the nuclear region, and either a Sersic, a Core-Sersic or a Nuker+Sersic profile to model the rest of the galaxy, i.e., a total of 6-11 parameters.
  
  While the overall rms of the fits with 6-11 parameters appear small, the residuals in many radial sections remain large, $\gtrsim$ 0.1 and sometimes $\gtrsim$ 0.2 $\mgasc$; considerably larger than the 0.01-0.05 $\mgasc$ uncertainty of HST quality data. For illustration we refer the reader to the fits in \cite{F06} using a combination of Sersic, Core-Sersic and King models where large, and sometimes divergent, residuals can be found in some regions. A comparison of fits with Core-Sersic, King and Nuker profiles in the central regions of some galaxies are also shown in \cite{L07}, clearly revealing the lack of a good fit in the central-most regions. More detailed discussion is provided in \S \ref{comparison}.

  Further, none of the above power-law based functions are defined in the limit $R\to0$. While this may seem to be of academic interest, since the density at $R$ $=$ 0 does not contribute much to the light enclosed, it can introduce uncertainties in interpreting deprojections needed to extract 3D intrinsic profiles, and also for fitting procedures using psf-convolved models, where one needs to specify a finite value for the SB at $R$ $=$ 0 for the convolution.
  \subsection{Multi-component modified exponentials}\label{modexp}
  All of the above power-law and Core-Sersic parametrizations are well-guessed but ad-hoc empirical fitting functions in 2D, in the sense that they are not a result of well established theoretical models of galaxy formation. However, despite the existence of residuals larger than measurement errors, the parameters of the fits are used to draw inferences on galaxy structure and evolution, which are intrinsically 3D phenomena. This could have been meaningful if the 2D models were deduced from an underlying physically motivated 3D distribution. In order to draw such inferences from functions that are merely fitting functions, one needs, at the very least, to have residual profiles consistently comparable to measurement errors over a large dynamic radial range, and not just a low $rms$.
  
  \cite{KFCB09}, hereafter KFCB09, addresses this issue partly by fitting a single modified exponential function, the Sersic profile, over a rigorously tested range where the function produces residuals comparable to measurement errors. However, this range is chosen to ensure that the Sersic profile produces a good fit and specifically excludes the entire domain within the break radius. Since the Sersic profile is an ad-hoc fitting function, parameters deduced from fits within a limited radial range can lead to misleading interpretations of the physics involved and in estimating properties of the region inside the break radius.
  
  Since the Sersic profile produces good fits over a large radial range and lenticulars have been modelled with a Sersic + exponential profile, a plausible alternative is to use a double-Sersic profile to model the entire galaxy. This approach had not received much attention until recently when \cite{Gonz03} showed that a double-Sersic profile provides much better fits to twelve Brightest Cluster Galaxies (BCG). An example of such a fit for the BCG in Abell 2984 is shown in their figure 2. \cite{CoteDS} also arrive at a similar conclusion for some galaxies in the ACSVCS and Fornax cluster surveys. A comprehensive study of fits with a double-Sersic profile is provided in Hopkins et al. (2009a,b) where they obtain low residuals over a large dynamic radial range. While they do not provide a residual profile, the {\it rms} of their fits are often larger than the measurement errors.
  
  From a mathematical stand point, the Sersic profile presents another difficulty: its deprojection, that can provide insights in to the 3D structure of the galaxy, is not very well analytically tractable. Asymptotic limits of deprojection can be found in \cite{Ciotti91} and approximate expressions are given in \cite{PS97} and \cite{LGM99} (hereafter, PS97 and LGM99). However, the PS97 and LGM99 approximations are not accurate at small $R$$\leq$$10^{-2}$$R_E$, and the 3D density becomes undefined as $r$$\to$$0$ for Sersic indices $m$$>$$1$ \citep{Ciotti91}, while all galaxies observed to date have $m$$>$$1$.
  
  \cite{BG11} (BG11) provide an exact analytical expression for the deprojection of Sersic profiles for all $m$ in terms of the Fox H function, which is extremely difficult, if not impossible to compute, even numerically. However, for rational values of $m$ they show that the deprojection can be expressed in terms of the Meijer G function. Rational $m$ requirement is not too stringent because for practical purposes any  $m$ can be well approximated by a rational number. The singularity in the deprojected central density for $m>1$ is, however, inherent to the form of the Sersic function.  We also note that the deprojections discussed above are assumed to extend to infinite 3D radius. However, just because a deprojection is analytically difficult, does not, by any means, suggest that the true 2D light distribution of galaxies is not described by a Sersic profile, and the 3D distribution is not a deprojected Sersic profile. What might suggest that the Sersic profile is not an optimal function over very large radial ranges is that even a $2$-component model,  i.e. a double Sersic (as in Hopkins et al. (2009a,b)), often yield residuals larger than measurement errors.
  
  Motivated by the finding that the Einasto profile provides better fits to pure dark matter high resolution N-body simulations of dwarf, galaxy and cluster sized haloes, \cite{DW10} presented, for the first time an extremely accurate -- fractional deviations of $\sim 10^{-4}$ to $10^{-2}$ -- analytical approximation to the surface mass density of a 3D Einasto profile. This function is valid for $n \gtrsim 0.2$ (see section (\ref{DWfunction}) below), and is expressed in terms of the 3D Einasto profile parameters, $\rho_{s}$, $r_{s}$, and $n$. Given the issues described above with the existing forms of the fitting function, in this paper we explore the quality of fits to the surface brightness profiles of ellipticals with a multi-component DW-function, which has the interesting property that the intrinsic 3D luminosity density is a multi-component Einasto profile. 
  
  \section{Data}\label{data}
  Since our primary goal is to explore how well a multi-component DW-function describes the surface brightness profiles of ellipticals, we restrict ourselves to ellipticals for which a large dynamic radial range of high resolution data is available. We hence looked at the well studied Virgo Cluster, for which KFCB09 provides, for the first time, an excellent composite compilation of many ground and space based observations, spanning up to five decades, i.e. the largest available radial range.
  
  In order to be able to probe the central regions, $\sim$ 0.1 arcsec, we select galaxies from the KFCB09 sample with published psf-deconvolved profiles for their central regions; primarily data from Lauer et al. (1992,~1995 \& 2005). The intermediate regions of all galaxies are supplemented by the high resolution HST ACSVCS data, while the extensive outer regions come from a wide range of ground based observations allowing for accurate sky subtraction. We refer the reader to KFCB09 for details on how many datasets have been used for each galaxy and the details of averaging between data sets to create the composite.

  From the sample of ellipticals in KFCB09 we exclude NGC4374 and NGC4515 since they do not have psf-deconvolved profiles, NGC4486A since its profile is strongly contaminated by dust, and NGC4486B whose profile is affected by its proximity to M87 (NGC4486). 
  
  NGC4261 does not have psf-deconvolved profile, but we include this galaxy in our sample. This is because a) KFCB09 uses HST NICMOS 1.6$\mu$ H-band data for its central regions (transformed to V-magnitudes) to account for dust absorption in {\it V}; b) the 1.6$\mu$ images in \cite{Quillen2000} show that dust absorption in near-IR is weak and restricted to the very central regions, and the {\it V-H} colour image shows that any nuclear point source is shielded by the dust; c) it has a shallow cusp and the effects of a psf are not as strong as in steep-cusp galaxies; and d) during fitting with a multi-component DW-function, discussed in sections \ref{MCfits} and \ref{fitresults}, we varied the lower end of the fit range from 0.07 to 0.3 arcsec (4$\times$0.075 arcsec, the NICMOS pixel scale) and found that our best fit parameters were robust within $10$ per cent.

  We thus obtain a sample of 23 ellipticals in and around Virgo, comprising of 22 ellipticals from the dataset presented in KFCB09 and a composite profile of NGC4494 from \cite{Nap4494}.

  \subsection{Uncertainties in the dataset}\label{uncertainties}
  Critical to our modelling of multiple components is the requirement that the fit-residuals be consistent with measurement errors. It is therefore important to discuss the various uncertainties reported by KFCB09 for their dataset. 
  
  Zero-points in KFCB09 are reported to have systematic uncertainties $\leq$ 0.05 $\mgasc$, and random errors of $\sim$ 0.03 $\mgasc$. While the authors do not provide quantitative values for profile measurement errors (either for every data point or an $rms$ for each galaxy) they do state that fits with a Sersic function are considered good when the resulting $rms$ of fit is comparable to profile measurement errors, which are of the order of a {\it ``few hundredths of a $\mgasc$''}. KFCB09 also report that the median $rms$ of fits with a Sersic profile, over a restricted radial range, is 0.04 $\mgasc$ with a dispersion of 0.02 $\mgasc$. We thus conclude that individual galaxies in the sample can have profile measurement errors in the range of (0.02-0.06) $\mgasc$, with an $rms$ of random errors $\sim$ 0.03 $\mgasc$.
  
  For the central most regions \cite{L98} report psf deconvolution errors of around 0.07 $\mgasc$, which along with the random errors of 0.03 $\mgasc$ imply that the central-most data points, at $\lesssim$0.1 arcsec, for our selected sample may have uncertainties $\sim$ 0.1 $\mgasc$. KFCB09 also report uncertainties in sky subtraction and errors due to matching profiles at large radii are around 0.1 $\mgasc$.

  \section{Surface Brightness with the DW-function}\label{DWfunction}
  In \cite{DW10} the authors derive an extremely accurate analytical approximation for the line-of-sight 2D projection of the 3D Einasto profile, equation~\eqref{einasto}. In terms of a scaled radius $X$$=$$\frac{R}{r_s}$ the surface density at a projected radial distance $R$ is 
  \begin{align}\label{sigmae}
    \Sigma_{DW} (R) &=\frac{\Sigma_{0}}{\Gamma(n+1)}\Bigg\{n\;\Gamma \left[n,
      b\left(\zeta_2 X\right)^{\frac{1}{n}}\right] \notag \\
    &  \!+\frac{b^n}{2}\sqrt{\frac{2n}{b}} X^{\left(1-\frac{1}{2n}\right)}\gamma\left[\frac{1}{2}, 
      \zeta^2_1\frac{b}{2n} X^{\frac{1}{n}}\right]
    e^{-b X^{\frac{1}{n}}} \\
    &  \!-\delta b^n X
    e^{-b\left(\sqrt{1+\epsilon^2}X\right)^{\frac{1}{n}}}
    \Bigg\}      \notag
  \end{align}     
  where, 
  \begin{align}\label{sigmanot}
    \Sigma_{0} = \Sigma(0)=\frac{2e^{b}r_s\rho_s n \Gamma(n)}{b^n},
  \end{align}
  with $\Gamma$$(n)$ as the Gamma function, $\Gamma$$[n,x]$ and $\gamma$$[n,x]$ as the upper and lower incomplete gamma functions, respectively, and $b$$=$$b(n)$ is related to $r_s$. The parameters $\zeta_2$, $\epsilon$, $\delta$, and $\mu$ are functions of $n$, as derived in \cite{DW10} (assuming $b$$=$$2n$ and $\zeta_1$$=$1). 
  \begin{align}\label{zeta2param}
    \zeta_2= 1.1513 +\frac{0.05657}{n} -\frac{0.00903}{n^2}
  \end{align}  
  \begin{align}
    \zeta_1=1
  \end{align}
  \begin{align}\label{epsilonparam}
    \epsilon = \frac{\zeta_2 + \zeta_1}{2}
  \end{align}  
  \begin{align}\label{deltaparam}
    \delta(X) = (\zeta_2-\zeta_1)\left\{1-\exp\left[- X^{\mu}\right]\right\}
  \end{align} 
  with,
  \begin{align}\label{muparam}
    \mu    = \frac{1.5096}{n}+\frac{0.82505}{n^2}-\frac{0.66299}{n^3}.
  \end{align}
  As described in Paper III (in preparation), generalizing the $\delta$ term \eqref{deltaparam} allows equations \eqref{sigmae}-\eqref{muparam} to describe the surface density of a projected Einasto profile for any choice of scale radius $r_s$ and an associated $b(n)$. However, these equations simplify considerably for the N+04 parametrization of $b$$=$$2n$ (DW10). This parametrization also has another nice feature that in a log-log plot of the surface density, the slope of the profile is $-1$ at $R$ $\approx$ $r_{-2}$ which is easily identified visually as the point where the profile begins to flatten out in log-space. In this paper we therefore adopt $b=2n$ and $r_{s}$$=$$r_{-2}$ and use the parametrizations derived in Paper III to infer the {\it 3D half-light radius} ($r_{3E}$, valid for $n$$\gtrsim$0.17) and {\it 2D half-light radius} ($R_{2E}$, valid for $n$$\gtrsim$0.5) of a single component through 
  \begin{align}\label{r3dr2}
    r_{3E} = r_{-2}\left(1.5 - \frac{0.1665}{n}+\frac{0.0035}{n^{2.2}}\right)^n
  \end{align}
  \begin{align}\label{R2dr2}
    R_{2E} = r_{-2}\left(1.5 - \frac{0.6328}{n} + \frac{0.145}{n^{1.704}}\right)^n.
  \end{align} 
  It should be evident from geometry that $R_{2E}$$<$$r_{3E}$.
  
  Instead of fitting to estimate $r_{-2}$ and $\Sigma_0$, it may be more desirable in some cases to fit directly for either $R_{2E}$ or $r_{3E}$ and the corresponding surface brightness at $\Sigma(R_{2E})$ and $\Sigma(r_{3E})$. In that case one will need to use the appropriate parametrizations for $b$$=$$b(n)$ given in Paper III, which presents a number of useful results on the Einasto profile, and a better parametrization to measure the half-light radius of the 2D Sersic profile.
  
  The above set of equations, \eqref{sigmae}-\eqref{muparam}, were derived and tested in DW10 for 1 $\leq$ $n$ $\leq$ 10. The fractional residuals with respect to a numerically projected Einasto profile were very small, $\sim$ $10^{-2}$ to  $10^{-3}$. In order to model the light of ellipticals, we compared equation \eqref{sigmae} to the numerical projection of \eqref{einasto} for a wider range of $n$. For 10 $\leq$ $n$ $\leq$ 50 the residuals are $\sim$ $10^{-4}$, and $\sim$ $10^{-2}$ for 0.5 $\leq$ $n$ $\leq$ 1. The function works very well even for 0.2 $\leq$ $n$ $<$ 0.5 with larger residuals in a small region around $r_s$, but with an overall $rms$ $\leq$ 0.05. However, none of the galaxies require an $n$ in this latter range. We also found that for the entire range 0.2 $\leq$ $n$ $\leq$ 50, the uncertainties in recovering the 3D Einasto profile parameters from 2D fits with \eqref{sigmae} to a numerical projection of an Einasto profile are better than $10^{-3}$. 
  
  The tests were conducted over a large dynamic range in radii, corresponding to the domain within which the surface density drops from $\Sigma_0$ to $\sim$ $10^{-8}$ $\Sigma_0$, which translates to a difference in magnitude $\Delta \mu$$=$20 $\mgasc$.  The set of equations \eqref{sigmae}-\eqref{muparam} can thus be used to estimate the 3D parameters of a projected Einasto profile for a very wide range in radii, and shape parameter 0.2 $\leq$ $n$ $\leq$ 50, corresponding to 0.02$\leq$ ($\alpha$$=$$1/n$) $\leq$ 5.  

  \subsection{Multi-component fits with the DW-function}\label{MCfits}
  As with other $3$-parameter functions discussed in section (\ref{MCstruc}), a single DW-function could not fit the SB profiles over the entire dynamic radial range of our galaxies. Hence, we start with the assumption that a minimum of two DW-functions, each with three parameters, are required to describe the light of ellipticals.   
  
  The decision to add a third component must be based on the level of measurement errors in the data (section \ref{uncertainties}), as well as on the available degrees of freedom, i.e. the addition of the third component must be statistically justifiable.
  
  We decide whether to fit a $3$-component model after considering the following factors.

  (i) Overall $rms$ of residuals: The random errors in zero-points of the KFCB09 data are $\sim$$\pm$0.03$\mgasc$, and result primarily from matching profiles of different filter magnitude systems. If the $rms$ of residuals are much less than this level, it may indicate over-fitting. In that case $3$-component models are not considered. On the other hand, if a $2$-component model has a larger $rms$, we explore a $3$-component model.
  
  (ii) Examination of the fit residuals: A low $rms$ of residuals does not necessarily imply a good or reliable model. One can obtain a low $rms$ due to very small residuals over a large radial range, and large residuals over a small range. We therefore examine $2$-component model residuals over the entire radial range and consider the fit to be good if it has consistently low residuals ($\lesssim$ 0.05 $\mgasc$), except possibly at very large $R$, and at the smallest $R$ $\lesssim$ 0.1 arcsec $\approx$ $2$ HST-WFPC2 pixels, where residuals up to 0.1 $\mgasc$ are considered acceptable (section \ref{uncertainties}). If the overall $rms$ is low, but there are systematically high residuals in some regions, other than at very large and very small $R$, we explore a $3$-component model.
  
  (iii) F-test: Having fit a $3$-component model, we employ the F-test to ensure that the reduction in $rms$ is statistically significant at $>3\sigma$, or $99.7$ per cent, and is not merely due to an increase in the number of parameters. 
  
  Note that a failure of the F-test does not indicate that an extra component is certainly not present. It just means that the number of degrees of freedom do not justify a statistically significant detection of the extra component. Better resolution, observations with a different filter and an increase in the number of independent data points may lead to a significant detection of an additional component from the surface brightness profile.
  
  The resulting surface brightness is then given as:
  \begin{align}\label{MCDWfunc}
    \Sigma_{DW}(R)= \sum_{i=1}^N \Sigma_{i} (R)
  \end{align}
  with $N$ $=$ $2$ or $3$ indicating the number of components and $\Sigma_{i}(R)$ as the density of the $i^{th}$ component given by equation \eqref{sigmae}-\eqref{muparam}, with each component uniquely characterized by the set \{$\Sigma_{0i}$, $r_{si}$$=$$r_{-2i}$, $n_i$\}. As indicated later in section \ref{spcases}, some galaxies may have $N$$>$$3$ but we do not explore this option in this paper.
  
  For a $2$-component model we shall refer to a central component, which typically describes the region within the break radius, and an outer component, which typically identifies the main body of the galaxy. For a $3$-component model there is an additional intermediate component, which describes a transition region between the central and outer components, except in two cases, NGC4621 and NGC4434, where this component indicates the presence of a weak system embedded within the outer component.
  
  It should be noted that in our models of a galaxy as a superposition of components, the central and intermediate DW-components are in excess to an inward extrapolation of the outer DW-component.
  \section{Results of fits}\label{fitresults}
  Figures~\ref{SB1}-\ref{SBlast} show fits to the SB profiles of the $23$ Virgo ellipticals with a multi-component DW-function along with the best-fit residuals. For all galaxies we present a $2$-component model. For fourteen galaxies a $3$-component model could be justified, while a $2$-component model is sufficient for the other nine galaxies. The figure captions identify whether the 2- or $3$-component model is statistically significant. The results of fits are summarized in Table~\ref{Virgosample}.

    The residuals of our models are consistent with measurement errors (section \ref{uncertainties}) over large dynamic ranges $\sim$ $10^5$ in radius for the largest shallow-cusp galaxies down to the resolution limit of the HST and $\sim$ $10^6$ in surface brightness for the smaller steep-cusp galaxies. The $rms$ is often as low as $\sim$ 0.025 $\mgasc$, with a median sample $rms$ of 0.032 $\mgasc$.
   
  The multi-component fits were carried out through a non-linear least squares Levenberg-Marquardt minimization using {\it GNUPLOT}. During the fitting process all components were allowed the entire dynamic radial range and no pre-defined restricted range in $R$ was imposed. While convergence does depend on a reasonable initial guess in any non-linear fitting, we did not find any strong degeneracies between the fit parameters especially for models where residuals are consistent with measurement errors.

  As noted in section \ref{MCfits} we seek a $3$-component model when the $rms$ of our $2$-component model is greater than the 0.03 $\mgasc$ $rms$ of random errors in our sample. We then perform an F-test to either accept or reject the $3$-component model at $>3\sigma$, or $99.7$ per cent. However, in the figures we present two cases where we do not rely on the F-test alone, and consider other factors. 
 
  (i)~ In NGC4649 (Fig.\ref{NGC4649}) an F-test indicates there is a 27 per cent chance that the reduction in $rms$ due to a $3$-component model is not statistically significant; hence for the rest of the paper we use the $2$-component model. However, from our understanding of errors in the central regions (section \ref{uncertainties}) it can be seen that the $3$-component model certainly improves the fit near the centre and is acceptable at $1.2$ $\sigma$. Hence, a failure of the F-test does not necessarily mean that a physically distinct system is not present. Its existence can be verified with more information, say spectroscopic, about the central region. In section \ref{Npdeproj} we will show that the $3$-component model may be necessary to infer the intrinsic 3D density.
  
  (ii)~ We show a $3$-component model for NGC4636 (Fig.\ref{NGC4636}) as an illustration where an F-test does not reject the $3$-component model (at $> 3\sigma$), but we do. This galaxy has a large dynamic radial range and a $3$-component model may well be admissible. We however reject it since our first criteria to admit a $3$-component model is that the $2$-component model must have an $rms$ $>$ 0.03 $\mgasc$ while its $2$-component $rms$ is 0.029 $\mgasc$. Hence, we do not feel confident in accepting this $3$-component model and emphasize that reliance on statistics and physical interpretations of models must be made with respect to the level of measurement errors in data.
  \begin{table*}
    \begin{center}
      \begin{minipage}{2.3\columnwidth}    
        \resizebox{1.0\columnwidth}{!}{
          \begin{threeparttable}
            \caption{Multi-component DW model properties of Virgo Ellipticals}\label{Virgosample}         
            \begin{tabular}{*{20}{c}}
              \toprule\noalign{\smallskip}
              & & & & \multicolumn{6}{c}{Component Parameters} & \multicolumn{2}{l}{Effective Radii} & & & & & & & \multicolumn{2}{c}{RMS (mag)} \\ \noalign{\smallskip}\cmidrule(lr){5-10}\cmidrule(lr){11-12}\cmidrule(lr){19-20}
              Name & Type & D & Scale & $n_{c}$ & $n_{int}$ & $n_{o}$ & $r_{3Ec}$ & $r_{3Eint}$ & $r_{3Eo}$ & $r_{3E}$ & $R_{2E}$ & $\mu_{0}$ & $q_c$ & $A_{v}$ & $V_T$ & $M_{VT}$ & $L_{VT}$ & Multi- & Double-\\
              ~ & ~ & (Mpc) & (pc) & ~ & ~ & ~ & (pc) & (kpc) & (kpc) & (kpc) & (kpc) & ($mag^{\square}$) & ~ & ~ & (mag) & (mag) & ($\times 10^{9} L_{\sun}$) & DW & Sersic \\
              (1)  &  (2) &   (3) &   (4)   &  (5) & (6) & (7) & (8) & (9) & (10) & (11) & (12)  & (13) & (14) & (15) & (16) & (17) & (18) & (19) & (20)\\ \midrule\noalign{\smallskip}
              NGC4472 & E2 & 17.14 &  83.10 & 0.717 & 1.860 &  5.359 & 382.75 &  2.09 &   29.87 &  25.59 &  18.88 & 15.92 & 0.806 & 0.072 &  7.88 & -23.29 & 176.5 & 0.031 & 0.07\\
              NGC4486 & E1 & 17.22 &  83.48 & 1.090 & 2.619 &  6.451 &  30.81 &  3.96 &   75.26 &  50.98 &  37.29 & 16.21 & 0.722 & 0.072 &  8.04 & -23.14 & 153.9 & 0.050 & 0.09\\
              NGC4649 & E2 & 17.30 &  83.87 & 0.997 & ----- &  5.444 & 951.50 & ----- &   14.96 &  13.78 &  10.20 & 15.64 & 0.828 & 0.086 &  8.46 & -22.73 & 106.2 & 0.032 & 0.08\\
              NGC4406 & E3 & 16.83 &  81.59 & 0.895 & 1.488 &  3.084 & 240.85 & 32.58 &    4.14 &  24.61 &  17.96 & 15.66 & 0.709 & 0.096 &  8.39 & -22.74 & 106.9 & 0.049 & 0.13\\ 
              NGC4365 & E3 & 23.33 & 113.11 & 0.970 & 3.720 &  7.233 & 549.05 &   5.07 &  43.98 &  28.07 &  20.70 & 15.58 & 0.717 & 0.068 &  9.09 & -22.75 & 107.5 & 0.036 & 0.09\\
              NGC4261 & E2 & 31.60 & 153.20 & 0.896 & 1.446 &  5.023 & 533.37 &   2.35 &  22.45 &  16.29 &  11.92 & 16.11 & 0.794 & 0.059 &  9.90 & -22.60 & 93.45 & 0.024 & 0.08\\
              NGC4382 & E2 & 17.86 &  86.59 & 1.545 & 0.903 &  3.330 & 186.71 &   0.70 &  12.21 &  11.14 &   8.25 & 14.65 & 0.761 & 0.101 &  8.78 & -22.48 & 83.74 & 0.090 & 0.11\\
              NGC4636 & E3 & 14.70 &  71.27 & 1.282 & ----- &  6.064 & 632.50 & ----- &   26.25 &  25.44 &  18.85 & 16.41 & 0.760 & 0.090 &  8.54 & -22.30 & 70.82 & 0.029 & 0.04\\
              NGC4552 & E1 & 15.85 &  76.84 & 0.754 & 3.306 &  6.528 &  89.56 &   1.09 &  20.34 &  12.51 &   9.14 & 14.48 & 0.873 & 0.133 &  9.29 & -21.71 & 41.22 & 0.049 & 0.09\\
              NGC4621 & E4 & 14.93 &  72.38 & 3.714 & 0.985 &  9.561 &  16.13 &   4.67 &   8.77 &   7.62 &   5.72 & 10.21 & 0.742 & 0.107 &  9.29 & -21.58 & 36.75 & 0.027 & 0.05\\
              $^{\dagger}$NGC4494 & E1 & 15.85 &  76.84 & 1.701 & 2.057 &  4.171 &  10.95 &   0.19 &   5.54 &   5.15 &   3.81 & 12.61 & 0.838 & 0.067 &  9.90 & -21.10 & 23.50 & 0.025 & -----\\
              NGC4459 & E2 & 16.07 &  77.91 & 4.091 & ----- &  3.835 & 257.37 & ----- &    4.84 &   4.44 &   3.29 & 13.16 & 0.804 & 0.149 & 10.09 & -20.94 & 20.36 & 0.050 & 0.05\\
              NGC4473 & E4 & 15.28 &  74.08 & 2.300 & ----- &  5.649 & 532.50 & ----- &    5.48 &   4.26 &   3.13 & 14.44 & 0.607 & 0.092 & 10.00 & -20.92 & 20.03 & 0.031 & 0.05\\
              NGC4478 & E2 & 16.98 &  82.32 & 0.941 & 1.217 &  2.641 &   4.25 &   0.07 &   1.49 &   1.45 &   1.08 & 13.55 & 0.822 & 0.080 & 11.37 & -19.78 &  6.99 & 0.035 & 0.10\\
              NGC4434 & E0 & 22.39 & 108.55 & 1.238 & 0.576 & 5.532 &  15.50 &   2.78 &   1.47 &   1.73 &   1.27 & 13.61 & 0.928 & 0.072 & 12.18 & -19.57 &  5.74 & 0.032 & 0.07\\
              NGC4387 & E4 & 17.95 &  87.02 & 3.562 & ----- &  2.621 &  82.31 & ----- &    1.68 &   1.65 &   1.23 & 14.18 & 0.633 & 0.107 & 12.14 & -19.13 &  3.83 & 0.037 & -----\\
              NGC4551 & E3 & 16.14 &  78.25 & 1.634 & 1.214 &  2.446 &  23.12 &   0.15 &   1.67 &   1.59 &   1.18 & 14.47 & 0.734 & 0.125 & 11.95 & -19.09 &  3.69 & 0.026 & 0.05\\
              NGC4458 & E1 & 16.37 &  79.36 & 1.502 & 3.078 &  3.149 &  23.05 &   0.12 &   2.07 &   1.74 &   1.28 & 12.93 & 0.879 & 0.077 & 12.12 & -18.95 &  3.24 & 0.026 & 0.06\\
              NGC4464 & E3 & 15.85 &  76.84 & 1.994 & 1.222 &  3.094 &  13.66 &   0.08 &   0.78 &   0.69 &   0.51 & 12.74 & 0.749 & 0.071 & 12.59 & -18.41 &  1.97 & 0.021 & 0.06\\
              NGC4467 & E3 & 16.53 &  80.14 & 2.673 & ----- &  2.402 &  31.67 & ----- &    0.54 &   0.51 &   0.38 & 14.38 & 0.813 & 0.074 & 14.18 & -16.92 &  0.499 & 0.020 & 0.02\\
              VCC1440 & E0 & 16.00 &  77.57 & 1.378 & ----- &  4.976 &   8.79 & ----- &    0.91 &   0.90 &   0.67 & 14.36 & 0.965 & 0.088 & 14.14 & -16.88 &  0.484 & 0.032 & 0.04\\
              VCC1627 & E0 & 15.63 &  75.78 & 2.002 & ----- &  2.907 &  17.47 & ----- &    0.38 &   0.37 &   0.27 & 14.63 & 0.928 & 0.127 & 14.53 & -16.44 &  0.324 & 0.037 & 0.04\\
              VCC1199 & E1 & 16.53 &  80.14 & 1.896 & ----- &  2.389 &  13.49 & ----- &    0.23 &   0.21 &   0.16 & 14.22 & 0.869 & 0.071 & 15.55 & -15.54 &  0.140 & 0.023 & 0.02\\
              \bottomrule\noalign{\medskip}
	      \end{tabular}        
              \begin{tablenotes}
		\small
              \item NOTES.---  Galaxy properties from the multi-component DW models best-fitting the surface brightness (SB) profiles compiled from KFCB09, except that of NGC4494 (marked with a $\dagger$) whose data is from {\protect \cite{Nap4494}}; we have, however, transformed their intermediate-axis profile to major-axis so as to have uniformity with the KFCB09 sample. Columns contain: (2) Galaxy type as defined in KFCB09. (3) Distance in Mpc from KFCB09 and references therein. (4) shows the physical size of $1~arcsec$ in pc at the distance given in column 3. (5-10) Einasto shape parameter $n$ and the intrinsic (3D) effective or half-light radius $r_{3E}$ of the central-, intermediate- (when applicable) and outer components, deduced from the multi-component DW fits to the SB profile. $r_{3E}$ of the components have been computed from the respective best fitting $r_{-2}$ using equation \eqref{r3dr2}. (11-12) Total effective or half-light radii of the galaxy, intrinsic ($r_{3E}$) and projected ($R_{2E}$), deduced by numerically integrating the best fitting multi-component DW model to infinity. (13) V-band Central ($r$$=$$0$) surface brightness in $\mgasc$ deduced from the best fitting multi-component DW model and corrected for Galactic extinction (column 15). (14) characteristic axis-ratio of the galaxy (see section~\ref{lumhalflight}). (16-17) Galactic extinction corrected total V-band apparent ($V_T$) and  absolute ($M_{VT}$) magnitude. (18)  V-band extinction corrected luminosity assuming $M_{V\sun}$$=$4.83. (19-20) Comparison of $rms$ of residuals between the best-fitting multi-component DW models (this paper) and double-Sersic models in Hopkins et al. (2009a,b), for the same SB dataset of KFCB09. 
              \end{tablenotes}       
            \end{threeparttable}
        }
	  \end{minipage}	  
      \end{center} 
    \end{table*}
    \subsection{Special cases}\label{spcases}
  In this section we detail peculiarities observed while modelling some of the galaxies and in section \ref{exceptions} we discuss three galaxies where interpretation of their components may require detailed modelling using additional (for example, spectroscopic) information.

  a)~For some of the larger galaxies like NGC4486 (M87, Fig.\ref{NGC4486}), even a $3$-component model leaves systematic patterns $\sim$ 0.1 $\mgasc$ in the fit residuals. Hence, although the fits look very good and {\it rms} is better than what is typically achieved in the literature for such a large dynamic radial range, we believe that this could be an indication of an underlying fourth component. We do not explore four components in this paper.
  
  b)~NGC4459 (Fig.\ref{NGC4459}) has well known embedded dust features which are clearly evident in the SB profiles. Its outer $n$ of the $2$-component model is fairly robust with respect to whether we include or exclude the radial range affected by the dust. This is, however, not true of its inner $n$, which changes appreciably based on the inclusion or exclusion of the dusty region. We adopt the fit that includes the dust region.

  c)~NGC4473 (Fig.\ref{NGC4473_87}) is quite an interesting case. The SB is fit extremely well with an $rms$ of 0.031 $\mgasc$ using only a $2$-component DW-function. The outer component has an Einasto shape parameter $n$$=$5.65 and the central region shows a density profile flattening typical of shallow-cusp galaxies. It has $M_{VT}$$=$-20.92, $L_{VT}$$=$$2$$\times$$10^{10}$$L_{\sun}$ and rather elliptical (E4/E5) isophotes, typical of the steep-cusp galaxies. The KFCB09 composite data extends to 300 arcsec (22 kpc), an extent that is larger than that of all steep-cusp ellipticals (except the unusual NGC4621, section \ref{exceptions}), but smaller than that of the smallest shallow-cusp elliptical NGC4552 (Fig.\ref{NGC4552}). It has a Galactic extinction corrected central surface brightness of $14.44~\mgasc$ which is also found in both families of galaxies (Table\ref{Virgosample}. However, its intrinsic 3D central density as well as the overall intrinsic density profile appears to be similar to that of shallow-cusp galaxies. 

  d)~NGC4387 (Fig.\ref{NGC4473_87}) is well fit with a $2$-component model but may well have a third component. Since the $2$-component $rms$ is already reasonably low at 0.0367, we do not explore a $3$-component model. However the SB profile is very similar to that of NGC4551 (Fig.\ref{NGC4551}) which clearly shows a $3$-component structure. The central component for NGC4387 is also spatially more extended than that of NGC4551 and has a larger $n$ compared to its outer component; a feature that is different from most galaxies in the sample. The $n$ and $r_{-2}$ of the outer component for NGC4387 and NGC4551 are also very similar. This suggests that the central component of NGC4387 could well be a sum of two components, which a SB analysis with the present data is not able to distinguish.

  Note that the fading light of the giant elliptical NGC4406 (M86) beyond 575 arcsec (Fig.\ref{NGC4406} and section \ref{exceptions}), affects the entire region of NGC4387. The centres of these two galaxies are separated by 668 arcsec. Their SB profiles extend through 800 and 93 arcsec for NGC4406 and NGC4387, respectively. The high SB central regions of NGC4387 are unlikely to be affected by the low signal ($>$ 25 $\mgasc$) from NGC4406, but the outer regions beyond $\sim$ $50$ arcsec could be. However, the SB of NGC4387 show no detectable features in its outer profile.
  
  e)~VCC1440 (Fig.\ref{NGC4467_VCC1440}) is unusual in that despite being a fairly low luminosity steep-cusp galaxy ($M_{VT}=-16.88$, $L_{VT}$$=$3.2$\times$ $10^{8}$ $L_{\sun}$) it has a fairly large $n$$=$4.98 for its outer component, similar to the massive shallow-cusp galaxies. 
    \subsection{Galaxies with uncertain components}\label{exceptions}
    The analysis of the SB profiles of NGC4406, NGC4382 and NGC4621 do not lead us to an unambiguous conclusion of which component characterizes the main body of the galaxy. We hence exclude these galaxies from studies involving specific aspects of individual components, for example, trends involving shape parameters and luminosity of components. The figure captions identify them as {\it exceptions}. The details are explained below:
  
  a)~The SB of NGC4406 (Fig.\ref{NGC4406}) is fit well with a $3$-component model with an rms of 0.049 $\mgasc$. However, it is not clear what constitutes the main body of the galaxy. The $n$$=$3.084 component contains 23 per cent of the total light, while the outer component contains 76 per cent, but has an unusually low $n$$=$1.488. Note that while NGC4406 has overlapping profiles with NGC4387 and NGC4374, they overlap only beyond 550 arcsec. The third component gains prominence around $100$ arcsec and continues through $800$ arcsec. It is hence intrinsic to NGC4406 and not a feature due to incorrect subtraction of the light of NGC4387 and NGC4374.

  This galaxy is streaming into Virgo at 1400 km s$^{-1}$ and it is possible that it has also gone through a recent interaction or a merger. Chandra images and the presence of large plumes of HI gas around this galaxy do indicate such a possibility. Therefore it is not surprising that the $n$ of its outer component is different from the typical values of other galaxies. Same is true of the intermediate component. 
  
  Our $2$-component fit over four decades in radius (with an upper limit of 153 arcsec, as in KFCB09) is not a good fit, although its outer $n$ is consistent with that of other large shallow-cusp galaxies. However, as mentioned earlier estimating fit parameters and structural properties from a limited radial range can be misleading. For example, the half-light estimated with the $2$-component model is 31 kpc, much larger than the $3$-component half-light of 18 kpc.
  
  In Fig.\ref{NGC4406} and Figs.(\ref{complum},\ref{rEMVT} and \ref{nMVT}) we identify the component with $n$$=$3.084 as the one characterizing the main body of the galaxy (which for most galaxies, is the outer component) and the component with $n$$=$1.488 as a perturbing or transitional component (which for most galaxies, is the intermediate component).

  b)~In the $3$-component model for NGC4382 (Fig.\ref{NGC4382b}), the outer component contains 94 per cent of the total light. However, the rather abnormal bump in the profile around $100$ arcsec and the large residual patterns may indicate that the system has not relaxed since a recent interaction or merger event (refer to KFCB09 for an image from http://www.wikisky.org showing signatures of a recent interaction).
    
  Hence the outer $n$ of its $3$-component model may not be representative of well relaxed systems, although its central and intermediate components are similar to those of other shallow-cusp galaxies analyzed in this paper. In fact, if the bump around $100$ arcsec is excluded (Fig.\ref{NGC4382a}), we find an outer $n$ of a $2$-component model more consistent with those of other shallow-cusp galaxies. However, we do not want to draw inferences on components obtained after excluding some radial ranges from the fit, especially in the absence of a theoretical motivation for doing so. We hence adopt the $3$-component model in Fig.\ref{NGC4382b}.
  
  c)~NGC4621 (Fig.\ref{NGC4621}) is an interesting case. With $M_{VT}$$=$-21.58, it is the most luminous steep-cusp galaxy in our sample. In general, galaxies with $M_{VT}$$<$-21.5 appear to have shallow-cusps, and are spatially more extended, while those with $M_{VT}$$>$-21.5 have much smaller spatial extent and usually have steep cusps. However, NGC4621 has the steepest central cusp and has the highest central surface brightness of all galaxies in our sample and is physically almost as large, and as luminous as the smallest shallow-cusp galaxy NGC4552 (Fig.\ref{NGC4552}). 
  
  The residuals of a $2$-component fit and an F-test justifies a $3$-component model, which also gives a far more reasonable $n$$=$3.71 for its central component than the unusually large $n$$=$8.79 of a $2$-component model. However, the outer $n$ of its $3$-component model is much larger than that of any galaxy in our sample, including the shallow-cusp galaxies that generally have large outer $n$. This could be due to the embedded intermediate component with $n$$=$0.985. NGC4621 is a E4 galaxy and its elliptical isophotes show sharp pointed features. It could be that the ellipticity of the embedded intermediate component is quite different from that of the outer component and accounting for this ellipticity may be necessary to obtain a better estimate of $n$ values. At the same time this galaxy could just be a special case, as mentioned above.
    
  NGC4434 (Fig.\ref{NGC4434}) may be a similar system. Its intermediate component, which is nearly gaussian, with $n$$=$0.576, is embedded within its outer component. The galaxy however is fairly round (E0) with regular isophotes. We thus do not exclude this galaxy but note that like NGC4621 and NGC4473 and VCC1440 (section \ref{spcases}) its outer $n$ is larger than that of all other steep-cusp galaxies, and is are more consistent with that of the shallow-cusp galaxies. However, like NGC4621, the embedded component may have altered the shape of the outer component. Its outer $n$ may therefore have large uncertainties. 
  \subsection{Estimating luminosity and half-light radii}\label{lumhalflight}
  In order to calculate the total luminosity, the cumulative luminosity within $R$ and the half-light radii one has to, in principle, account for the varying ellipticity $\epsilon$$=$$1-q$; where $q$$=$$b/a$ is the axis ratio. Numerically, this can be done by expressing the area element $dA(R)$ in terms of an axis ratio, $q(R)$, as 
  \begin{align}
    dA(R) = 2\pi q(R) R + \pi R^2 \frac{dq}{dR}
  \end{align}
  and the projected luminosity is then given by
  \begin{align}\label{sigmatot}
    L_p(R_a) = \int^{R_a}_{0} \Sigma(R) dA(R)
  \end{align}
  where $R$ and $R_a$ are along the major axis. However, this involves taking derivatives of axis ratios which show large and some times abrupt variations. Further, in addition to real variation in $q$ there could also exist artificial variations due to the limitations of the ellipse-fitting process.
  
  To avoid the above difficulties, during the ellipse-fitting process one can add up the projected light non-parametrically in elliptical isophotes, $L^{NP}_p$$(R_a)$, through the last data point, $R_a$, and define a characteristic axis ratio, $q_c$, such that
  \begin{align}\label{characq}
    q_c = \frac{L^{NP}_p(R_a)}{2 \pi \int^{R_a}_{0} \Sigma_{AN}(R) R dR},
  \end{align}  
  where $\Sigma_{AN}(R)$ is the spherically symmetric analytical function; here a multi-component DW-function. 
  
  To estimate the total luminosity, we use this value to integrate through infinity, even though this ellipticity seldom represents the axis ratio of the outer most isophotes. Doing so can be justified for profiles with large dynamic radial range, as in this paper, since by definition (equation \eqref{characq}) it is a weighted axis ratio and the extremely low luminosity regions extending through infinity are unlikely to change this weight. Further, since we do not know the true axis ratio beyond the last data point we would not like to assume that the axis ratio through $\infty$ is equal to that of the last data point. 
  
  The total luminosity is then given by
  \begin{align}\label{lumqtot}
    L_{pT} = 2 \pi q_c \int^{\infty}_{0} \Sigma_{AN}(R) R dR
  \end{align}  
  and the projected half-light radius, $R_{2E}$, can be estimated from
  \begin{align}\label{lumqhalflight}
    L_{pT} = 4 \pi q_c \int^{R_{2E}}_{0} \Sigma_{AN}(R) R dR.
  \end{align}
  
  Note that the characteristic axis ratio defined above, \eqref{characq}, is neither the mean nor a luminosity weighted axis ratio in the usual sense. Nevertheless, it is a useful definition that reduces the error in estimating the total light and half-light radii, and also ensures that they are weakly dependent on the specific choice of parametrization of the SB profile and assumptions of ellipticity; provided of course that the fit parameters have been deduced after modelling the entire dynamic radial range and not from a limited range.
  
  Analytical expressions for the integrals in equation \eqref{characq} and \eqref{lumqtot} may not exist. However, for the projection of an Einasto profile, analytical forms are given in Paper III, which accurately models a numerical integration with the DW-function as well.
 
  In this paper, we use the apparent magnitudes estimated non-parametrically by KCFB09 through the last data point to calculate the numerator in equation \eqref{characq}. As a cross-check one can verify that the characteristic axis ratio estimated through equation \eqref{characq} is usually consistent with the E-type of the galaxies listed in Table~\ref{Virgosample} and the ellipticity values in the literature. The {\it V}-band total (integrated through infinity) magnitudes are listed as $V_T$ and the corresponding total absolute magnitudes are listed under $M_{VT}$.
  \begin{figure}
    \begin{minipage}{2.0\columnwidth}
      \centering
      \includegraphics[width=142mm]{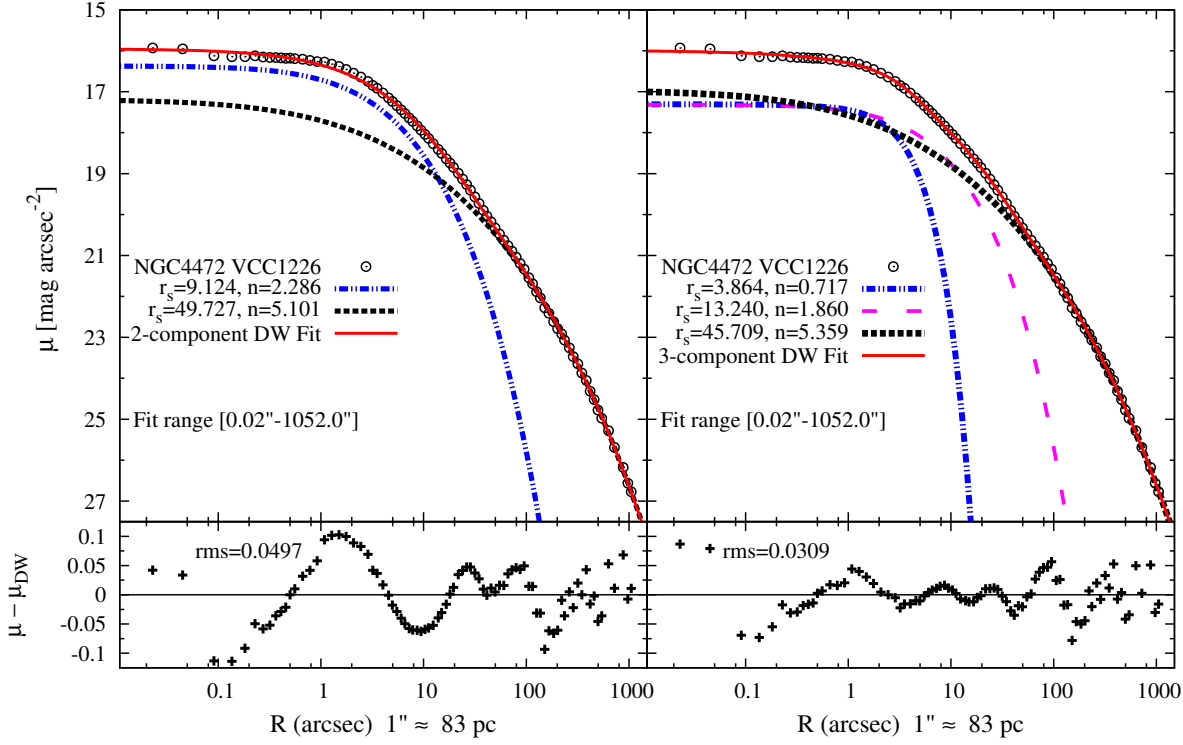}
      \caption{Multi-component DW models of the surface brightness (SB) profiles of $23$ galaxies in the Virgo Cluster. For each galaxy, the figure keys show the best-fitting Einasto index $n$ and scale radius $r_{s}$$=$$r_{-2}$ of the components. The $rms$ of residual is shown in the adjoining residual profile. The fit range is listed in the figure panels and in a few cases when some points are excluded from the fit, the excluded region is also marked on the SB profile. A $2$-component model is always shown, while a $3$-component model is shown when it is found to be statistically significant through an F-test (section \ref{MCfits}). The model accepted is marked '{\it adopted}' in the caption.  Circles indicate data from KFCB09, solid (red) line shows the total multi-component DW model, dot-dashed (blue) marks the central component, short-dashed (black) the outer component or the component resembling the main body of the galaxy and long-dashed (magenta) shows an intermediate or embedded component in some galaxies. Note that the central and intermediate DW-components are in excess to an inward extrapolation of the outer DW-component in those regions. \newline {\bf Above} NGC4472 (VCC1226)  Left: $2$-component, Right: $3$-component (adopted). The large residuals around 1.5 arcsec in the $2$-component model indicates that a $3$-component model may be necessary, which not only has a much lower $rms$ of 0.0309 but also has consistent low and non-divergent residuals $\lesssim$ 0.05 over a large dynamic radial range $\sim$ $10^{5}$. Also note the occurrence of an Einasto index of $n$ $\lesssim$ 1 in the central component of the statistically significant (adopted) model; as in all massive shallow-cusp galaxies. (Colour versions of these figures are available in the online edition.)}
      \label{SB1}
      \end{minipage}
  \end{figure}    
  \begin{figure*} 
    \includegraphics[width=142mm]{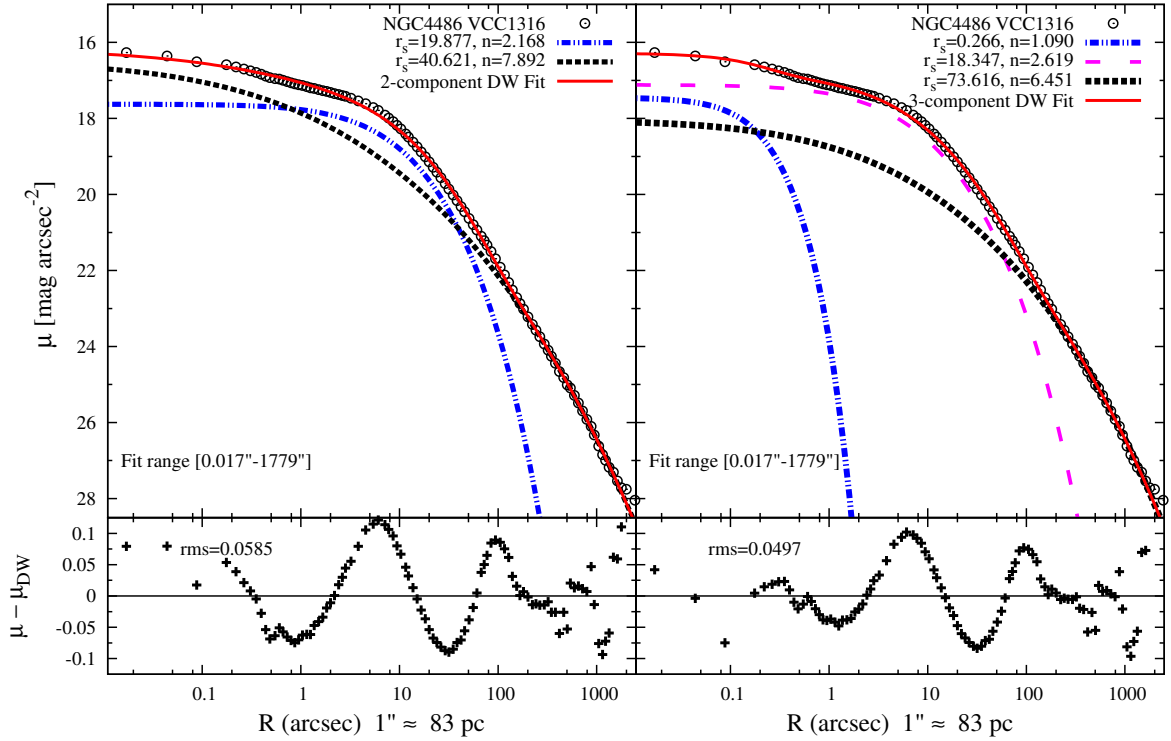}
    \caption{NGC4486 (VCC1316 or M87) Left: $2$-component; Right: $3$-component (adopted). The $3$-component model shows an overall improvement and a reasonable $rms$=0.0497 with non-divergent residuals over an extremely large radial range $\sim$ $10^{5}$. However, the presence of ($\sim$ 0.1 mag) systematic patterns between 3-150 arcsec, greater than the $\sim$ 0.03 mag measurement errors,  suggests the possible presence of a fourth component. This is not surprising given the large spatial extent of this galaxy. We do not explore 4-component models in this paper. Refer to caption of Fig.\ref{SB1} for further details.}
    \label{NGC4486}
  \end{figure*}
  \begin{figure*} 
    \includegraphics[width=142mm]{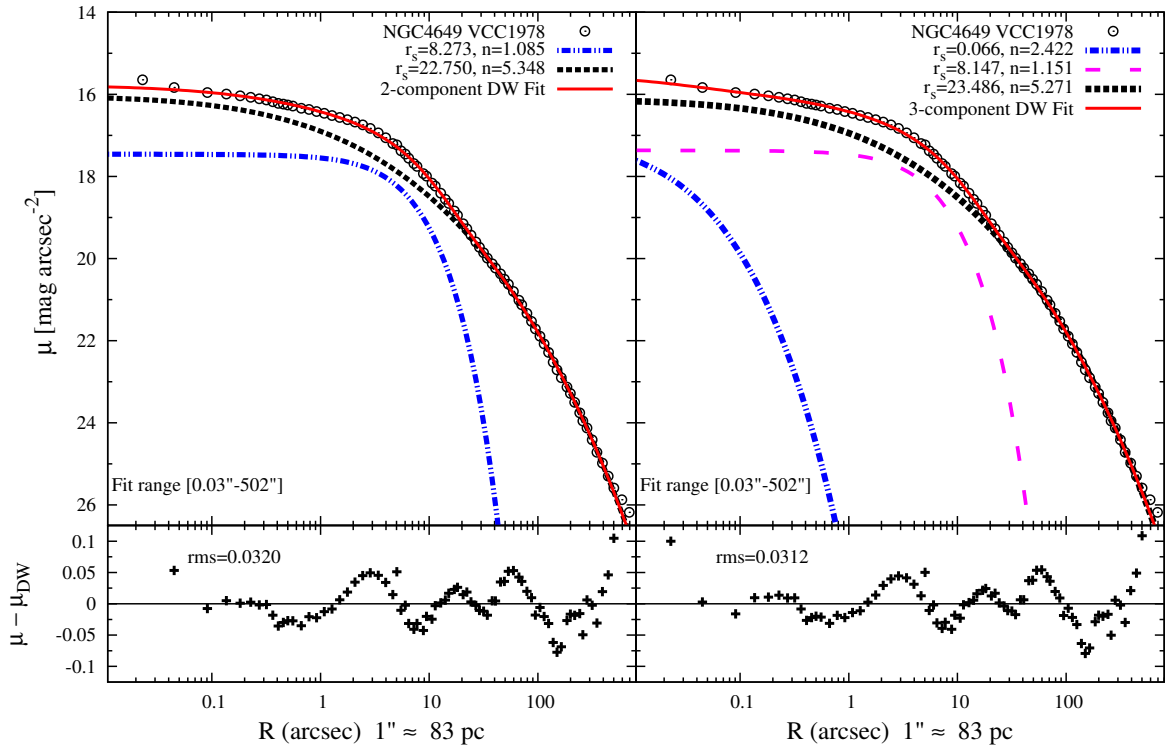}
    \caption{NGC4649 (VCC1978) Left: $2$-component (adopted); Right: $3$-component (significant at 1.2$\sigma$). The $3$-component model could be used to compare with non-parametric deprojections (section\ref{Npdeproj}). Refer to caption of Fig.\ref{SB1} for details.}
    \label{NGC4649}
  \end{figure*} 
  \begin{figure*} 
    \includegraphics[width=142mm]{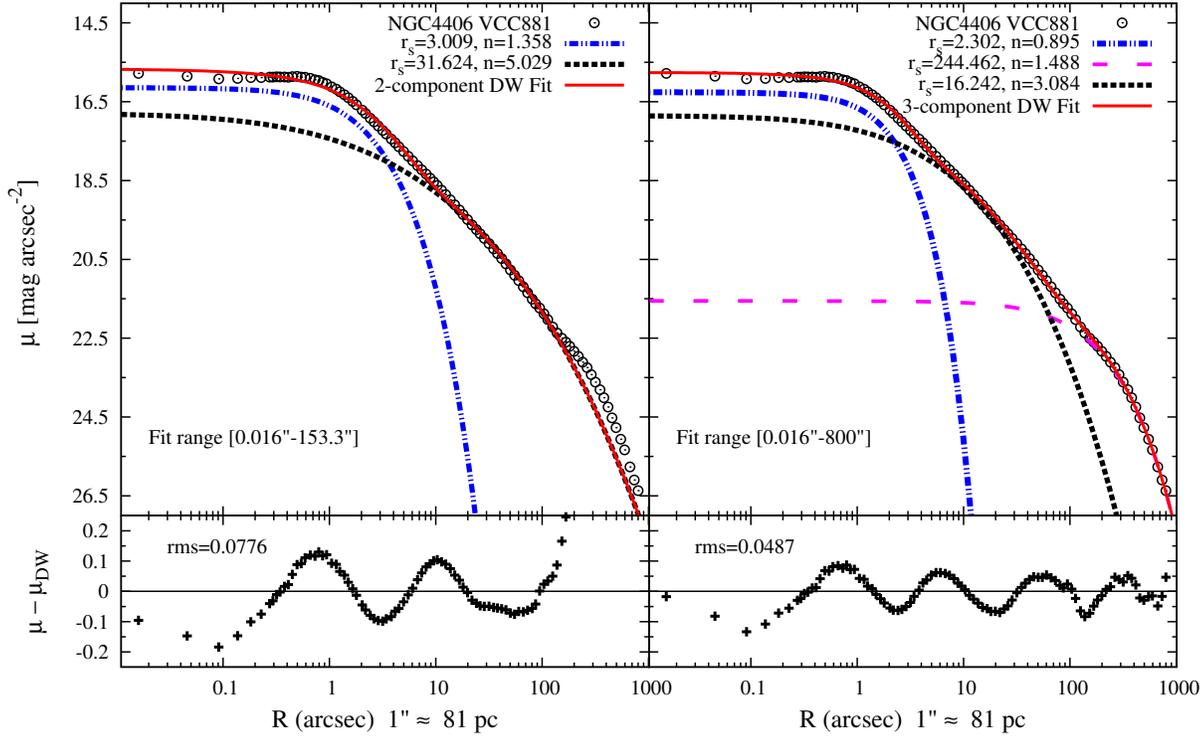}
    \caption{NGC4406 (VCC881) Left: $2$-component fit upto 153.3 arcsec - the upper-limit of fit with a Sersic profile shown in KFCB09; Right: $3$-component (adopted) fit over the entire radial range through 800 arcsec. However, which component forms the main body of the galaxy is not clear and hence this galaxy is an {\it exception} (section \ref{exceptions}). Refer to caption of Fig.\ref{SB1} and section \ref{exceptions}.}
    \label{NGC4406}
  \end{figure*} 
  \begin{figure*}
    \includegraphics[width=142mm]{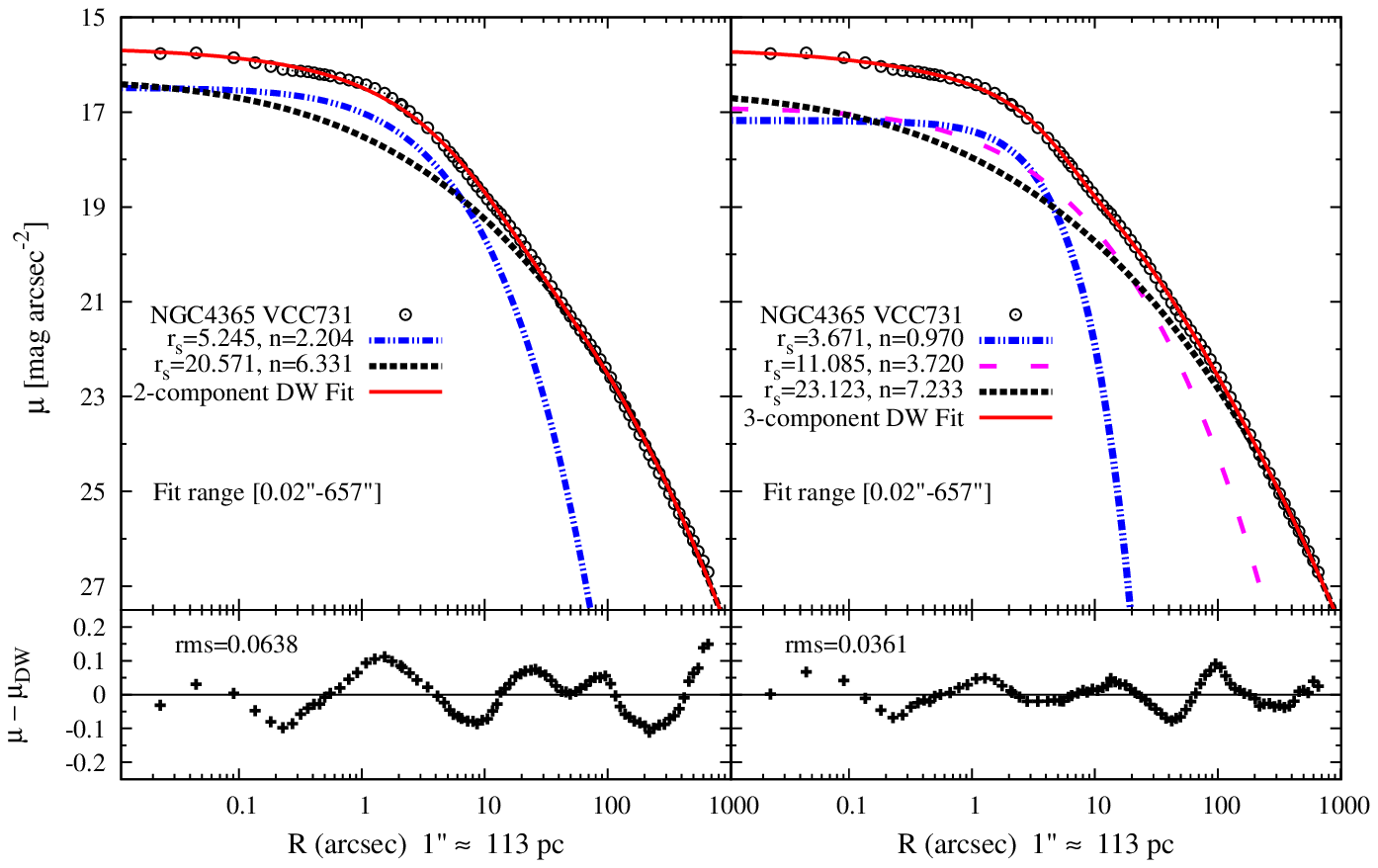}
    \caption{NGC4365 (VCC731) Left: $2$-component; Right: $3$-component (adopted). Refer to caption of Fig.\ref{SB1} for details.}
    \label{NGC4365}
  \end{figure*} 
  \begin{figure*}
    \includegraphics[width=142mm]{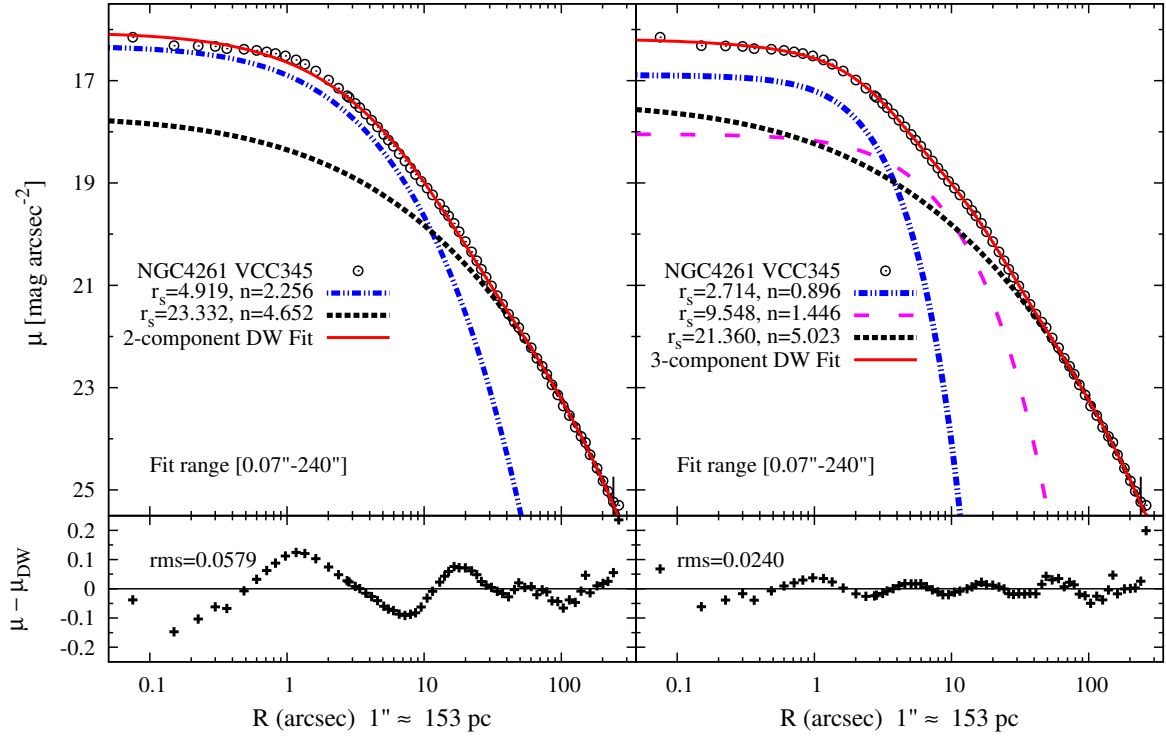}
    \caption{NGC4261 (VCC345) Left: $2$-component; Right: $3$-component (adopted). The improved fit with a $3$-component model, around the transition radius, and the near vanishing of large scale systematic patterns along with a significant reduction in $rms$ is clearly visible. Refer to caption of Fig.\ref{SB1} for details.}
    \label{NGC4261}
  \end{figure*} 
  \begin{figure*}
    \includegraphics[width=142mm]{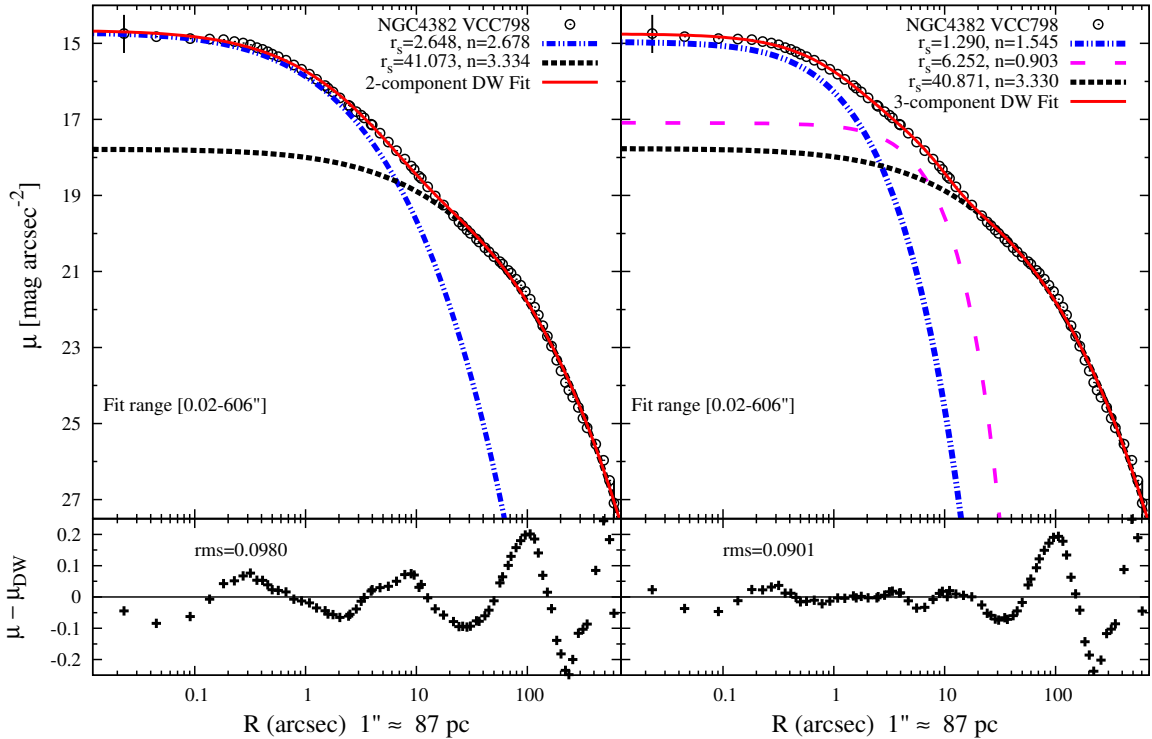}
    \caption{NGC4382 (VCC798) Left: $2$-component; Right: $3$-component (adopted). Although this galaxy shows a visibly shallow-cusp, note the lower outer-n and an inner $n=1.55$ similar to that of some of the less massive steep-cusp galaxies. The $3$-component fit is very good, except in a region around $100~arcsec$. Hence the best-fitting parameters of its outer component may not represent the true values. The outer regions of this galaxy also show indications that it may not have relaxed from a recent interaction and we therefore mark this galaxy as an {\it exception} (see section \ref{exceptions}). Refer to caption of Fig.\ref{SB1} for further details.}
    \label{NGC4382b}
  \end{figure*} 
  \begin{figure*}
    \includegraphics[width=142mm]{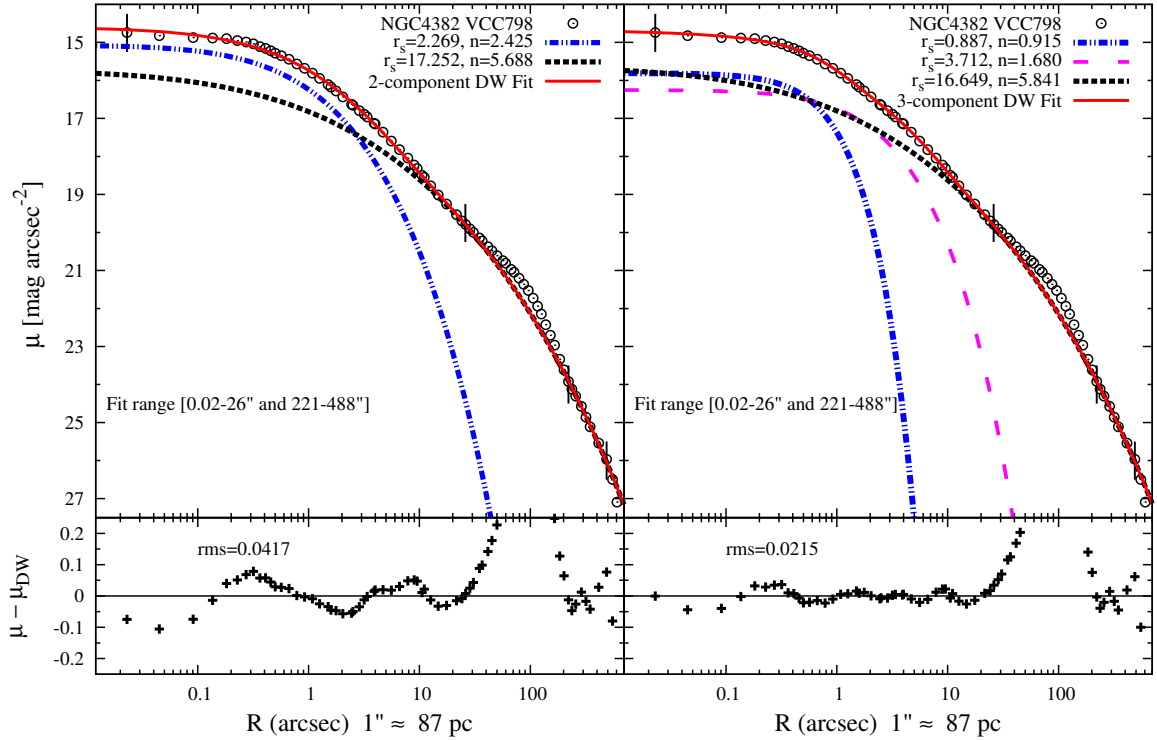}
    \caption{NGC4382 (VCC798) Left: $2$-component; Right: $3$-component. As noted in Fig.~\ref{NGC4382b}, this galaxy is an {\it exception}. Here we show a fit by excluding a region around the bump from $26-221$ arcsec, as in KFCB09. Note the larger outer-n and the $n\sim1$ for the inner component, similar to the pattern seen in other massive shallow-cusp galaxies. However, we adopt the $3$-component fit in Fig.~\ref{NGC4382b} since excluding domains of fit with ad-hoc fitting functions, may not reveal the true structure of these regions. Refer to caption of Fig.\ref{SB1} for further details.}
    \label{NGC4382a}
  \end{figure*}  
  \begin{figure*}
    \includegraphics[width=142mm]{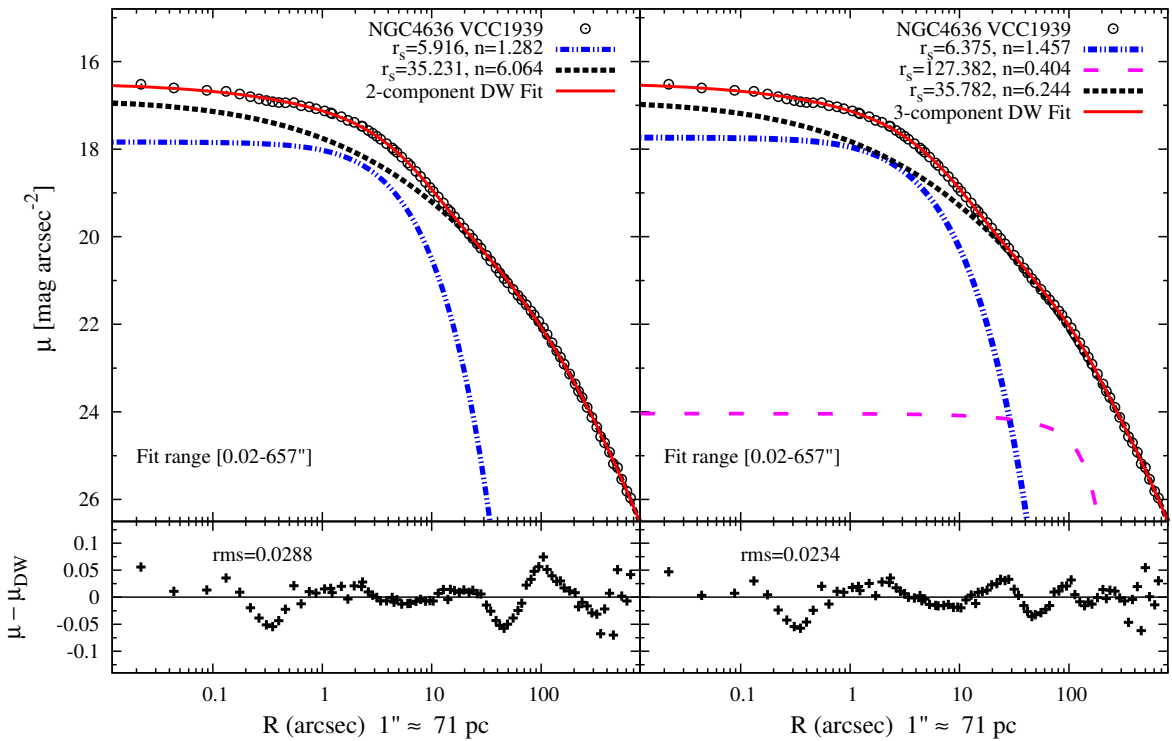}
    \caption{NGC4636 (VCC1939) Left: $2$-component (adopted); Right: $3$-component. The $3$-component model is shown as an illustration of a case where an F-test does not reject it but we do, since the $2$-component model has a rms$\sim$0.03 and the residuals are non-divergent. Refer to caption of Fig.\ref{SB1} for details.}
    \label{NGC4636}
  \end{figure*}  
  \begin{figure*}
    \includegraphics[width=142mm]{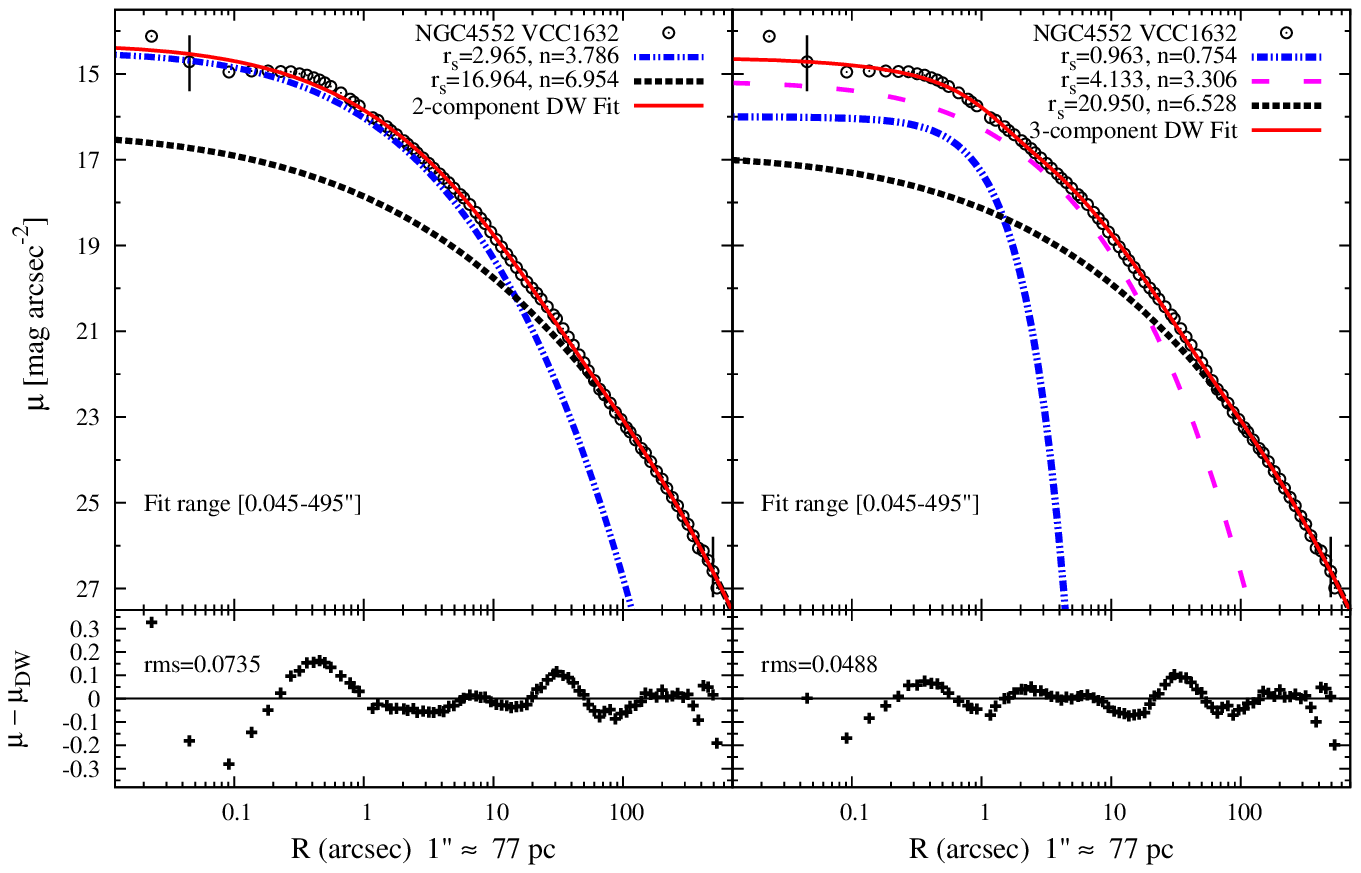}
    \caption{NGC4552 (VCC1632) Left: $2$-component; Right: $3$-component (adopted). At $M_{VT}$$=$$-21.71$, the smallest shallow-cusp elliptical in Virgo is more luminous than all other steep-cusp ellipticals. As in most shallow-cusp ellipticals, it also has an $n$$<$$1$ for its central component which is not seen in any of the steep-cusp ellipticals. HST observations ({\protect \cite{Renzini95}}, {\protect \cite{Cap99}}) reveal variable UV-flare activity in the centre which is interpreted to arise from a low-level AGN. Contributions from such a point source to the central-most data point is excluded from the fit. See caption of Fig.\ref{SB1} for details.}
    \label{NGC4552}
  \end{figure*}
  \begin{figure*}
    \includegraphics[width=142mm]{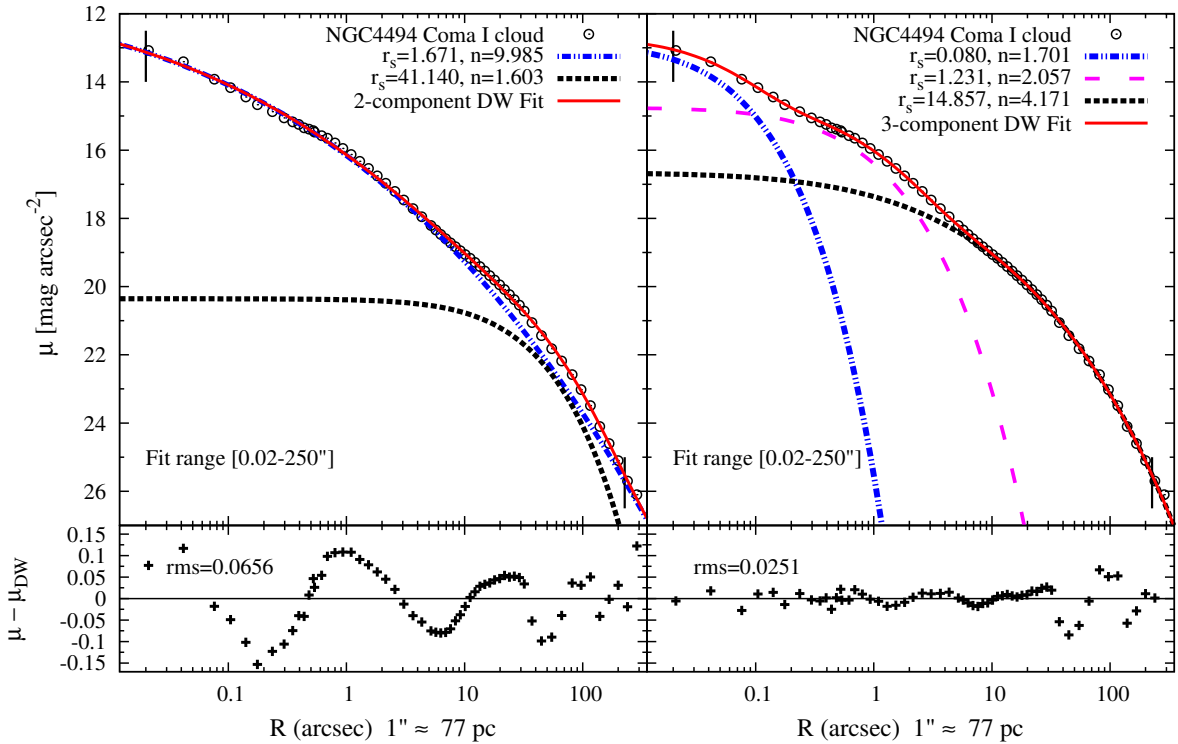}
    \caption{NGC4494. Left: $2$-component; Right: $3$-component(adopted). This galaxy is in the Coma-I cloud around Virgo and comparable to the more luminous steep-cusp ($> 2\times 10^{10} L_{V\sun}$) Virgo ellipticals. The composite data from {\protect \cite{Nap4494}} is plotted in terms of the major axis radius, as in other galaxies from the KFCB09 sample. This is the only galaxy that is not in the KFCB09 sample. Refer to caption of Fig.\ref{SB1} for details.}
    \label{NGC4494}
  \end{figure*}   
  \begin{figure*}
    \includegraphics[width=142mm]{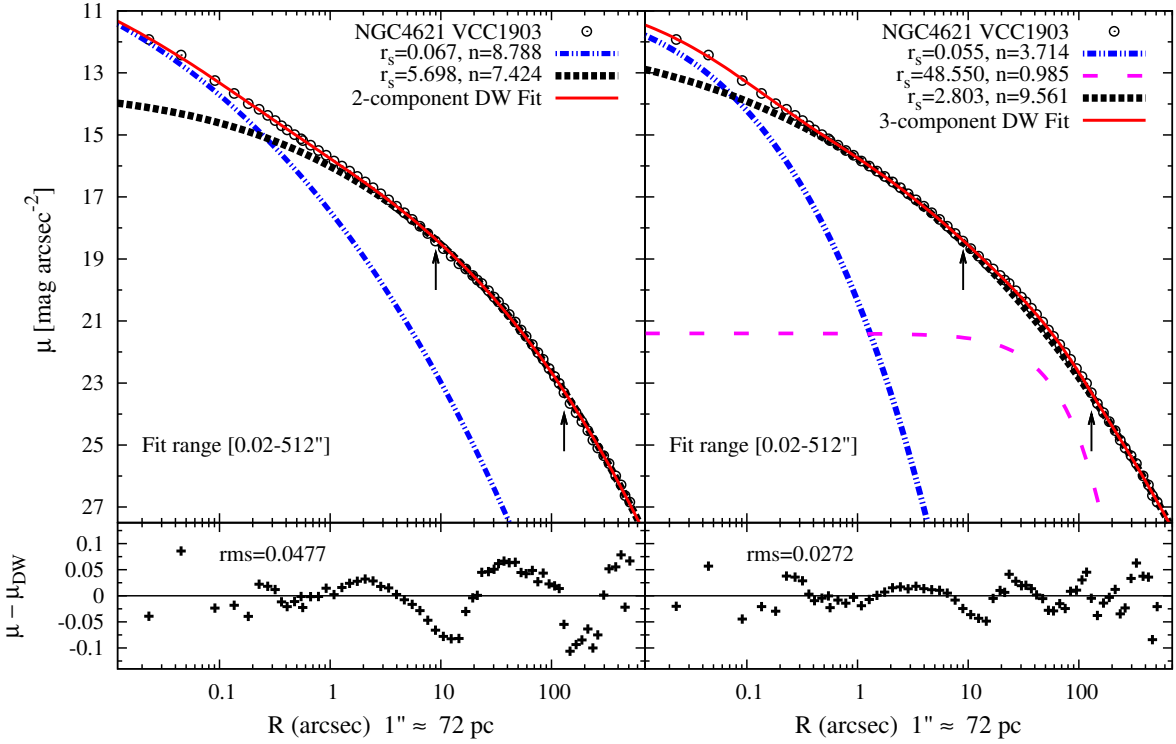}
    \caption{NGC4621 (VCC1903) Left: $2$-component; Right: $3$-component (adopted). The most luminous ($M_{VT}$$=$-21.58, $L_{VT}$$=$3.68$\times$$10^{10}$$L_{V\sun}$) steep-cusp elliptical in Virgo. The $3$-component model improves the fit in the region from 9-130 arcsec and also gives a more reasonable central-n. However, it also has a much larger outer-n. The SB has sharp pointed isophotes and it is not clear how much the embedded component and its ellipticity influences our determination of outer-n. We hence mark this galaxy as an {\it exception} (see section \ref{exceptions}). Also see caption of Fig.\ref{SB1}.}
    \label{NGC4621}
  \end{figure*}
  \begin{figure*}
    \includegraphics[width=142mm]{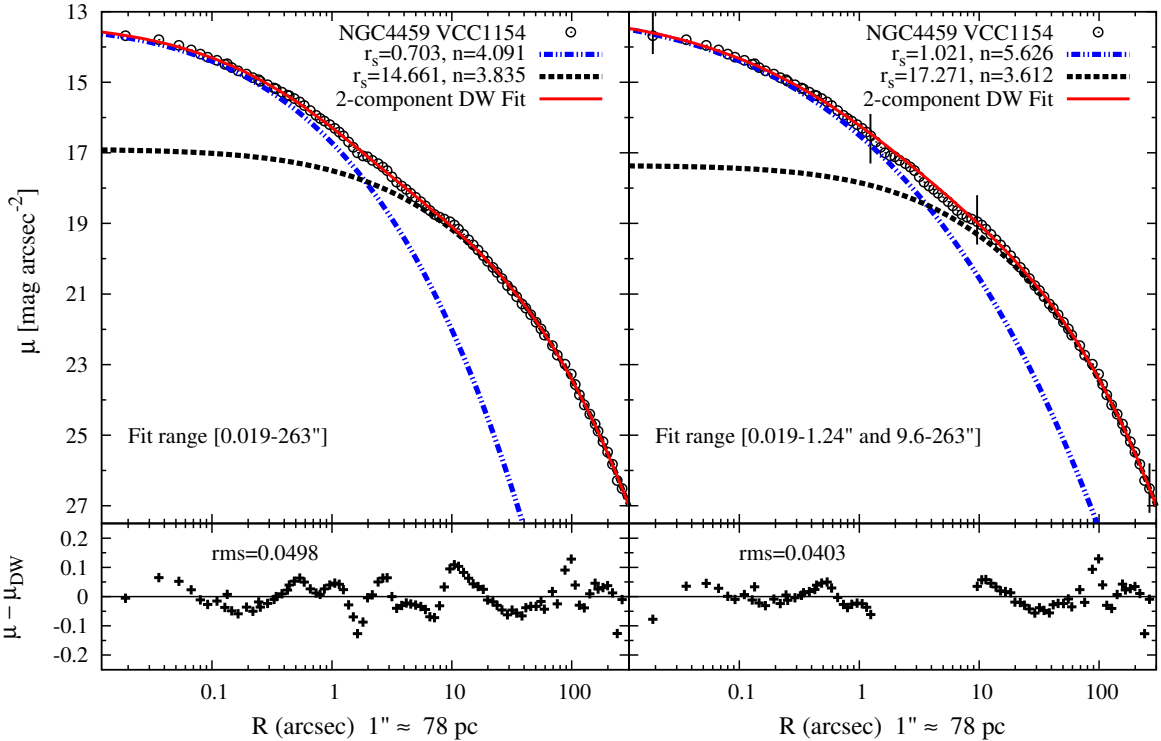}
    \caption{NGC4459 (VCC1154) Left: $2$-component(adopted); Right: $2$-component excluding the region affected by dust from $1.24-9.6$ arcsec, as in KFCB09. It is one of three galaxies in our sample that has an $n_{central} > n_{outer}$. Refer to caption of Fig.\ref{SB1} for details.}
    \label{NGC4459}
  \end{figure*}
  \clearpage
  \begin{figure*}
    \includegraphics[width=142mm]{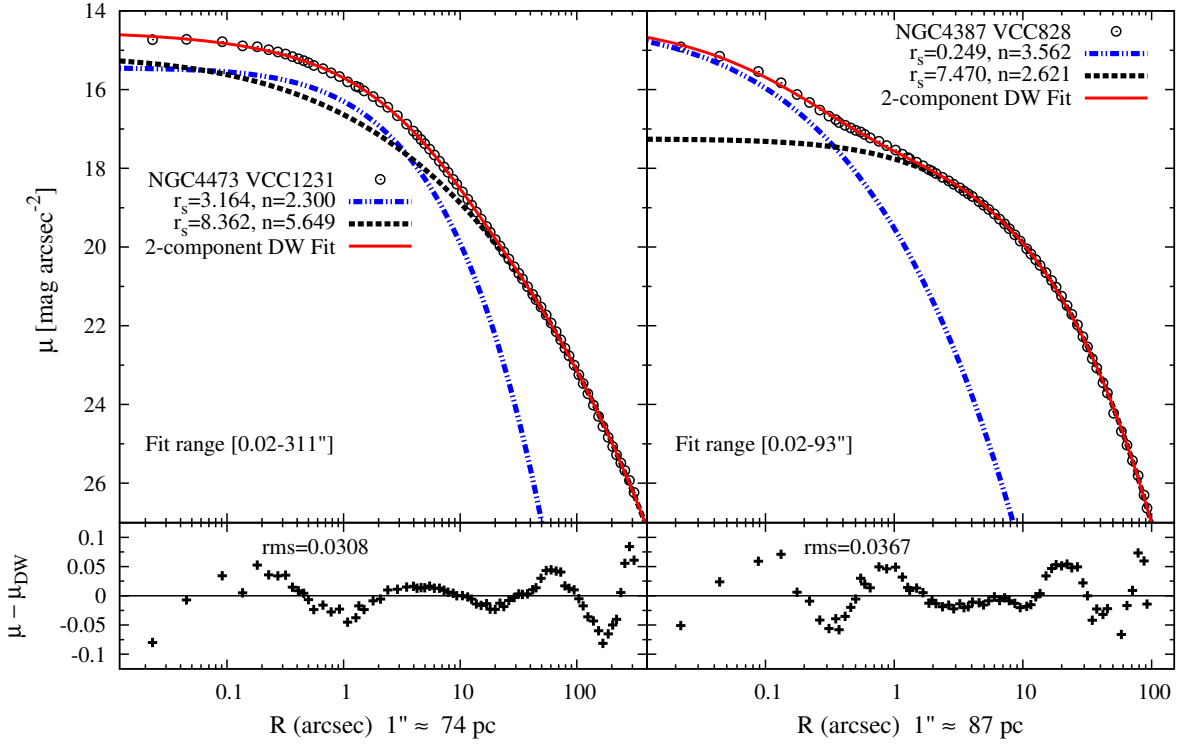}
    \caption{Left NGC4473 (VCC1231): Its $2$-component model has an rms of only 0.031 over a dynamic radial range $\sim$$10^{4}$ and $12$ mags in SB. It is the least luminous of $\sim$$10^{10}$$L_{V\sun}$ ellipticals in Virgo with properties similar to both the steep and shallow cusp families (section \ref{spcases}). However, unlike the massive shallow cusps that have central-$n$$\lesssim$$1$, it has $n$$=$2.23 leading to a much gradual central flattening than the sharper transition seen in the former (section \ref{outcomp}); Right NGC4487 (VCC828): It is one of three steep-cusp ellipticals with $n_{central}$$>$$n_{outer}$ and has a luminosity of 3.83$\times$$10^{9}$$L_{V\sun}$. Its profile could be affected by the light of NGC4406 (section \ref{spcases}). Note the similarity of this profile with that of NGC4551 (Fig.~\ref{NGC4551}) where $3$-components could be justified.}
    \label{NGC4473_87}
  \end{figure*}
  \begin{figure*}
    \includegraphics[width=142mm]{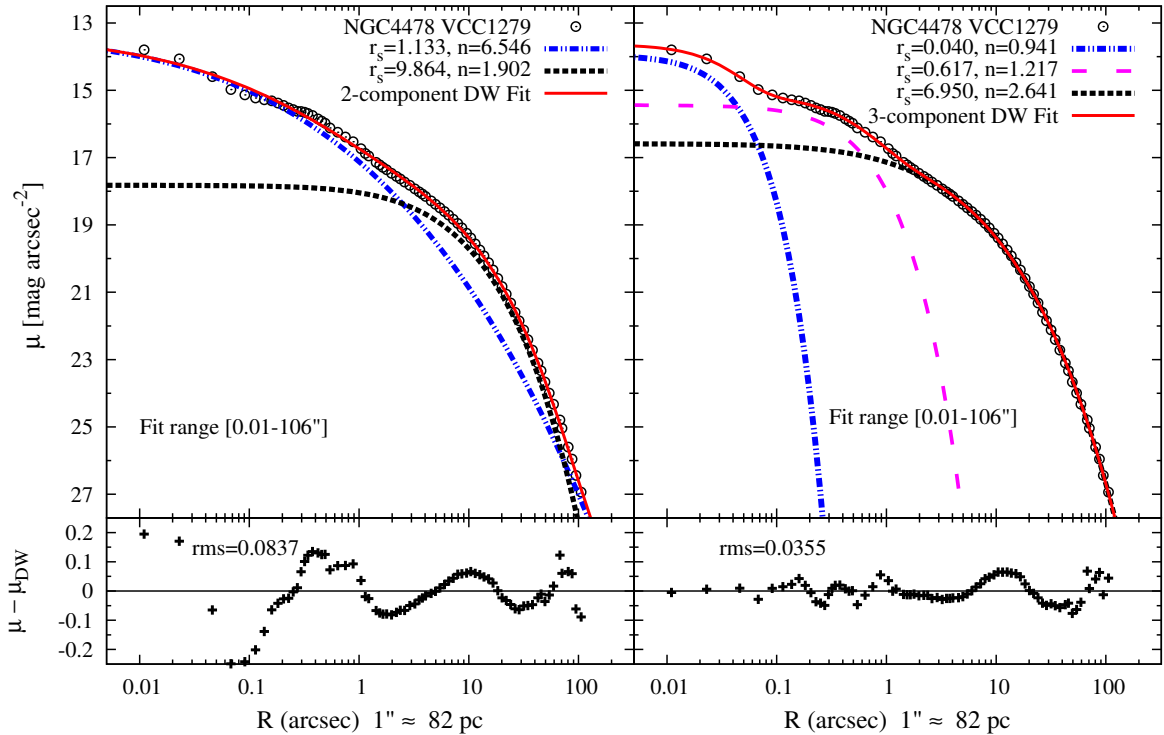}
    \caption{NGC4478 (VCC1279) Left: $2$-component; Right: $3$-component (adopted). As in NGC4494, three components are clearly visible. It is the most luminous of $\sim$$10^{9}$$L_{V\sun}$ ellipticals in Virgo. Along with NGC4434, they are the only two steep-cusp ellipticals with a central-n$\sim$$1$. See Fig.\ref{SB1} for details.}
    \label{NGC4478}
  \end{figure*}  
  \begin{figure*}
    \includegraphics[width=142mm]{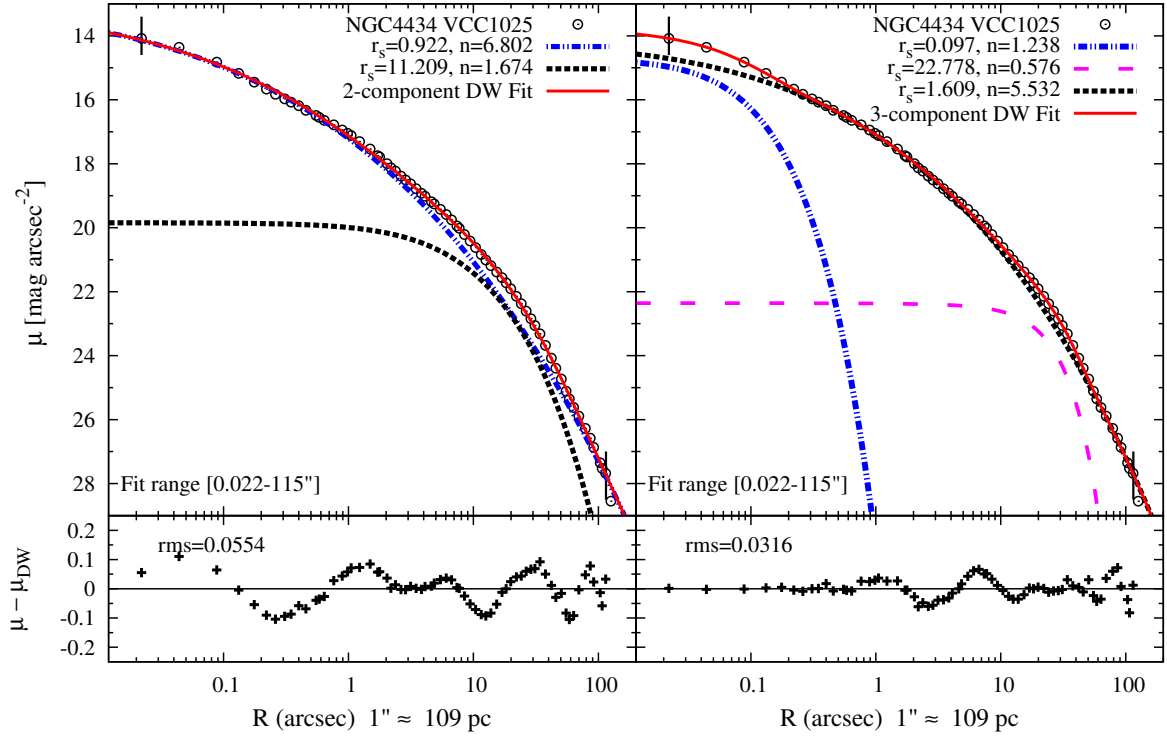}
    \caption{NGC4434 (VCC1025) Left: $2$-component; Right: $3$-component (adopted). A $3$-component model is clearly needed. However, it has an unusually large outer-n typical of the more massive shallow-cusp galaxies. Refer to caption of Fig.\ref{SB1} for details.}
    \label{NGC4434}
  \end{figure*}
  \begin{figure*}
    \includegraphics[width=142mm]{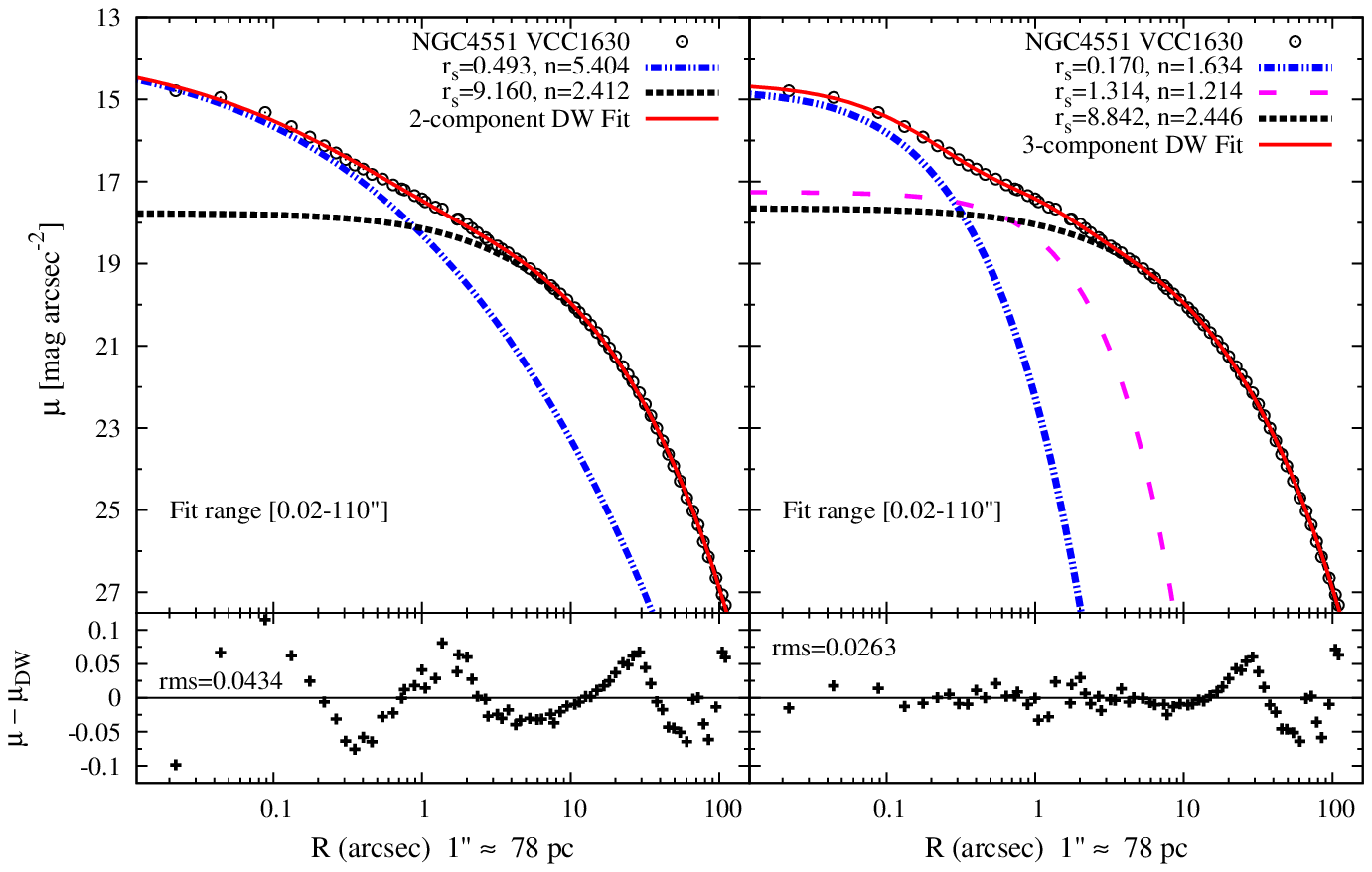}
    \caption{NGC4551 (VCC1630) Left: $2$-component; Right: $3$-component (adopted). Refer to caption of Fig.\ref{SB1} for details.}
    \label{NGC4551}
  \end{figure*}
  \begin{figure*}
    \includegraphics[width=142mm]{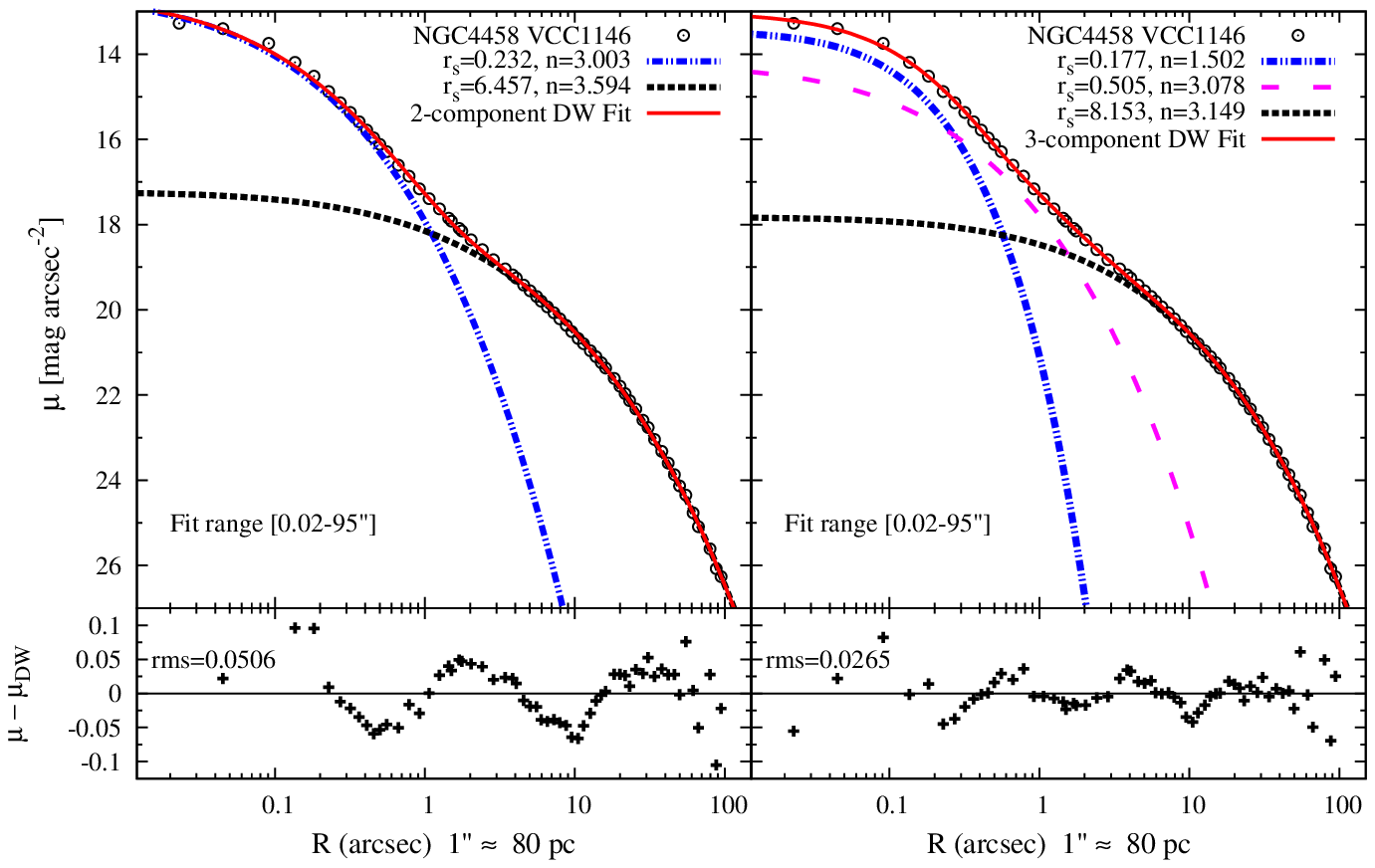}
    \caption{NGC4458 (VCC1146) Left: $2$-component; Right: $3$-component (adopted). Refer to caption of Fig.\ref{SB1} for details.}
  \end{figure*}
  \label{NGC4458}
  \begin{figure*}
    \includegraphics[width=142mm]{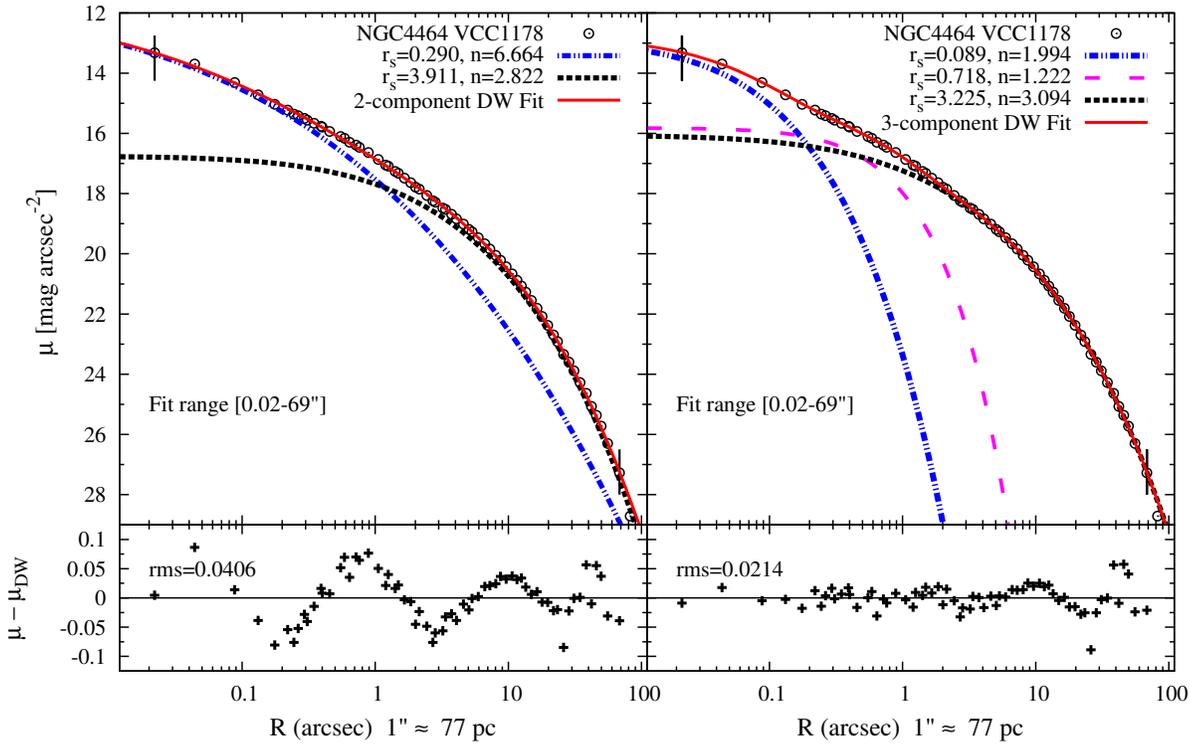}
    \caption{NGC4464 (VCC1178) Left: $2$-component; Right: $3$-component (adopted). It is the least luminous of $\sim 10^{9} L_{V\sun}$ ellipticals in Virgo. Refer to caption of Fig.\ref{SB1} for details.}
    \label{NGC4464}
  \end{figure*}
  \begin{figure*}
    \includegraphics[width=142mm]{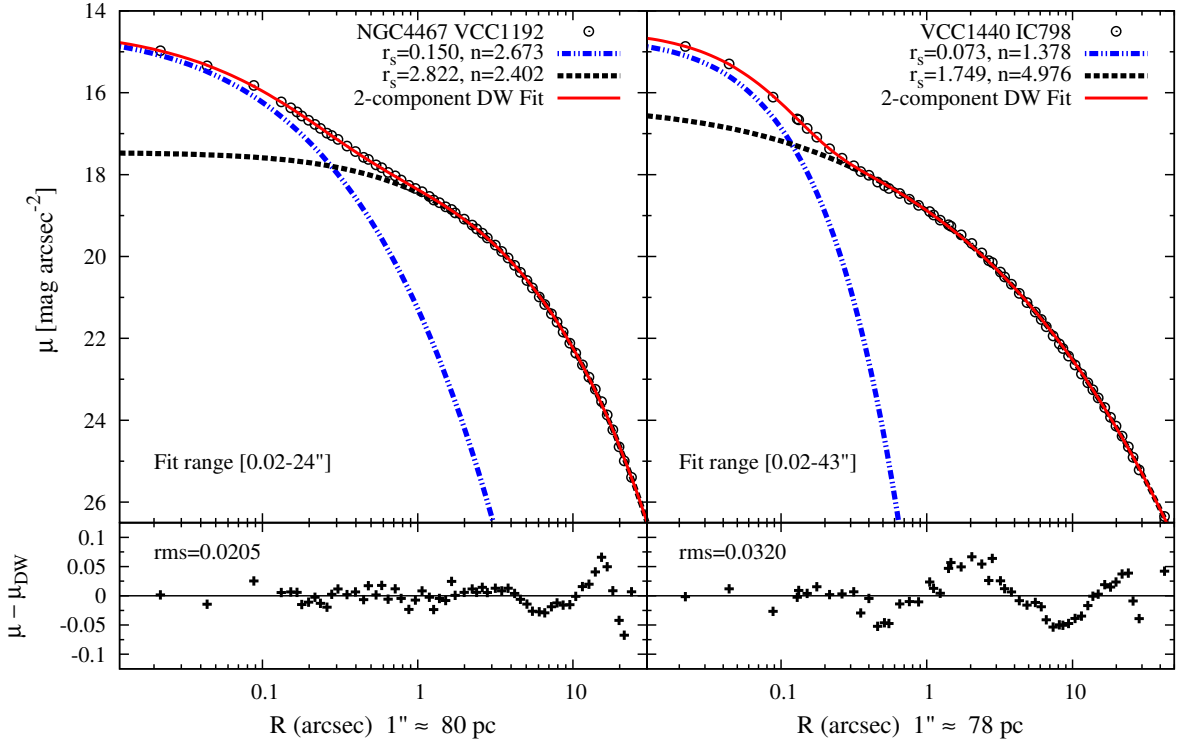}
    \caption{Left NGC4467 (VCC1192); Right: VCC1440. Low luminosity ellipticals in Virgo ($\sim 10^{8} L_{V\sun}$) bordering the dwarf elliptical population. Both galaxies show a steep-cusp, but note the large outer-n for VCC1440 typical of the more massive shallow cusps. Refer to caption of Fig.\ref{SB1} for details.}
    \label{NGC4467_VCC1440}
  \end{figure*}
  \begin{figure*}
    \includegraphics[width=142mm]{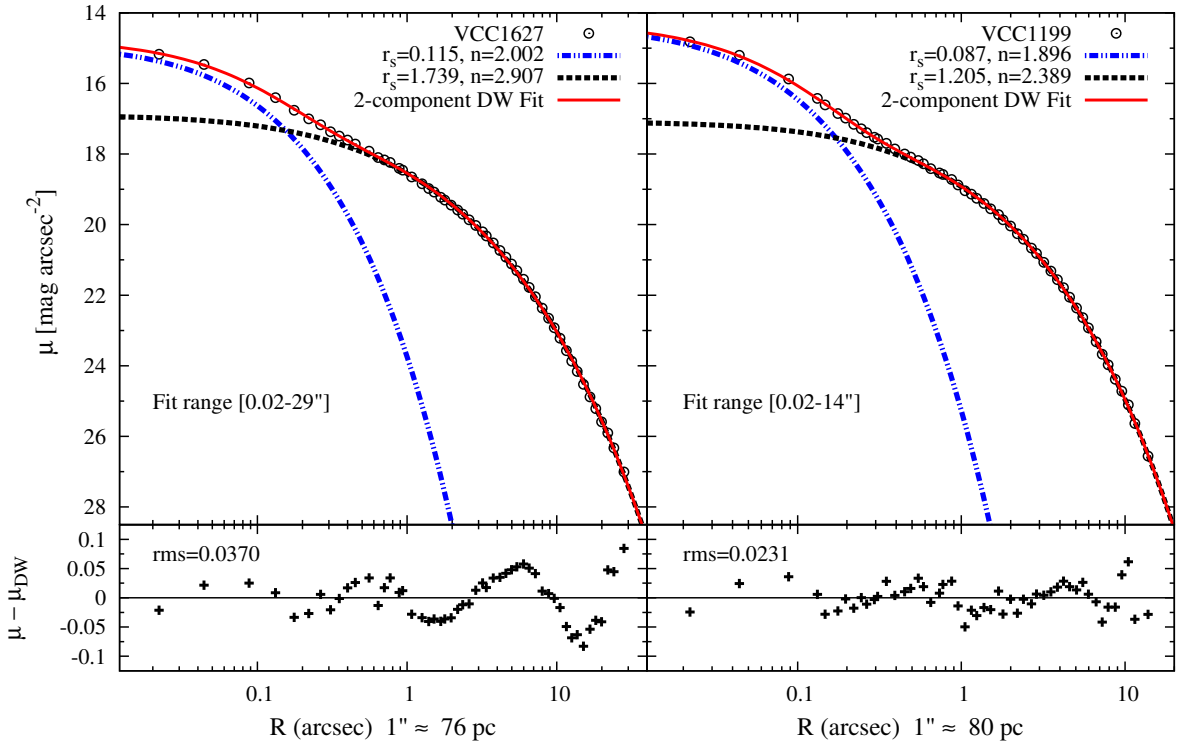}
    \caption{Left: VCC1627; Right: VCC1199. The least luminous ellipticals in Virgo ($\sim 10^{8} L_{V\sun}$). Refer to caption of Fig.\ref{SB1} for details.}
    \label{SBlast}
  \end{figure*}  
  \section{Comparison with other parametrizations}\label{comparison}
  The goal of this paper is to investigate how well can a multi-component DW-function model the SB and consequently whether the intrinsic 3D density can be described with a multi-component Einasto model. A detailed comparison with other parametrizations for every galaxy is beyond the scope of this paper. However, these parametrizations often show residuals larger than measurement errors (sections \ref{powerlaw} and \ref{modexp}). Since the literature has a few examples of fits with other parametrizations, for the galaxies we model here, we present a comparison in sections \ref{NukCSKing}, \ref{SSer} and \ref{dSer}.
  \subsection{Core-Sersic, Sersic+King and Nuker models}\label{NukCSKing}
  All galaxies in our sample have been fit with a combination of psf-convolved Sersic, Core-Sersic (hereafter, CS) and King models to the ACSVCS Sloan g- and z-band profiles in \cite{F06}. While the residuals are difficult to ascertain from their figures, fit residuals of King+Sersic, Core-Sersic and Nuker models in the central $0.02-20~arcsec$, or $\sim 3$ radial decades, are presented in \cite{L07} (L+07); see their fig.~9 and 10. These are the same psf-deconvolved profiles that were used in KFCB09 and the present paper where the data in the central regions are from psf-deconvolved WFPC V-band and $1.6~\mu$ NICMOS images of Lauer et.al. (1992,~1995 \& 2005). Note that L+07 fit the same models used in F+06 (for a given galaxy) to psf-deconvolved WFPC/NICMOS images and then compare the F+06 models to fits with a Nuker model. L+07 also notes that differences in choice of psf, camera and observing band are not significant to prevent a comparison.
  
  The Nuker fits usually exclude fitting 'nuclei'. These regions are fit by F+06 with a King and a Sersic ($6$-parameters) or CS profile ($9$-parameters); the CS+King galaxies of F+06 are not shown in L+07. Further, the fits in L+07 are compared within the central $10-20~arcsec$ domain of validity of the Nuker profile. All other parametrizations have been fit over larger radial ranges. Fits in this paper, which use KFCB09 data, use the largest range. The $19$ galaxies that overlap with our sample are discussed below:
  
  (i)~Steep-cusp with a King model for 'nuclei': For most steep-cusp galaxies F+06 required a King profile for the central region and identified them as 'nuclei'. Note that King models have a truncation radius and a flat core in 3D, and hence in 2D as well. The Nuker fits exclude these regions. For galaxies that we have in common with F+06 and L+07 -- NGC4387, NGC4467, NGC4551, NGC4458, VCC1199, VCC1440 and VCC1627 -- we show that they can be well fit with DW-functions, often with just two components over the entire dynamic radial range, and with much better residuals than with either the King+Sersic or the Nuker profiles. 
  
  (ii)~Steep-cusp with a single Sersic model: For some galaxies, F+06 showed that the entire profile can be described with a single component Sersic model. The comparison plots in L+07 show that a single Sersic profile can not fit the central-most regions of NGC4478, NGC4473, NGC4621, NGC4434 and NGC4464 where the Nuker performs better. We show that it is possible to easily quantify these regions using the DW-function with better residuals than with the Nuker profile.
  
  (iii)~Shallow-cusp: NGC4365, NGC4382, NGC4406, NGC4552, NGC4649, NGC4472 and NGC4486 have been fit with the Core-Sersic and Nuker profiles. Although the residuals are smaller than those in the case of the steep-cusp galaxies, we show that fits with a multi-component DW-function produce even smaller residuals. It should also be noted that the Nuker fits apply to a limited range, and as shown by \cite{Graham03} the parameters depend strongly on the selected radial domain of fit.
  \subsection {Single-Sersic models}\label{SSer}
  KFCB09, whose composite SB profiles we use in this paper, presents fits to the SB with a single Sersic $3$-parameter function for each galaxy. Since such a 1-component Sersic profile cannot adequately fit the full radial range of 4-5 decades of KFCB09 profiles, the authors estimate the largest radial range over which a single Sersic profile produces robust fits and residuals comparable to measurement errors. This range is typically 1-2.5 radial decades, and excludes regions interior to the transition radius, where the slope changes rather abruptly. 
  
  KFCB09 also had to invoke certain constraints to estimate the total luminosity and half-light radii, for example, a limiting magnitude up to which to integrate the light.  This was usually done for galaxies where the Sersic profile failed to model the SB profile over a large radial range. While as noted in KFCB09 the existence of a physically justified limiting magnitude is possible -- for example, a tidal truncation radius or incorrectly subtracted light of a neighbour -- such limits should not exist due to the failure of an ad-hoc fitting function to model the entire SB profile. Further, since the fits were obtained from a limited range, they give a biased estimate of the Sersic index (for example, in the case of NGC4406 and other massive galaxies in table 1 of KFCB09).
  
  We agree with KFCB09 that fits must be consistent with the measurement errors, and show that this can be achieved using a multi-component DW-function, which also avoids systematic deviations between the data and the fits.  Our estimates of structural properties are therefore likely to be more meaningful, and can be used to obtain direct estimates of the intrinsic 3D structural properties of galaxies.
  \subsection{Double-Sersic models}\label{dSer}
  Hopkins et al. (2009a,b) fit a $2$-component Sersic profile to all galaxies in the KFCB09 sample that we use here. While they do not show the residual profiles, we note that the $rms$ of their fits with a $2$-component Sersic profile is usually $40$ per cent larger than the $rms$ of our $2$-component DW models, and nearly $200-300$ per cent larger than the $rms$ of our $3$-component DW models. In Table~\ref{Virgosample} we list the $rms$ of fits of our best-fitting DW models and that of the double-Sersic models in Hopkins et al., for comparison. 

  From their double Sersic models Hopkins et.al. conclude that steep-cusp and shallow-cusp galaxies are two disjoint populations, and that the outer Sersic index $m$ does not depend on the mass or luminosity of these galaxies. 
  
  Our fits lead us to a different conclusion. We observe that the outer Einasto index $n$ does increase with luminosity (and consequently size, $r_{3E}$ - the 3D half-light radius) in a seemingly continuous manner. A similar trend was also noted by \cite{Graham96} and F+06 based on their fits with Sersic profiles.
  \subsection{The form of the 2D structure}  
  An examination of the SB profiles show that the central regions have distinct variations in slope. This is more true for the steep-cusp galaxies than shallow-cusp ones. This deviation from pure power-laws are reflected in the large fit residuals in the central regions as discussed in section \ref{NukCSKing}.

  Fig.~\ref{SB1}-\ref{SBlast}, when compared to the fits in the literature described above, demonstrate that the 2- or $3$-component DW-function provides a better fit, over a larger radial range, than other existing functional forms. Our overall $rms$ are comparable to, or lower than that of other models, and our residuals are consistently low over the 4-5 radial decades of the available composite observations (KFCB09).  Furthermore, the DW-function is a very accurate 2D projection of the 3D Einasto profile, and is expressed in terms of the 3D Einasto profile parameters. This means that if the 2D fits are good, the intrinsic 3D luminosity structure is that of superimposed Einasto profiles, and can be inferred directly, with no further modelling.

  We hence propose that the light of ellipticals that was believed to be well fit with a Sersic profile in 2D, is instead better described by a multi-component form of a similar function (the Einasto profile) in 3D, whose 2D projection is given by a multi-component DW-function. 
  \section{Component properties: Luminosity, half-light radius and Einasto index $n$}\label{compprops}
  In this section we investigate the structural properties of the components deduced from the multi-component DW fits. We shall be referring to the statistically significant best-fitting models only, listed in Table~\ref{Virgosample}, and not all $2$-component and $3$-component fits shown in Figs.\ref{SB1}-\ref{SBlast}. Nine galaxies are described with two DW-components (two shallow cusps and seven steep cusps), and fourteen galaxies are described with three DW-components (seven each of shallow and steep cusp galaxies).
  
  In our modelling of a galaxy as a linear superposition of components, the central and intermediate DW-components described in the following sections \ref{centcomp}-\ref{outcomp} are in excess to an inner extrapolation of the outer DW-component; and do not contain all of the light in the central and intermediate regions. This is consistent with similar decompositions in the literature. Superposition of components, comprising of a Sersic profile for the central bulge superimposed on an underlying exponential model (a Sersic profile with $m=1$), is often applied to the case of lenticulars and spirals. \cite{CoteDS} have shown that a similar decomposition using a double-Sersic profile as a fitting function can also be applied to ellipticals with $M_B$ $\gtrsim$ $-19.5$. The resulting central Sersic-component is then used to evaluate physical properties of the central region.
  
  For the case of shallow-cusp galaxies, as in the spectroscopic and kinematic modelling of M87 in \cite{TEH91} and the fitting function (double Sersic) based modelling of 'core' galaxies in \cite{Hop09b}, we also find that these galaxies can be modelled as a linear superposition of components. While such fit components need not correspond to real physical systems, we show in section \ref{compsys}, three cases of shallow-cusp ('core') galaxies whose central DW-components coincide very well with spectroscopically identified systems.

  In order to estimate luminosities of components, a knowledge of a characteristic axis ratio (equation \eqref{characq}) over the extent of each component is required. Since the components are superimposed, it is very difficult to isolate their characteristic axis ratios using SB analysis; except maybe the outer-most dominant one. However, elliptical galaxies have axis ratios of 0.3 (E7) $\leq$ $q$ $\leq$ 1.0 (E0) and for any arbitrary choice of $q$ the uncertainty in estimating the luminosity can at most be off by a factor of $2-3$. In our sample, except 3 galaxies which are E4, the rest are between E1 and E3. Our worst errors in estimating component luminosities are therefore less than a factor of $2$. Additional spectroscopic or kinematic modelling may be used to constrain the ellipticities of individual components.

  With this understanding, for the purpose of computing luminosity and half-light radius of components, we assume that the axis ratio of all components are the same as the characteristic axis ratio $q_{c}$ for the entire galaxy (equation \eqref{characq}). Note that uncertainties in interpreting component luminosities do not affect our estimate of the total luminosity of the galaxy.
  \subsection{Central component}\label{centcomp}
  From the nine galaxies with two components and fourteen galaxies with three components, we observe that:

  1)~ In all galaxies, the central component has a lower $n$ than that of the outer component. Since Einasto functions of the form $e^{-x^{2}}$ ($n$$=$0.5) $\to 0$ faster than $e^{-x}$ ($n$$=$1), a lower $n$ implies a larger concentration. Hence, the central component is more concentrated than the outer component.

\begin{figure}
    \begin{minipage}{1.0\columnwidth}
    \includegraphics[width=84mm]{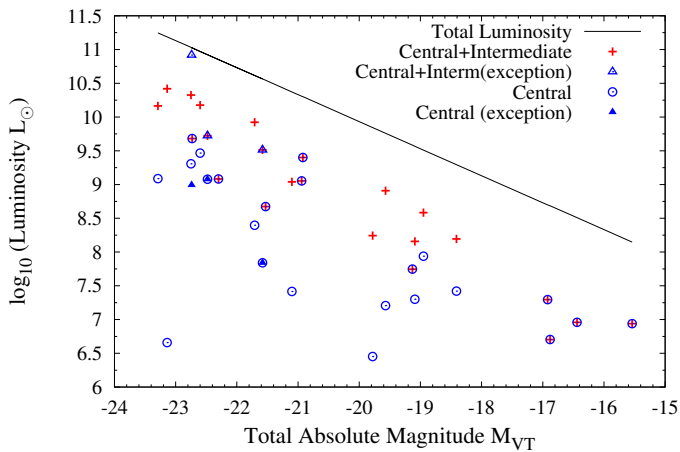}
    \caption{Luminosity of components resulting from our multi-component DW models as a function of total luminosity of the galaxy ($M_{VT}$). Galaxies whose component parameters are uncertain (see section \ref{exceptions}) are marked as {\it exceptions} in the figure key. The central component of massive (generally shallow-cusp) ellipticals are $\sim$ $10^{2}$ more luminous than that of the less massive (generally steep-cusp) ellipticals. This effect is more apparent for the total luminosity in the central+intermediate component. Note that luminosities of the central and intermediate DW-components are excess luminosities in those regions with respect to contributions from the outer DW-component. A colour version of this figure is available in the online edition.}
    \label{complum}
    \end{minipage}
  \end{figure}

  2)~ The central component of the shallow-cusp galaxies are one or two orders of magnitude more luminous than that of the steep-cusp galaxies (Fig.\ref{complum}). Recall that the uncertainty in $q$ is at worst a factor of $2$ while the luminosities differ by a factor of $10-100$. The shallow-cusp galaxies thus host an unambiguously larger luminosity in their central component. Note that more luminous galaxies, $M_{VT}$ $\lesssim$ $-21.5$ mag, typically have a shallow cusp, while fainter galaxies have steep central cusps.
 
\begin{figure}
    \begin{minipage}{1.0\columnwidth}
      \includegraphics[width=84mm]{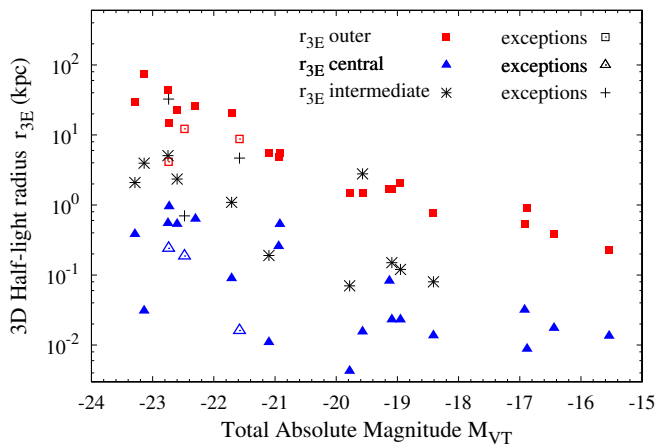}
      \caption{The intrinsic (3D) half-light radii $r_{E}$ of all components appear to be generally increasing with galaxy luminosity. The $r_{3E}$ of the massive shallow cusps are generally a factor of $10$ larger than that of the steep-cusp galaxies. This trend is stronger for the outer component. Note that, as shown in Table\ref{Virgosample}, $r_{3Eo}$ of the outer component is generally slightly larger than $r_{3E}$ of the galaxy which is computed using the total light from all components. Also refer to caption of Fig.~\ref{complum}. A colour version of this figure is available in the online edition.}
      \label{rEMVT}
    \end{minipage}
  \end{figure}

  3)~ The 3D half-light radius $r_{3E}$ of the central component of shallow-cusp galaxies is typically an order of magnitude larger than that of steep-cusp galaxies (Fig.\ref{rEMVT}). The same is also true for the scale radius $r_{-2}$ (listed in the keys of Figs.\ref{SB1}-\ref{SBlast}), which characterizes a radius inside which the logarithmic slope of the 3D profile falls below $-2$, and that of the SB profile falls below $-1$.
  
\begin{figure}
    \begin{minipage}{1.0\columnwidth}  
      \includegraphics[width=84mm]{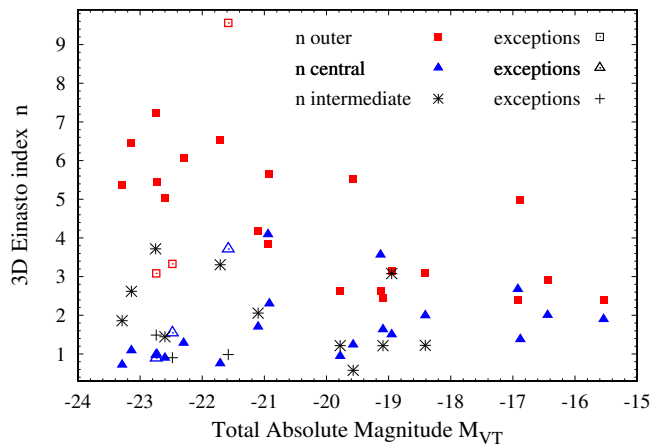}
      \caption{Einasto index $n$ of components. All massive shallow-cusp galaxies have $n$ $\lesssim$ 1 for their central components, while the steep cusps typically have $n$ $\gtrsim$ 2. The $n$ of the outer component shows a distinct trend increasing with the total luminosity of the galaxy. Also note the large difference in $n$ between the central and outer components of the shallow-cusp galaxies, which diminishes with decreasing luminosity for the steep-cusp family. Refer to caption of Fig.~\ref{complum} for {\it exceptions}. A colour version of this figure is available in the online edition.}
      \label{nMVT}
    \end{minipage}
  \end{figure}

  4)~ The shape parameter $n$ of the inner component of shallow-cusp galaxies is always significantly smaller than that of their outer component, with the difference reduced for the steep-cusp galaxies (Fig.\ref{nMVT}). Exceptions, with $n_{central}$ $\approx$ $n_{outer}$, are NGC4467 and NGC4459 whose central region is affected by a huge dust disk, and NGC4387 that has $n_{central}$ $>$ $n_{outer}$, whose light could be affected by NGC4406 or, as discussed in section \ref{spcases}, could be a $3$-component system being modelled with $2$-components.

  5)~ All of the nine massive shallow-cusp galaxies ($M_{VT}$$\leq$$-21.71$) have $n$ $\lesssim$ 1 for their central component (NGC4382 has $n$$=$1.55), while for the fourteen steep-cusp galaxies usually $n$ $>$ 2.0 and with a larger dispersion; except NGC4478 and NGC4434 where $n$ $\sim$ 1 and NGC4551, NGC4458 and VCC1440 with $n$ $\sim$ 1.5. Lower $n$ implies a larger concentration. Hence, the central components of shallow-cusp galaxies appear to be more concentrated than those of the steep-cusp galaxies.
  
  The above observations indicate a new trend with regards to the central components of galaxies: Even though steep-cusp galaxies have a higher central density, the central component of the shallow-cusp galaxies is far more luminous and massive, spatially more extended, and more concentrated than that of the steep-cusp galaxies.

  6)~ The large incidence of $n$ $\sim$ 1 (61 per cent) for the central component and especially in all the massive shallow-cusp galaxies is interesting. Disks are usually fit with an exponential, or $m$$=$1 Sersic profile. The central regions of these galaxies thus possibly support disk-like systems. While this conclusion is very tentative, because it is based solely on a SB analysis, in section \ref{compsys} we describe two cases, M87 and NGC4261, whose central regions have spectroscopically identified disks, and our $3$-component -- but not $2$-component -- fits show that they have $n$$\sim$1. We reiterate that while disks usually have $m$$=$1, not all $m$$=$1 systems should be called disks (and vice-versa) without spectroscopic verification.
  
  To compare with simulations we reviewed Hopkins et al. (2008, 2009a,b). While their simulations do not probe the very central regions, $\lesssim$ $50$ pc, they found a similar trend as mentioned above; all steep-cusp and shallow-cusp galaxies show a light excess in their central regions with respect to the inner extrapolation of an outer Sersic profile. They also note that, on an average, they could recover the total light in the true central component of steep-cusp galaxies by fixing, without fitting, the Sersic index of the inner component to $m$$=$1. However, whenever they fit for the Sersic index $m$ in the steep-cusp galaxies observed with the HST, they typically obtain a wide range of 0.6$\lesssim$$m$$\lesssim$5.75 for the central component and rarely $m$$\sim$1 (refer to the online version of table 3 in \cite{Hop09a}). 
  
  This is consistent with the large dispersion in the inner $n$ we get for our $2$-component DW models in cases where $3$-component DW models produce significantly better fits. Using three components in these cases, not only is the variance in the inner $n$ reduced, we also see a tendency of $n$$\sim$1 similar to what Hopkins et.al. find for the fits to their simulations with a Sersic profile. 

  We caution that only within a limited range of projected radii, systems with Einasto profiles of index $n$$\sim$1 can be modelled with Sersic profiles of index $m$$\sim$1 (from equations 3.2 and 3.3 in DW10). Over large radial ranges the Sersic profile fails to model a projected Einasto profile. The above comparison should thus be taken as qualitative. Nevertheless, this indicates that it may not always be meaningful to fit all galaxies with a predecided number of components which can further lead to misleading interpretations of other structural parameters. It is more meaningful to infer the existence of these components either from the residual patterns in the data (as in this paper) or motivate their existence through other (possibly spectroscopic) observations. 
  \subsection{Intermediate component}\label{intcomp}
  Our multi-component models indicate that $14$ of the $23$ galaxies have an intermediate DW-component. These include both steep cusp and shallow cusp galaxies. Only in two out of the $14$ galaxies (NGC4621 and NGC4434) it identifies features within the outer DW-component, but in the rest it is located between the inner and outer components, and thus forms a transition region. From Fig.\ref{complum} we also observe that within a factor of two, all galaxies contain a similar fraction of the total light in their central+intermediate DW-components. There appears to be an indication that the massive shallow-cusp galaxies may contain a larger fraction, although this is not very clear from our small sample.

  Without additional kinematic or spectroscopic data it is not possible to ascertain the physical origin of the intermediate component, but based on the results of existing galaxy formation models, we speculate that at least three scenarios are possible:

  We suggest that, the stars making up the intermediate component -- (i) may have formed as a result of local star formation (as in \cite{MH94}, \cite{Hop08} for the central regions), (ii) scattered into this region during mergers by a central supermassive black hole (SMBH) (\cite{BBR80}, \cite{BMQ04}) , or (iii) in the case of shallow-cusp galaxies only, stars could have been scattered to these radii by the central SMBH, or coalescing binary SMBHs which are believed to scour out few 100 pc regions in galaxy centres \citep{GM08}. 
  
  The physical interpretation of the central and outer regions of a galaxy depends on how the overall SB profile is modelled. In our models for shallow-cusp ('core') galaxies, we observe an excess luminosity due to the central DW-component with respect to an outer DW-component. In section \ref{compsys}, we show that for three 'core' galaxies, the central DW-component does correspond to spectroscopically identified real systems. Hence, the central DW-component is not necessarily a mere mathematical construct. Alternatively, as has been suggested in the literature, the mechanisms mentioned in the previous paragaraph are believed to have caused a deficit of mass in the central regions, with respect to an inward extrapolation of an outer Sersic profile which is fitted to the SB at radii beyond the transition, or break radius. In other words, the existence of a shallow cusp may imply -- (a) a deficit of mass in the central regions  or (b) an excess of mass with small $n$, which is superimposed on to the outer DW-component, as in this paper. The dynamical interpretation of the central region will probably be different depending on whether (a) or (b) are mostly correct, but either case is consistent with SMBH scouring out mass. However, SB analysis alone cannot resolve this issue; more data and dynamical modelling are required. Further discussion on mass deficits is provided in section \ref{massdeficit}.
  \subsection{Outer component}\label{outcomp}
  The multi-component DW models reveal a huge dominant outer component that usually contains a much larger fraction of the total light than any of the other components. Fig.~\ref{nMVT} indicates that the outer $n$ increases with luminosity. From the SB profiles (Fig.~\ref{SB1}-\ref{SBlast}) we note that for galaxies with outer $n$$\gtrsim$5, the outer component makes a significant contribution to the density in the central region.  This is usually the case for the massive shallow-cusp galaxies, but is also seen in smaller, less luminous galaxies, like NGC4473 and NGC4434, where the SB appears to indicate a shallow inner slope. On the other hand, the outer components of the steep-cusp galaxies do not contribute fractionally as much light to the central regions. 
  
  From the above discussion of components in sections \ref{centcomp}-\ref{outcomp} we observe that all shallow-cusp galaxies have a sharp transition to a shallow inner slope. These SB profiles are well modelled by a combination of -- (i) a central $n$$\sim$1 DW-component whose intrinsic half-light radius $r_{3E}$ (Table \ref{lumdensample}) is greater than about ten times that of steep-cusp galaxies, and (ii) a non-negligible contribution to the density in the central region from a $n$$\gtrsim$5 outer DW-component. The central slope flattening is especially pronounced because an Einasto profile with $n$$\sim$1 is quite concentrated, compared to that of larger $n$'s, and when projected it produces a step-function like sharper transition in a log-log plot of density versus radius.
  \subsection{Components and systems: Three examples}\label{compsys}
   The components obtained through fitting ad-hoc functions may not correspond to physically distinct kinematic systems or stellar populations, unless the form of our fitting function happens to be correct.  While a detailed analysis of components and systems for every galaxy is beyond the scope of this paper, we present three interesting connections between the components we deduce from fitting the V-band SB profiles, and spectroscopically identified systems.
   
   \cite{Jaffe93} and \cite{Ford94} reported the earliest detections of nuclear disks around supermassive black holes (SMBHs) in the centres of galaxies. We discuss structural similarities between such spectroscopic detections of systems and the components deduced from our multi-component DW models for three galaxies in our sample.
  
   1)~Images of M87 (NGC4486) show a prominent nuclear disk. The central DW-component of our $3$-component model (Fig.\ref{NGC4486}) from the broadband I-band (WFPC F785LP) images, suitably scaled to V-magnitudes in KFCB09, has a best fit Einasto shape parameter $n$$=$1.09 which is consistent with it being a disk. Further, at the scale length of $r_2$$=$0.266 arcsec the intensity of the central DW-component drops to $10$ per cent; a size consistent with the spectroscopic observations of \cite{Harms94}. \cite{Tsvet99} study the morphology of the disk in detail with the narrow band F658N filter and observe that a significant light excess is detected inside 0.5 arcsec. We note that at 0.5 arcsec the intensity of the central DW-component falls to $1$ per cent of maximum. 

   2)~NGC4261 also has large spectroscopically confirmed central systems (\cite{Jaffe93}, \cite{FFJ96}). FFJ96 fit a double exponential model with scale lengths of 1.83 and 8.73 arcsec. Our best fit $3$-component model (Fig.\ref{NGC4261}) show that the central and intermediate components have nearly exponential profiles. The central component has a shape parameter $n$$=$0.9 and $r_{-2}$$=$2.71 arcsec, while the intermediate component has $n$$=$1.488 and $r_{-2}$$=$9.548 arcsec. 

   3)~Another well studied galaxy is NGC4473 (see \cite{Pink03} and references therein). SAURON integral-field spectroscopic observations \citep{Emsel04} shows an inner spheroidal system extending through $20$ arcsec.  Our $2$-component DW model (Fig.\ref{NGC4473_87}) has a central component with $n$$=$$2.3$ and $r_{-2}$$=$3.16 arcsec which at $20$ arcsec contributes $\sim$ 10 per cent (2.5 magnitudes fainter) to the total SB at that radius.

  This is a fairly remarkable identification of physically distinct systems from a purely surface brightness analysis that had no information or priors about the spectroscopic properties of these regions. While identification of physical systems must not be made solely from identification of fit-components through a purely surface brightness analysis, the above examples indicate that a good fit consistent with measurement errors are unlikely to detect spurious artificial components. Further, all three galaxies exhibit a shallow central cusp. Our models thus allows one to account for structure within the shallow-cusp ('core'). This is not possible with either the Core-Sersic or Nuker profiles which cannot model deviations of the SB profiles from pure power-laws within the central region.
   \section{Uniqueness of deprojection}\label{uniqueness}
   Given that a multi-component DW-function fits the SB profiles extremely well, our next major goal is to explore the conditions under which the 3D intrinsic luminosity density profiles can be described with a multi-component Einasto model. An attractive feature of such an interpretation is that the Einasto profile can be a likely descriptor of both the baryonic, and $\Lambda$CDM N-body dark matter haloes, revealing an universality in their functional form. 
   
  Deprojections, however, are generally not unique. In this section we review some of the limitations in obtaining unique deprojections of surface density profiles that have been taken into consideration while providing the intrinsic 3D luminosity density profiles in the next section.
  \subsection{Konus and semi-konus densities}
  \cite{Ryb87} showed that based on the angle of inclination with the line of sight $i$ of an axisymmetric system, there exists a cone of ignorance ($\theta$$=$$90$$\degr$$-$$i$), such that a family of densities -- called {\it konus densities} \citep{GB96} -- that is non-zero only within $\theta$, can project to yield zero SB unless the system is seen edge-on ($i$$=$$90$$\degr$). The range of possible deprojections increases with decreasing $i$. For triaxial systems this non-uniqueness increases dramatically \citep{GB96}. \cite{KR96} further extend the range of possible functions through {\it semi-konus densities} and their linear combinations. 
  \subsection{Magnitude of effect of semi-konus densities}
  \cite{vB97} (hereafter, vdB97) highlight physical admissibility conditions that limit the range of possible semi-konus densities. Through a generalized set of semi-konus densities, he investigated how they may (or may not) have an effect on the dynamical and photometric properties of galaxies. We summarize here some of the key results from his work:
  
  1) For intrinsic axis ratio, $q\gtrsim0.8$ the effects of semi-konus densities are negligible for almost all inclination angles. This is similar to saying that spherically symmetric systems have unique deprojections.
  
  2) For oblate spheroids with a constant core, the presence of konus densities manifests as detectable wiggles along the minor axis, but not along the major axis, an effect seldom seen in real galaxies. Although real galaxies seldom have perfect constant density cores, this can be an important distinguishing feature for shallow-cusp galaxies. 

  3) Using a more representative, double power-law \citep{Qian95} parametrization of the intrinsic 3D density of the central regions of real galaxies, vdB97 showed that even for an intrinsic axis ratio of $q=0.5$, the maximum amount of semi-konus density that can be added is negligible for inclination angles $i\gtrsim 70\degr$. This is seen for a wide range of cusp steepness, $-2\le\alpha\le0$ inside the core radius, or break radius of the double power-law, say $R_b$ . For smaller inclination angles, $i\lesssim 70\degr$, the maximum semi-konus density that can be added increases as the ratio of the scale-length of the konus density (say $r_k$) to the core radius of the galaxy decreases (refer to his fig.~8).

  A consequence of this effect of decreasing $r_k/R_b$ is that the semi-konus density does not add significantly to the mass, or light. Furthermore, as $r\to 0$ the mass of the central supermassive black hole will completely overwhelm any contribution of the konus density. It should also be noted that the amplitude shown in his fig.~8 is the ratio of maximum konus-density at $r=0$ to the power-law galaxy density at $r=R_b$. Since galaxy densities continue to rise for $r<R_b$, this means the relative strength of konus to galaxy density is even smaller at comparable $r$.
  
  4) The generalized konus densities are not power-laws. However, to characterize their effect in terms of the slopes of the konus densities (not to be confused with slope of the galaxy SB profile), vdB97 approximates them as power-laws. He showed that if one can approximate $\rho_{konus}(r) \propto r^{-\alpha}$, then $\alpha$ must be $\leq 1$. i.e., the konus densities themselves can not be too cuspy. 
  \subsection{The effect of triaxiality}\label{triaxiality}
  The discussion in the previous section applies to axisymmetric systems. For triaxial systems the range of possible deprojections increases. However, this increase in non-uniqueness will also depend on the degree of triaxiality. 

  Triaxiality as well as axisymmetry can leave an imprint on the kinematic structure of galaxies. \cite{Emsel04} provides detailed kinematic maps of 48 E/S0 galaxies from the SAURON survey \citep{deZeeuw02}; eight of these galaxies are part of our sample. Using an estimate of their specific angular momentum $\lambda_R$, within one effective radius $R_E$, \cite{Emsel07} characterized galaxies as fast ($\lambda_{R}$$>$0.1) or slow ($\lambda_{R}$$<$0.1) rotators. \cite{Cap07} observe that the fast rotators are typically found to be oblate axisymmetric systems and have steep-cusps while the slow rotators generally have shallow-cusps and are triaxial. 

  Further, \cite{GB85}, \cite{MF96}, \cite{VM98} have shown that triaxial models with steep cusps or with a moderately sized supermassive blackhole ($M_{BH}\sim 0.005~M_{gal}$) are not able to sustain a triaxial shape. This indicates that the steep-cusp galaxies which are typically fast rotators can be very well approximated as axisymmetric systems.
  
  For galaxies with shallow cusps which typically rotate slowly, theoretically, triaxiality can not be excluded. However, \cite{Cap06} observe that significant triaxiality will lead to an increase in the otherwise relatively small scatter in the M/L-$\sigma$ relation of the SAURON sample deduced using axisymmetric models. It thus appears that the shallow cusps may not be strongly triaxial. In our sample the shallow-cusp galaxies also appear nearly spherically symmetric in projection and hence we model them as oblate axisymmetric systems as well.  
  \subsection{Non-parametric vs. Parametric deprojection}
  In addition to the above issues pertaining to konus densities, \cite{MT94} and \cite{Geb96} highlight the importance of non-parametric deprojections. This should be the approach of choice to reveal the range of possibilities in 3D since small deviations from fits to the SB with ad-hoc fitting functions can translate into larger deviations in the 3D distribution. 

  However, with very high precision data as used in this paper, large deviations, or features in 3D will manifest themselves as detectable, but possibly small features in the 2D SB profiles. Upon projection, smaller features in 3D may remain hidden within measurement errors.  Hence if the goal is to extract smaller, local features in the 3D distribution, then a non-parametric inversion must be performed. 

  Our goal in this paper is, however, to investigate whether the gross properties of the 2D and 3D distribution of light can be described with a multi-component Einasto model. Further, in section \ref{Npdeproj} we show that for two galaxies in our sample that have non-parametric deprojections in the literature, our parametric estimates are in good agreement with the non-parametric estimates. We hence do not take a non-parametric option and note that our inferred intrinsic profile is likely to fall within the confidence intervals of a non-parametric deprojection. 

  \section{Intrinsic 3D luminosity density profiles}\label{lumdensity}
  In this section we discuss the luminosity density profiles of $14$ galaxies in our sample where non-uniqueness of deprojection can be reasonably minimized. Based on the discussion in section \ref{triaxiality} we model all galaxies under the assumption of oblate axisymmetry.

  For oblate axisymmetric systems with intrinsic axis-ratio $q$ and with a minor axis inclined at an angle $i$ with the line of sight ($i=90\degr$ is edge-on), it can be shown that the surface brightness is given by: 
  \begin{align}\label{sigmaellip}
    \Sigma(R)=\frac{2\;q}{\sqrt{\cos^2(i)+q^2\sin^2(i)}} \int^{\infty}_{0} \rho(r) d\zeta
  \end{align}
  where, $R$ is a coordinate along the projected major axis, which for oblate axisymmetric systems is the same as the true major axis; $\zeta=z/q$ is the reduced coordinate along the line of sight and $r=\sqrt{R^2+\zeta^2}$.
  
  It can be further shown that the observed axis-ratio, $q'$, is given by: 
  \begin{align}\label{qobs}
    q'=\sqrt{\cos^2(i)+q^2 \sin^2(i)}
  \end{align}
  
  When the SB is described in terms of the semi-major axis using a spherically symmetric function, generally from equation \eqref{sigmaellip}, it can be seen that the inferred intrinsic central density $\rho_0 \propto \Sigma_0 (q'/q)$. An estimate of the light enclosed within an intrinsic radius $r$, measured along the semi-major axis, when expressed in terms of $\Sigma_0$, does not require knowledge of the true axis ratio $q$ and inclination angle $i$, and is given by $L(r) \propto \Sigma_0 q'$.
    
\begin{figure}
    \begin{minipage}{2.0\columnwidth}
      \centering
      \includegraphics[width=142mm]{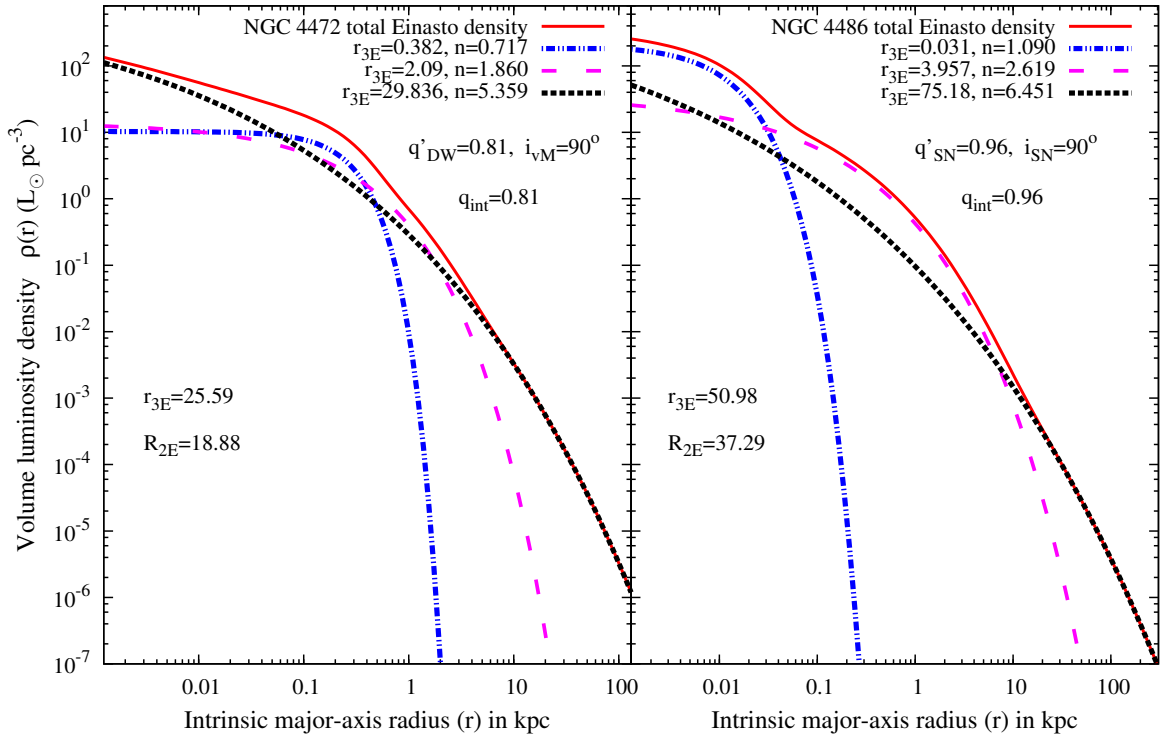}
      \caption{Multi-component Einasto models of the intrinsic (3D) volume luminosity density profiles for $14$ galaxies in our sample (see Table~\ref{lumdensample} and section \ref{lumdensity}) assuming oblate-axisymmetry. Only the statistically significant best-fitting models (Table~\ref{Virgosample}) are shown. The component profiles have been computed using equation \eqref{rho0q} in \eqref{einasto}. Colours and line types are as in Fig.\ref{SB1}. The figure keys list the Einasto shape parameter $n$ and the intrinsic (3D) effective or half-light radius of the components (in kpc), estimated from the best-fitting values of $r_{-2}$, using equation \ref{r3dr2}. The total half-light radii in kpc, intrinsic ($r_{3E}$) and projected ($R_{2E}$), are also shown separately in the figure panel. Also listed are the observed axis ratio $q'$, the inclination angle $i$ of the minor axis to the line-of-sight with suffixes labelling the references(see Table~\ref{lumdensample}) -- DW (this paper), vM (van der Marel et al. 1990), SN stands for SAURON \citep{Cap07}, SG\citep{SG10} and vB\citep{vBRCE08}. For some galaxies that are fairly round and for which we could not find an inclination angle in the literature, we have assumed an arbitrary $i=85\degr$. These cases are labelled as $i_{arb}$. Generally we use $q'$ from the same reference that contains the $i$ listed in Table~\ref{lumdensample}. However, if $q'_{DW}\approx q'$ of the reference, we use $q'_{DW}$. The intrinsic axis ratio $q_{int}$ is computed using equation \ref{qobs}.  The horizontal axis showing the intrinsic (3D) radius (in kpc) is up to $\sim 1.5 \times$ the projected radius of available data. \newline {\bf Above} NGC4472 (left) and NGC4486 (right). (Colour versions of these figures are available in the online edition.)}
      \label{lumden1}
    \end{minipage}
  \end{figure}

\begin{figure*}
    \includegraphics[width=142mm]{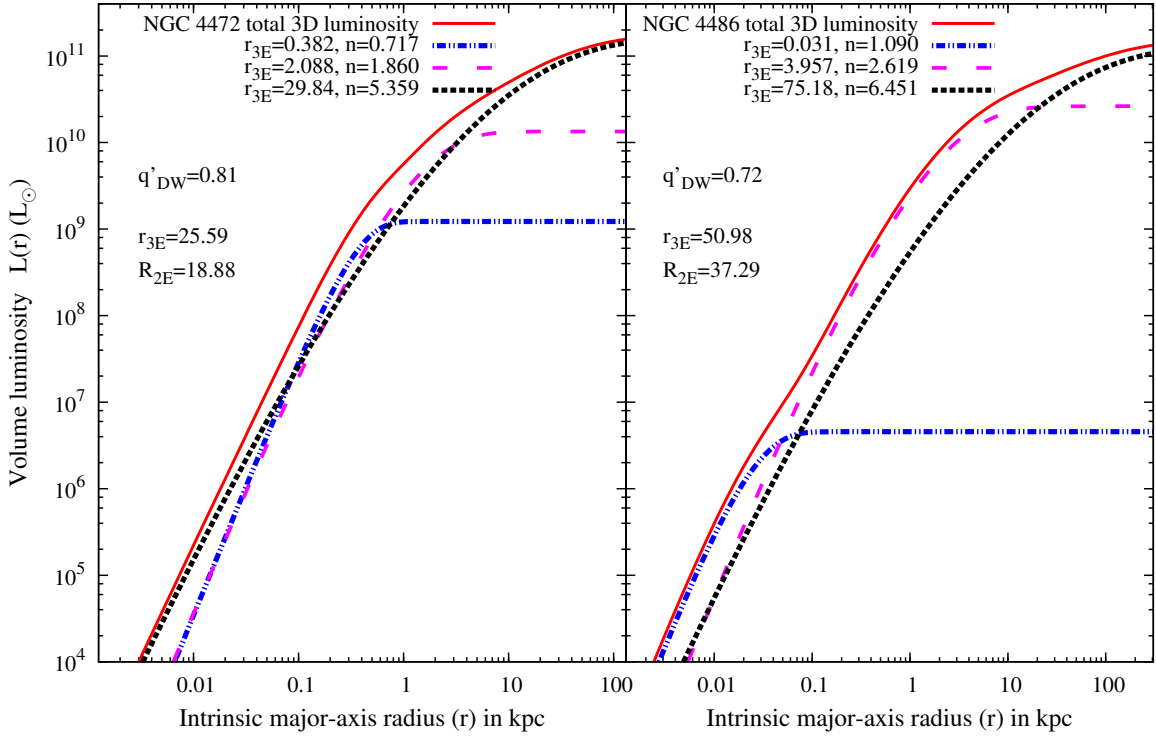}
    \caption{Cumulative light enclosed within an intrinsic (3D) radius $r(kpc)$ for the $14$ galaxies for which we infer the intrinsic luminosity density (see Table~\ref{lumdensample} and section \ref{lumdensity}). The component contributions have been estimated using \eqref{lightq}. Profiles are shown only for the statistically significant best-fitting models (Table~\ref{Virgosample}). The observed axis ratio $q'$ and the total half-light radii, intrinsic $r_{3E}$ and projected $R_{2E}$, are also shown separately within the figure panels. Colours and line types are as in Fig.\ref{SB1} and details on figure keys and labels are as in Fig.\ref{lumden1}. Above NGC4472 (left) and NGC4486 (right). (Colour versions of these figures are available in the online edition.)}
    \label{cumlight1}
  \end{figure*}

\begin{figure*}
    \includegraphics[width=142mm]{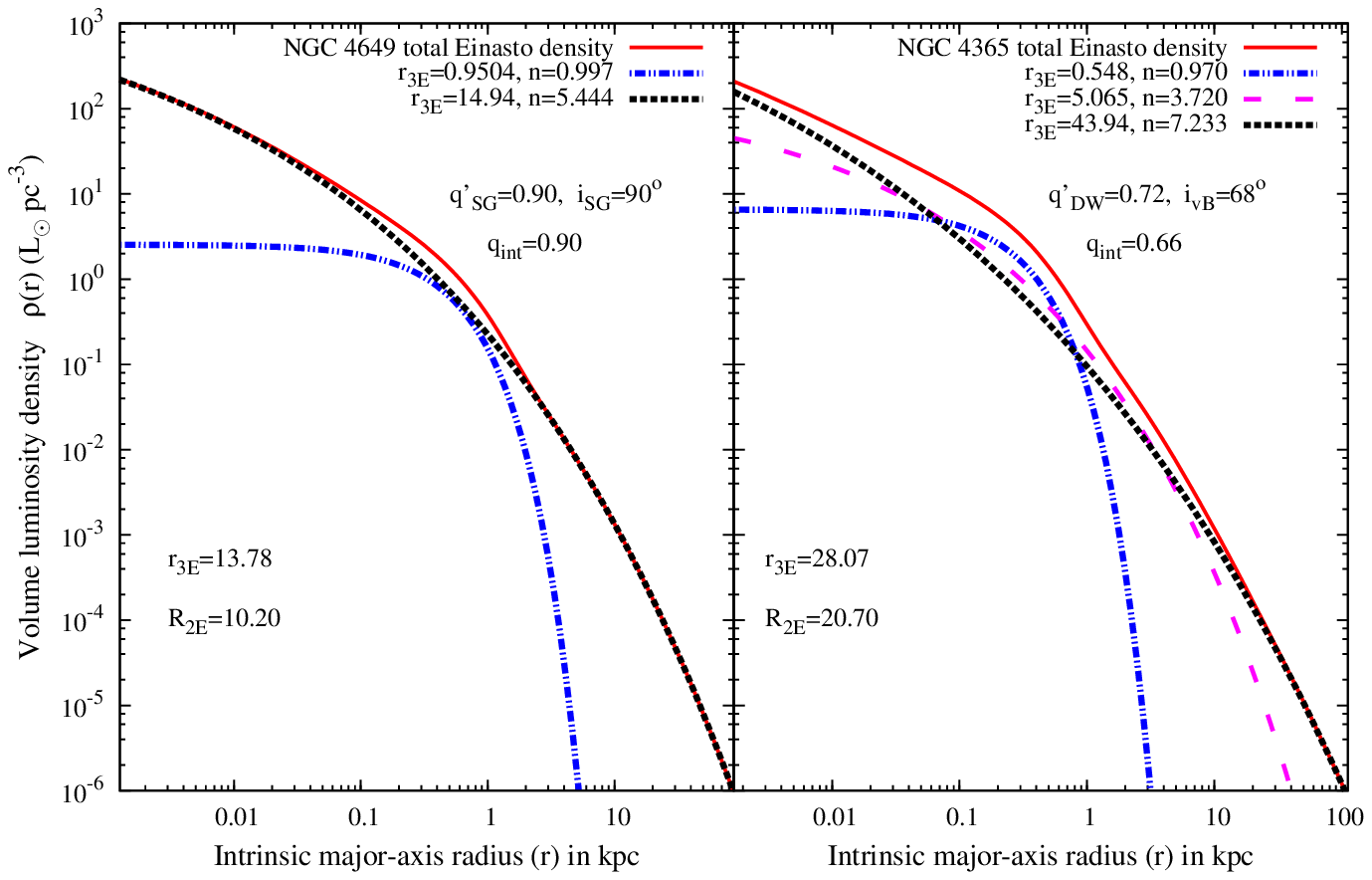}
    \caption{Intrinsic (3D) luminosity density for NGC4649 (left) and NGC4365 (right). Refer to caption of Fig.\ref{lumden1} for details.}
    \label{NGC4649_4365}
  \end{figure*}

  For the Einasto profile, equation \eqref{einasto}, using \eqref{sigmae} and \eqref{sigmanot} this leads to:
  \begin{align}\label{rho0q}
    \rho_0 = \frac{\Sigma_0\;b^n}{2\;r_s\;\Gamma(n+1)}\left(\frac{q'}{q}\right),
  \end{align}
  and,
  \begin{align}\label{lightq}
    L(r)= 2\pi\left(\frac{r^2_s}{b^{2n}}\right)\frac{\gamma\left(3n,b[\frac{r}{r_s}]^{\frac{1}{n}}\right)}{\Gamma(n)} \Sigma_0\; q'
  \end{align}  
  The $\rho(r)$ and $L(r)$ profiles in Figures ~\ref{lumden1}-\ref{lumdenlast} have been obtained using equation \eqref{rho0q} in \eqref{einasto} and \eqref{lightq}.

  Table~\ref{lumdensample} lists the inclination angles (and the corresponding references) used to obtain the intrinsic luminosity density profiles and the cumulative light enclosed as a function of intrinsic 3D radius $r$ in Figures~\ref{lumden1}-\ref{lumdenlast}. For galaxies with observed axis ratio $\gtrsim 0.8$ for which we could not find inclination angles in the literature, we infer their 3D intrinsic luminosity density by assuming an arbitrary inclination angle $i=85\degr$, to distinguish them from galaxies with $i=90\degr$ estimated from modelling.  
  \begin{table}
    \begin{center}
      \resizebox{1.0\columnwidth}{!}{ 
        \begin{threeparttable}
          \caption{Luminosity density sample}\label{lumdensample}
          \begin{tabular}{*{5}{c}}
            \toprule\noalign{\smallskip}
            Name    &  $q'_{DW}$ & $q'_{ref}$  & $i$ & Reference \\ 
            & & & (deg) & \\ \midrule\noalign{\smallskip}
            NGC4472 & 0.806	  & 0.83  & 90	&  \cite{vdM90}  \\
            NGC4486 & 0.722	  & 0.96  & 90	&  \cite{Cap07}  \\
            NGC4649 & 0.828	  & 0.90  & 90	&  \cite{SG10}   \\
            NGC4365 & 0.717	  & 0.75  & 68	&  \cite{vBRCE08} \\	  
            NGC4636 & 0.760	  & ----- & 85	&  Arbitrary assumed $i$ \\
            NGC4552 & 0.873	  & 0.96  & 90	&  \cite{Cap07} \\
            NGC4621 & 0.742	  & 0.66  & 90	&  \cite{Cap07} \\
            NGC4478 & 0.822       & ----- & 85	&  Arbitrary assumed $i$ \\ 
            NGC4434 & 0.928	  & ----- & 85	&  Arbitrary assumed $i$ \\
            NGC4473 & 0.607	  & 0.61  & 73	&  \cite{Cap07} \\
            NGC4458 & 0.879	  & 0.88  & 90 	&  \cite{Cap07} \\
            NGC4467 & 0.813	  & ----- & 85 	&  Arbitrary assumed $i$ \\
            VCC1627 & 0.928	  & ----- & 85 	&  Arbitrary assumed $i$ \\
            VCC1199 & 0.869	  & ----- & 85 	&  Arbitrary assumed $i$ \\
            \bottomrule
          \end{tabular}
      \begin{tablenotes}
        \small
      \item NOTES.--- Sample of $14$ galaxies whose intrinsic (3D) luminosity profiles are shown in Figs.\ref{lumden1}-\ref{lumdenlast}. Refer to sections \ref{uniqueness} and \ref{lumdensity}. In the columns above: Galaxy type are from KFCB09; $q'_{DW}$ is the characteristic axis ratio deduced using equation \eqref{characq}; $q'_{ref}$ and $i$ are the axis-ratio and inclination angle of the minor-axis with respect to the line-of-sight, from the reference listed in the last column. For some galaxies that are fairly round but for which we could not find an inclination angle in the literature, we have assumed an arbitrary inclination angle of $i=85\degr$. 
      \end{tablenotes}
      \end{threeparttable}
      }
    \end{center}
  \end{table}
  Given that the residuals of our fits to the 2D surface brightness (median $rms=0.032~\mgasc$), are similar to the $rms$ of random errors of the data $\sim 0.03~\mgasc$, we can be fairly confident that our parametric description of the 3D distribution as a multi-component Einasto profile will fall within the wider confidence limits of non-parametric inversion. This is especially true at large $R$ and for galaxies with uniformly low residuals over the entire dynamic radial range. This is however not true for cases like NGC4382 where our $3$-component model has systematic residuals significantly larger than measurement errors. We also do not provide luminosity density profiles for NGC4406 which has a strong central dip in its SB profile, as well as galaxies with large inclination angles, for example, NGC4261.

  \subsection{Comparison with non-parametric deprojections}\label{Npdeproj}
  Some of the galaxies for which we present the intrinsic luminosity density, were also deprojected by other authors. The $2$-component model for NGC4649, Fig.~\ref{NGC4649_4365}, can be compared to the non-parameteric inversion of the same data set (from KFCB09) presented in \cite{SG10} (SG10). Note that in this paper we use values for observed axis ratio $q$ and inclination angle $i$ quoted by SG10. The overall profiles are consistent once a correction is made for the distance to the galaxy; SG10 adopt $15.7$ Mpc, while we use $17.3$ Mpc. The central luminosity density is steeper in SG10. However, one must note that the central-most points in the surface brightness have larger errors, arising primarily from psf-deconvolution. 
  
  Also as mentioned in section \ref{fitresults}, including a third component improves the fit to the SB in the central regions (Fig.~\ref{NGC4649}), making it consistent with the $\sim 0.1~\mgasc$ errors in the central-most data points. We do not use the $3$-component model because the F-test rejects it at $1.2\sigma$. One can thus use the $3$-component model purely as a fitting function (without making strong inferences on the resulting components) and in doing so we find very good agreement with the non-parametric deprojection in SG10. This indicates that a subdued third component could well exist, as shown in Fig.~\ref{NGC4649}. 
  
  A similar comparison of our $3$-component parametric model (Fig.~\ref{NGC4486}) with a non-parametric deprojection of NGC4486 (M87) in \cite{GT09} also shows that they are consistent.
  \subsection{The form of the 3D intrinsic structure}
  Most of the galaxies for which we present luminosity density profiles are fairly round: ten are E0-E2, and four are E3/E4.  Twelve galaxies have large, $\sim$ $90\degr$ inclination angles, and two are at $i$ $\sim$ $70\degr$. None have any wiggles or dips in the central regions of their SB profiles. These considerations ensure that the contribution of semi-konus densities to the density profile is small, or non-existent. Further, our multi-component models have residuals consistent with measurement errors over a large dynamic radial range that allows for a parametric deprojection.
\clearpage
\pagebreak
\begin{minipage}{2.0\columnwidth}
  We conclude that for the $14$ galaxies presented in this section, the 3D intrinsic density profile can be described with a multi-component Einasto model. These galaxies span a wide range of luminosities $-24<M_{VT}<-15$ and belong to both the steep-cusp and shallow-cusp families. It is therefore likely that the intrinsic 3D baryonic density of other ellipticals can also be described with a multi-component Einasto model.
  \subsubsection{Considerations for detailed modelling}
  Here we note a few important factors that may affect our interpretation of the intrinsic luminosity profiles.
  
  i) We have assumed a constant ellipticity while galaxies seldom have constant ellipticity. A varying ellipticity can be incorporated in our multi-component models for a more accurate deprojection.

  ii) We show luminosity density profiles of $14$ galaxies which we could justify as oblate axisymmetric systems. However, some galaxies are prolate while some are triaxial and some have a combination of axisymmetric and triaxial regions. For such galaxies, the 3D intrinsic density should not be inferred from a SB analysis alone. One may assume a multi-component Einasto profile for the intrinsic density and use additional kinematic information to include triaxiality in the kinematics to constrain the Einasto profile parameters either in 3D or in 2D through the DW-function.

  iii) The luminosity density profiles are subject to uncertainties in estimating the absolute magnitude (which are subject to uncertainties in distance measurements) and Galactic extinction. These uncertainties however do not affect the overall shape of the density profile.
\end{minipage}
  \begin{figure*}
    \includegraphics[width=142mm]{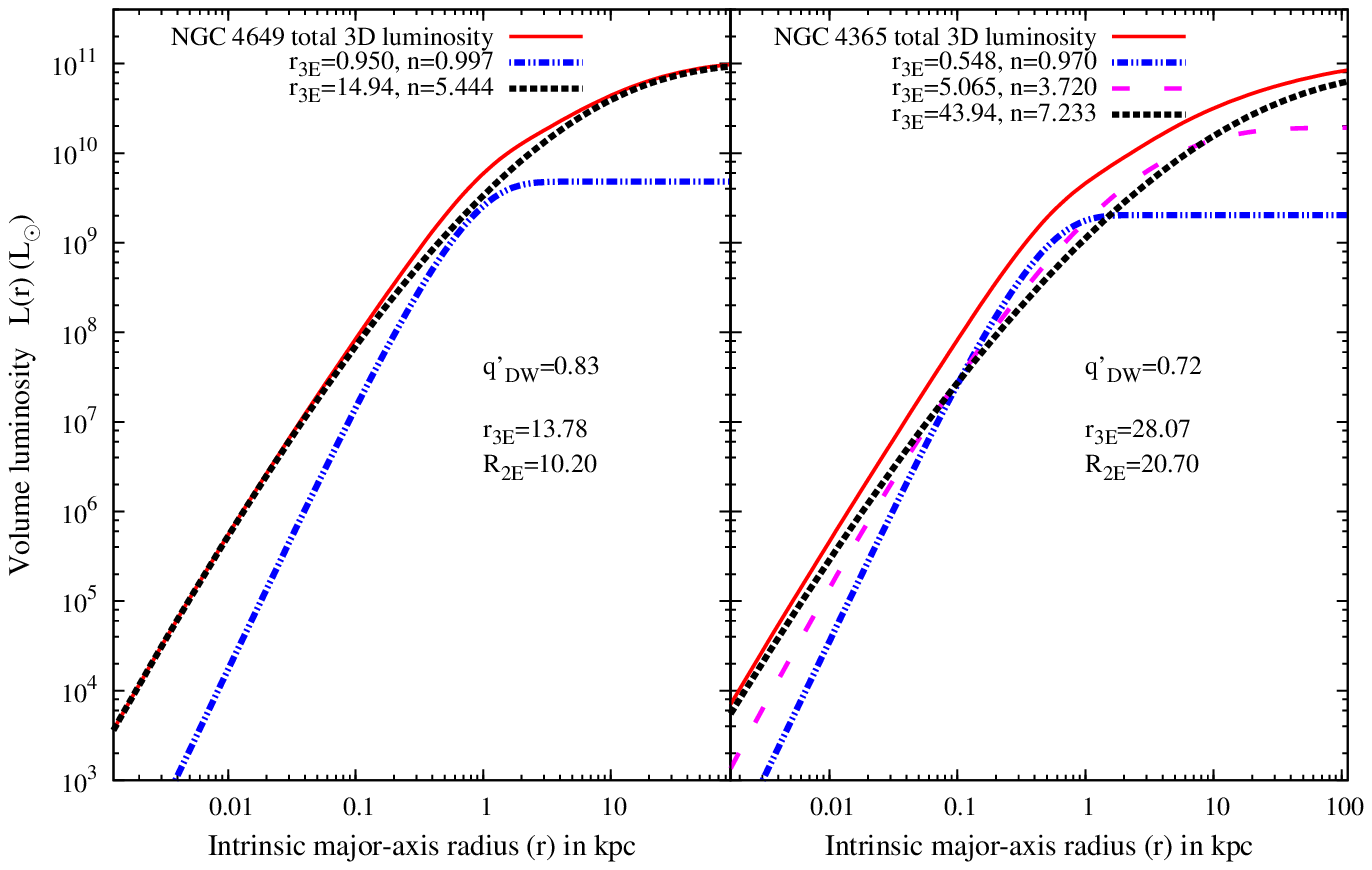}
    \caption{Cumulative volume luminosity profile for NGC4649 (left) and NGC4365 (right). Refer to caption of Fig.\ref{cumlight1} for details.}
  \end{figure*}
  \begin{figure*}
    \includegraphics[width=142mm]{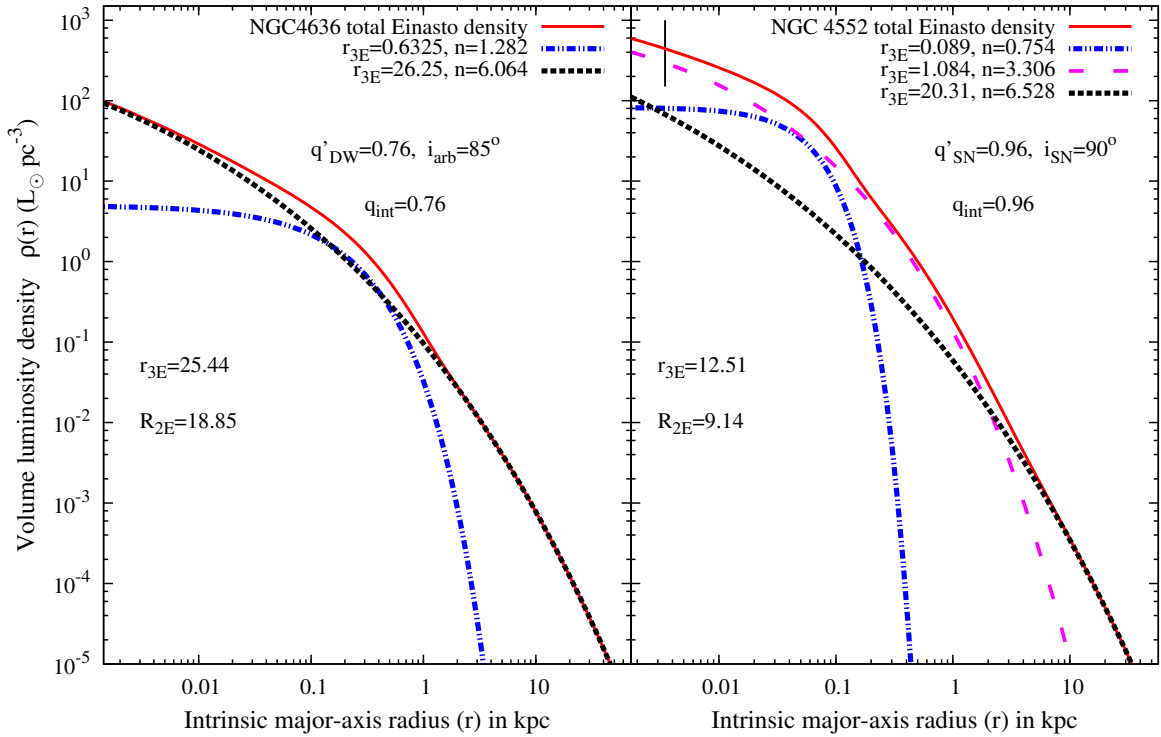}
    \caption{Intrinsic (3D) luminosity density for NGC4636 (left) and NGC4552 (right). The vertical marker on the profile for NGC4552 shows the inner radius beyond which the luminosity density profile shown, can be trusted. This is because the best fitting $3$-component DW model for this galaxy (Fig.\ref{NGC4552}) is not a good representation of the surface brightness profile within $0.045~arcsec$ where the light is affected by the variable UV flare activity interpreted to arise from a low-level AGN ({\protect \cite{Renzini95}}, {\protect \cite{Cap99}}). Also refer to caption of Fig.\ref{lumden1} for details.}
    \label{N4552lumden}
  \end{figure*}
  \begin{figure*}
    \includegraphics[width=142mm]{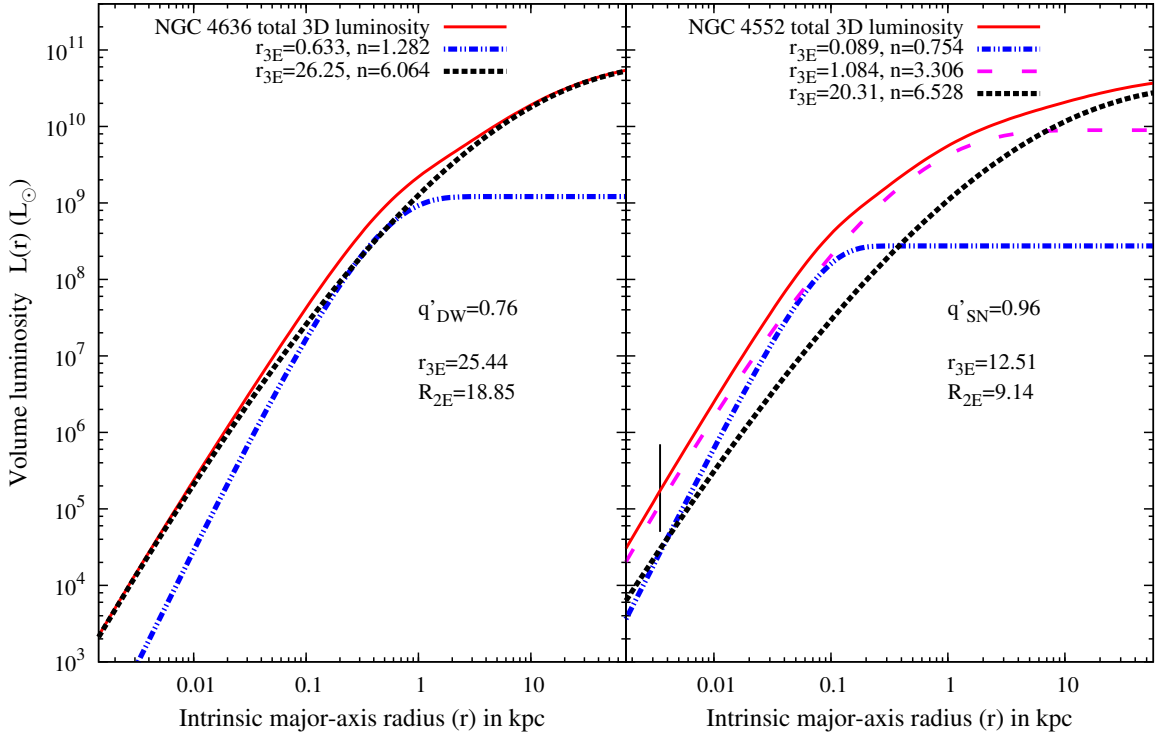}
    \caption{Cumulative volume luminosity profile for NGC4636(left) and NGC4552(right). The vertical marker on the profile for NGC4552 shows the inner radius beyond which the profile shown, can be trusted. This is because the best fitting $3$-component DW model for this galaxy (Fig.\ref{NGC4552}) is not a good representation of the surface brightness profile within $0.045~arcsec$. Refer to caption of Fig.\ref{cumlight1} and Fig.\ref{N4552lumden} for details.} 
  \end{figure*}
  \begin{figure*}
    \includegraphics[width=142mm]{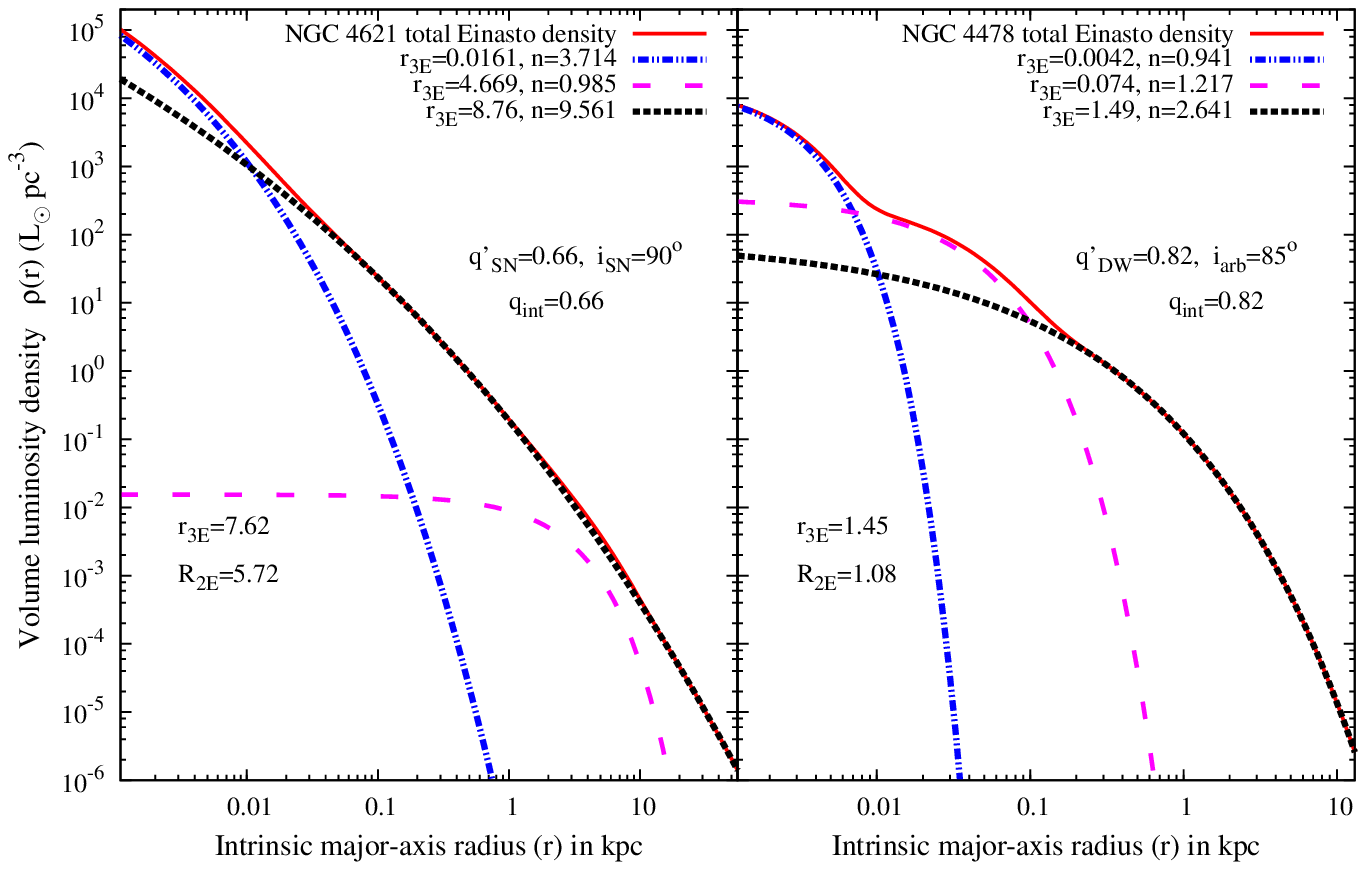}
    \caption{Intrinsic (3D) luminosity density for NGC4621 (left) and NGC4478 (right). Refer to caption of Fig.\ref{lumden1} for details.} 
  \end{figure*}
  \begin{figure*}
    \includegraphics[width=142mm]{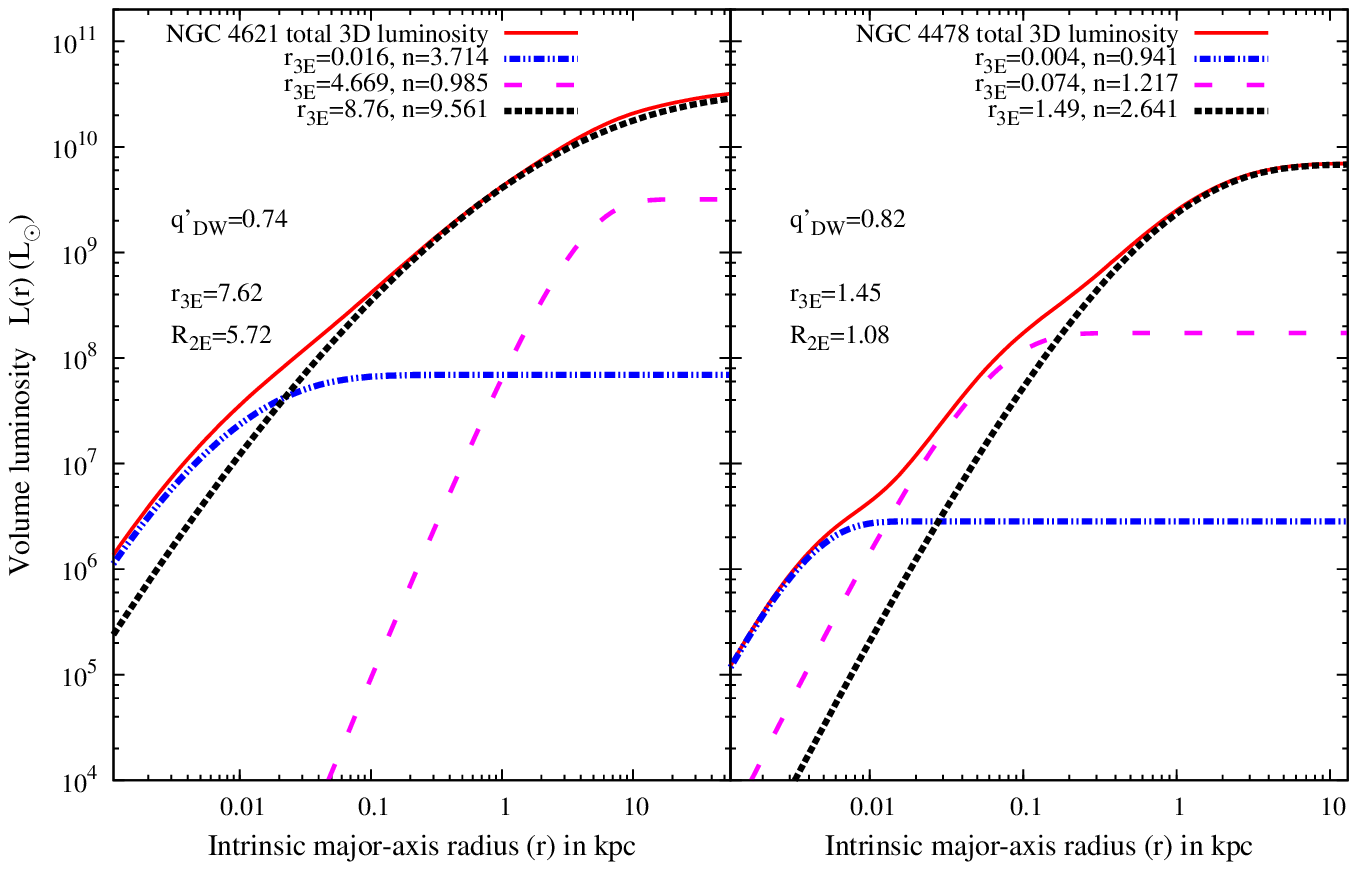}
    \caption{Cumulative volume luminosity profile for NGC4621 (left) and NGC4478 (right). Refer to caption of Fig.\ref{cumlight1} for details.}
  \end{figure*}
  \begin{figure*}
    \includegraphics[width=142mm]{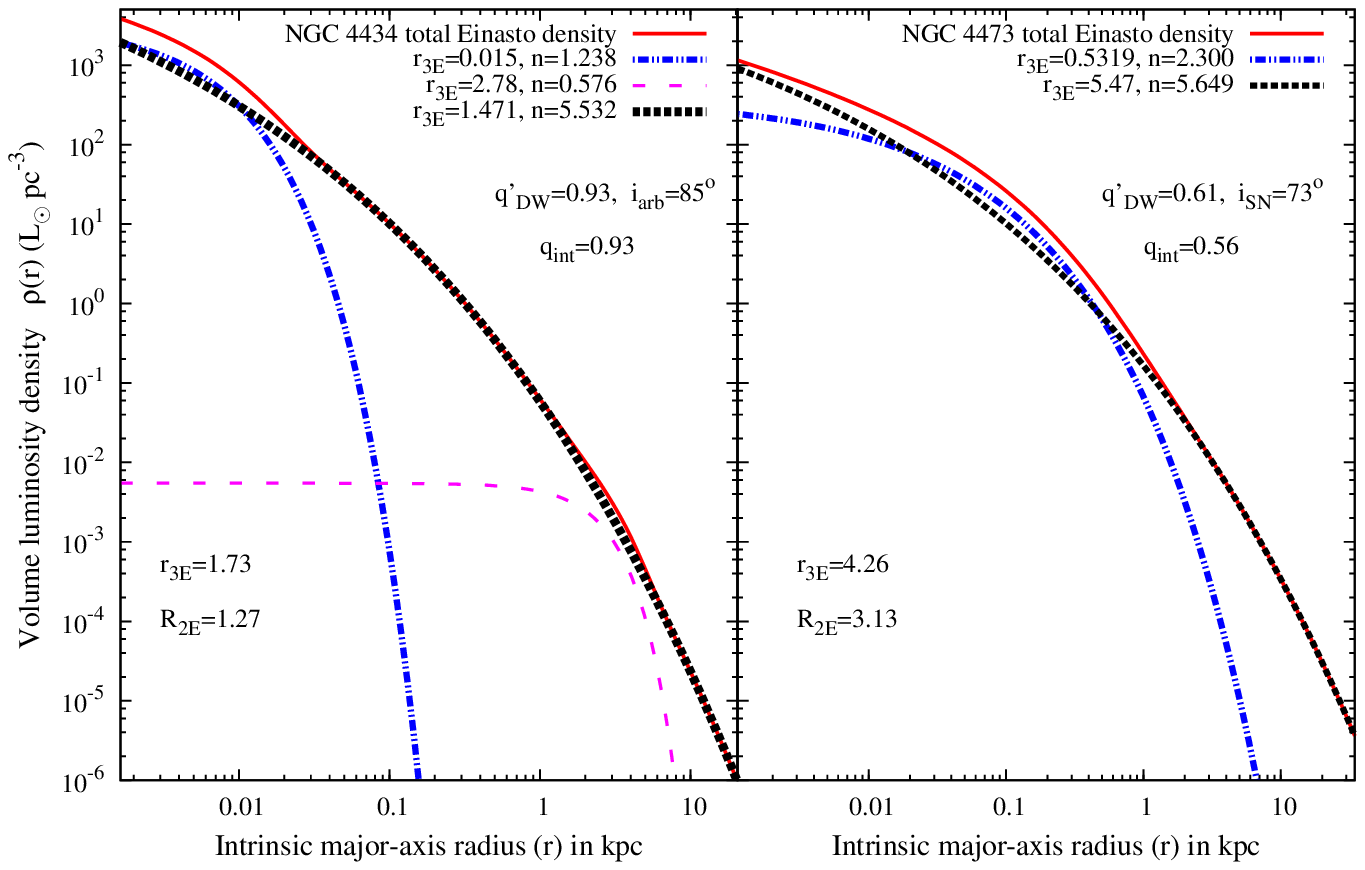}
    \caption{Intrinsic (3D) luminosity density for NGC4434 (left) and NGC4473 (right). Refer to caption of Fig.\ref{lumden1} for details.} 
  \end{figure*}
  \begin{figure*}
    \includegraphics[width=142mm]{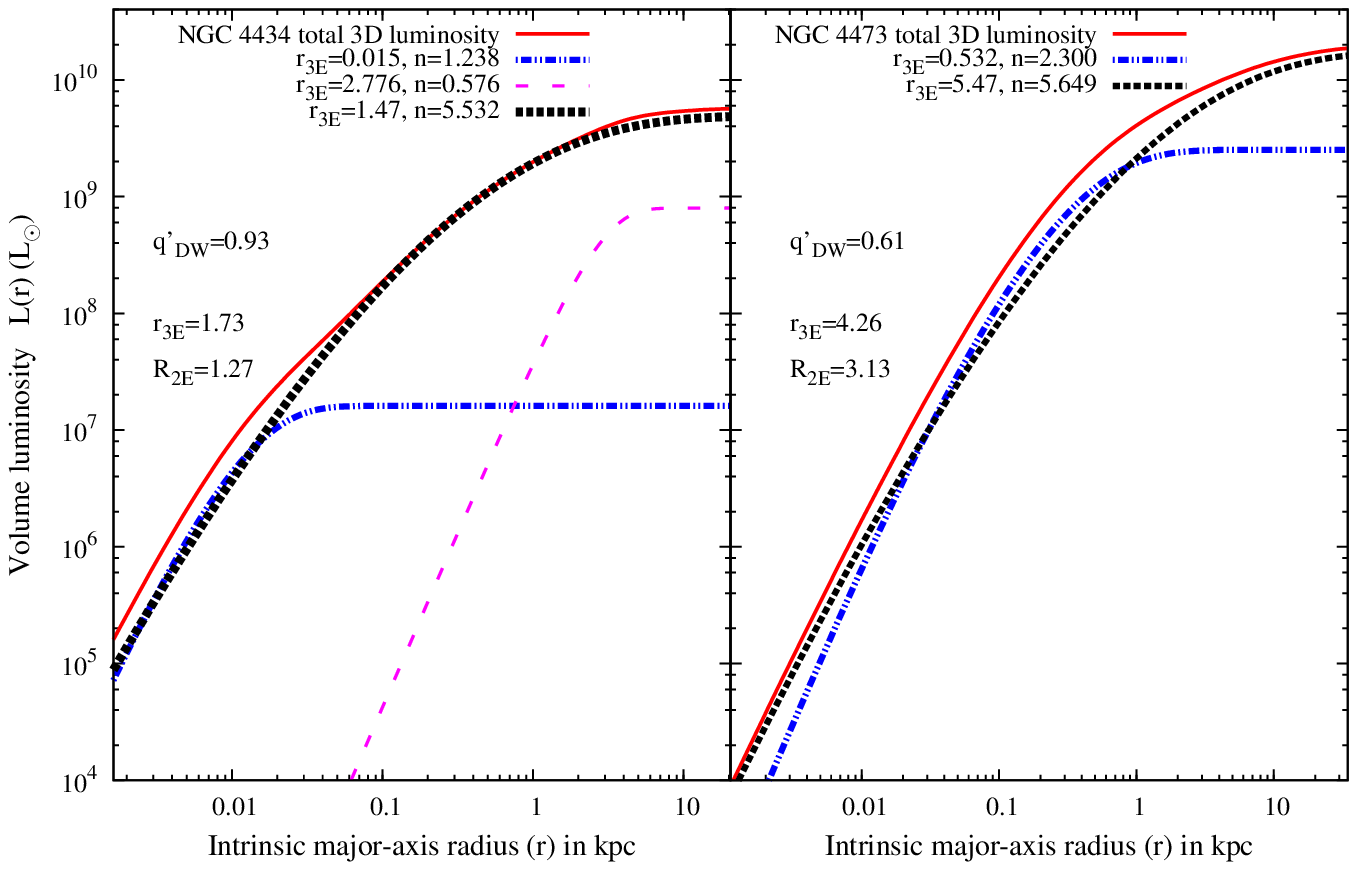}
    \caption{Cumulative volume luminosity profile for NGC4434 (left) and NGC4473 (right). Refer to caption of Fig.\ref{cumlight1} for details.} 
  \end{figure*}
  \begin{figure*}
    \includegraphics[width=142mm]{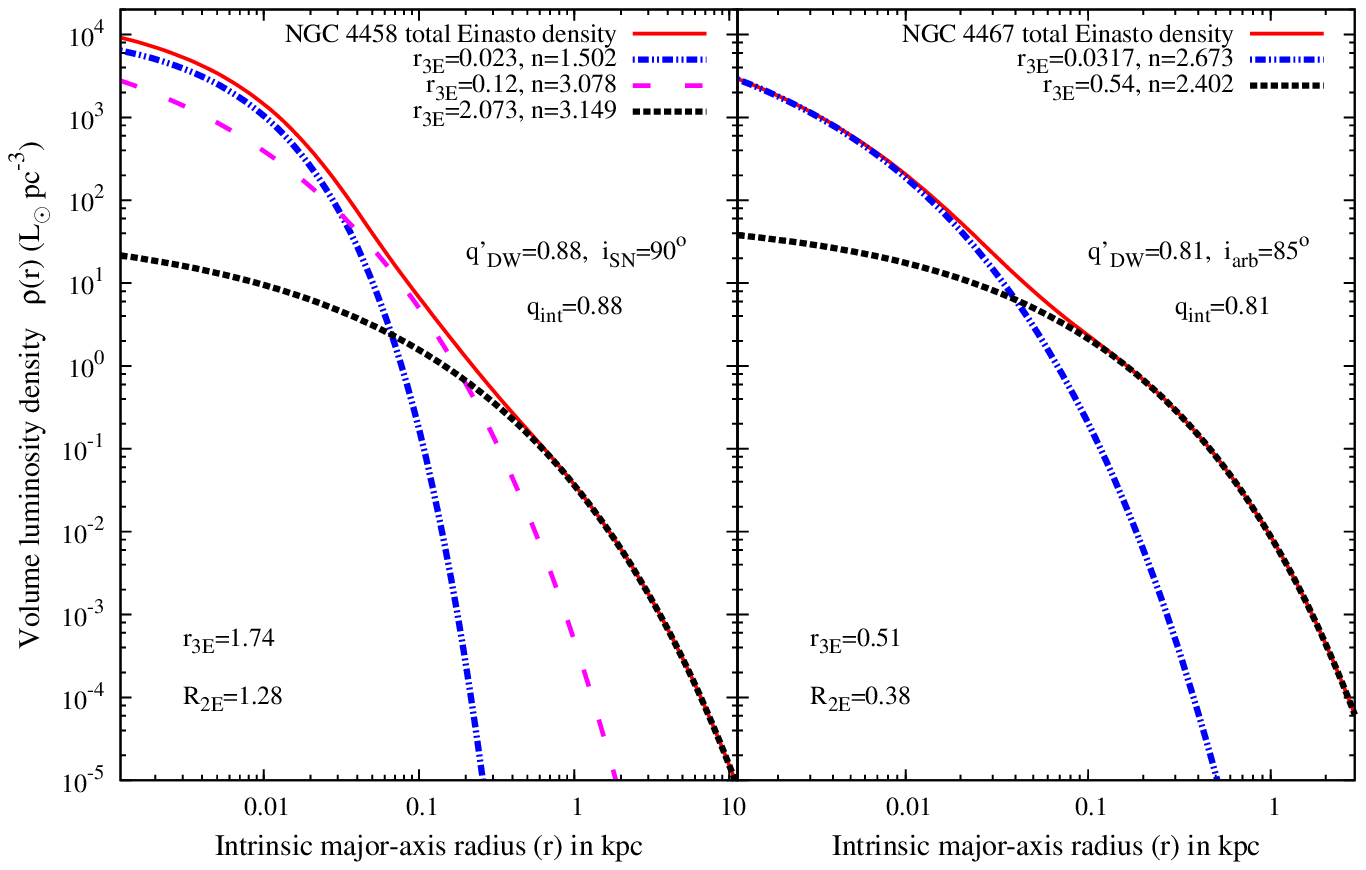}
    \caption{Intrinsic (3D) luminosity density for NGC4458 (left) and NGC4467 (right). Refer to caption of Fig.\ref{lumden1} for details.}
  \end{figure*}
  \begin{figure*}
    \includegraphics[width=142mm]{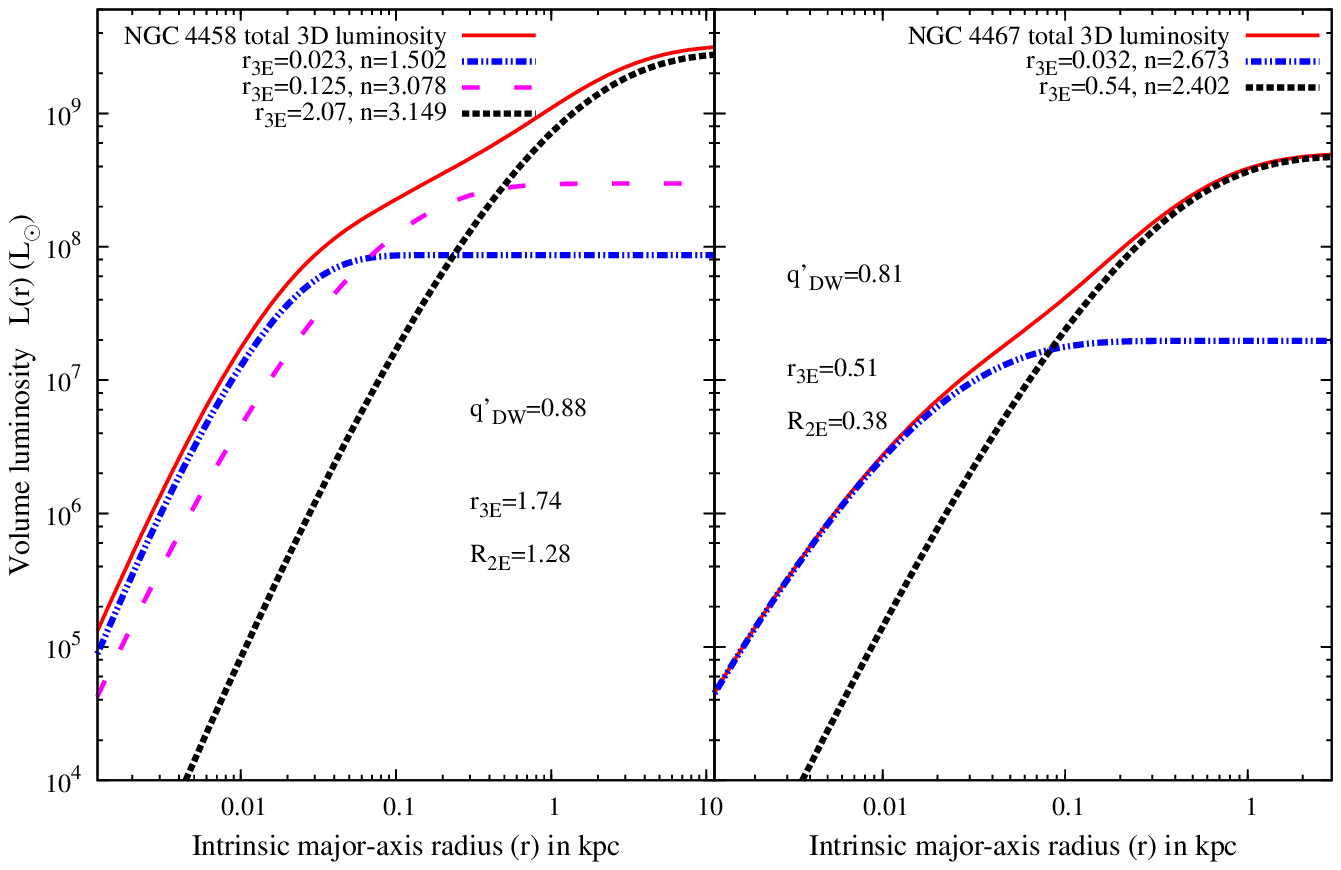}
    \caption{Cumulative volume luminosity profile for NGC4458 (left) and NGC4467 (right). Refer to caption of Fig.\ref{cumlight1} for details.} 
  \end{figure*}
  \begin{figure*}
    \includegraphics[width=142mm]{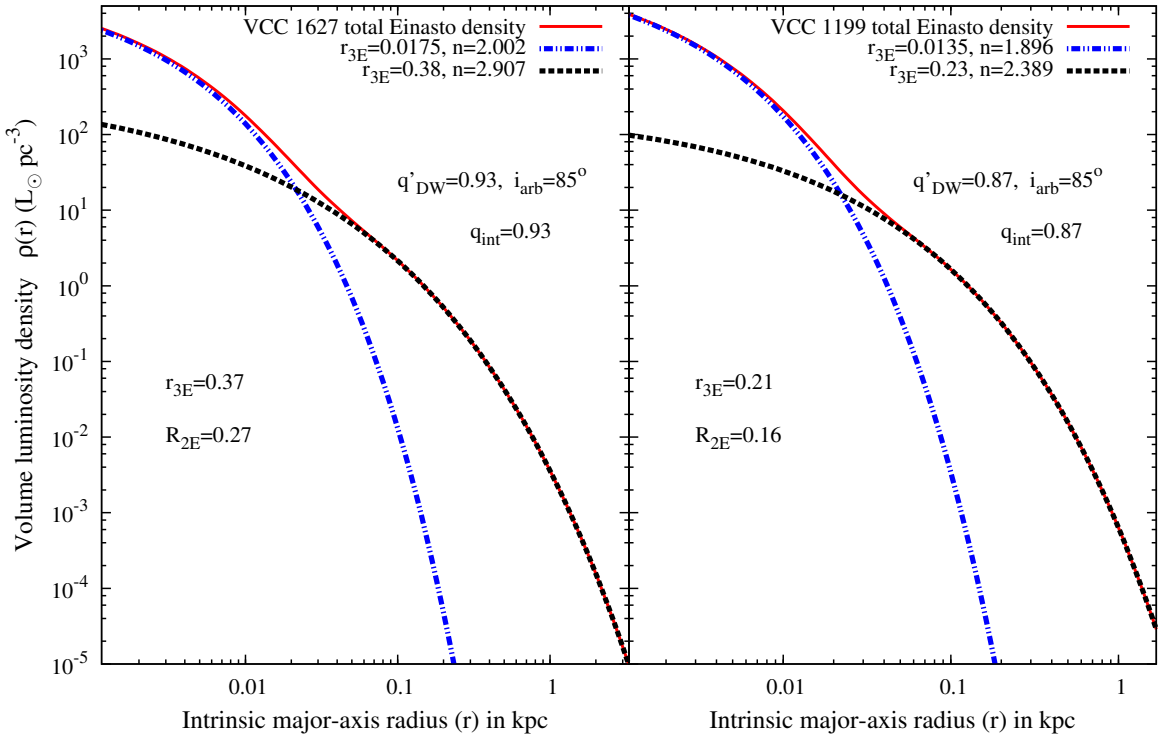}
    \caption{Intrinsic (3D) luminosity density for VCC1627 (left) and VCC1199 (right). Refer to caption of Fig.\ref{lumden1} for details.}
  \end{figure*}
  \begin{figure*}
    \includegraphics[width=142mm]{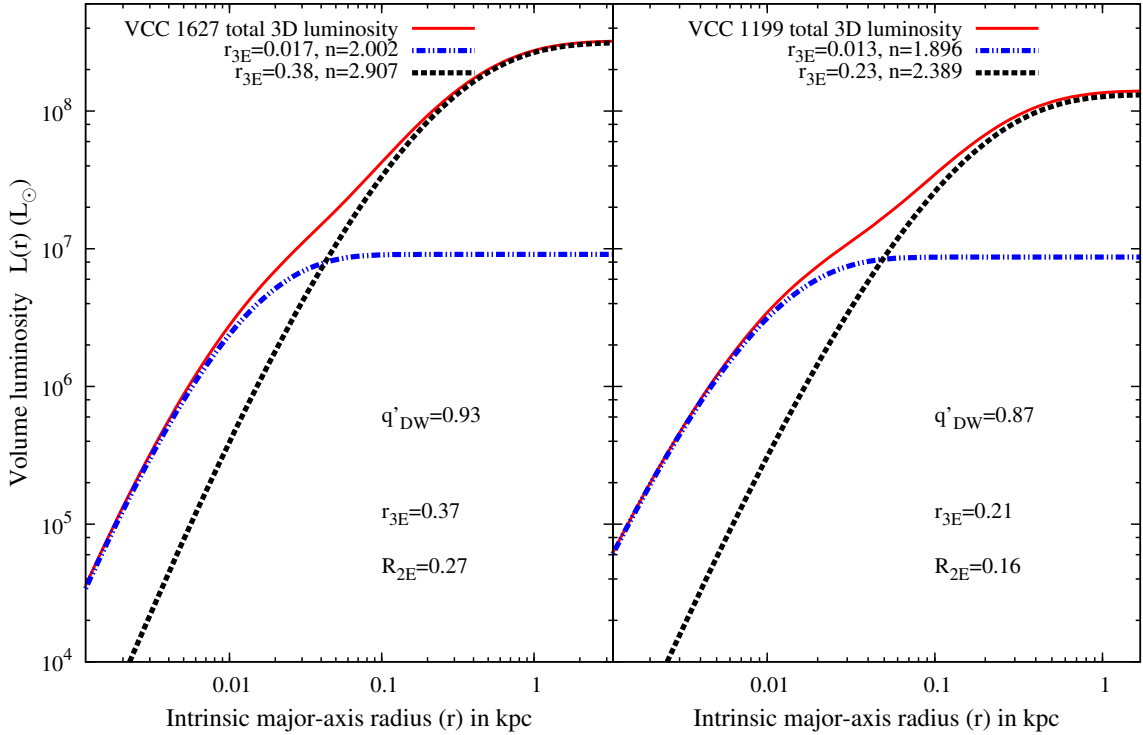}
    \caption{Cumulative volume luminosity profile for VCC1627 (left) and VCC1199 (right). Refer to caption of Fig.\ref{cumlight1} for details.}
    \label{lumdenlast}
  \end{figure*}
\pagebreak
\section{Mass deficit in massive ellipticals}\label{massdeficit}
\cite{Peebles72} predicted that a compact massive object, of mass $M$, can alter the density structure around its radius of influence,
\begin{align}\label{rhPeebles}
 r_{inf} = GM/\sigma_{*}^2
\end{align}     
such that for $r<r_{inf}$ a stellar density cusp (in 3D) $\propto$ $r^{-9/4}$ forms with a velocity dispersion $\sigma$$(r)$ $\propto$ $r^{-1/2}$; and a core, of constant density and velocity dispersion, forms within a core-radius $r_{c}$$>>$$r_{inf}$. Here $\sigma_{*}$ is the velocity dispersion at $r$$>>$$r_{inf}$.

\cite{BBR80} discuss the formation, evolution, eventual coalesence and possible recoil, of a binary SMBH, as a result of galaxy mergers. Since then, N-body simulations have shown that the evolution of the binary leads to a co-evolution of the density profile of the galaxy, in a region around the radius of influence, due to ejection of an amount of mass proportional to the mass of the binary.

 Our work, describing the structure of ellipticals as a superposition of DW-profiles, implies the presence of {\it excess light} in the centre of all ellipticals with respect to an inner extrapolation of the outer DW-components. We emphasize that our models describe the current structure, post any modification by the binary, and can be consistent with mass ejection by SMBHs.

 In this section we discuss a related concept of mass deficit in massive ellipticals, which has its genesis in - 

 i) the identification of massive ellipticals exhibiting a shallower surface brightness (SB) profile in their central regions, inwards of a projected break radius, with respect to the trend of SB outside the break radius - i.e., a light deficit. This is in contrast to less massive ellipticals that seem to contain steeper light profiles inside the break radius - i.e., a light excess; 

ii) the assumption that evolution of the binary SMBH is the sole factor responsible for the observed break radius; and 

iii) the assumption that in the absence of the binary SMBH, the trend (functional form) of SB outside the break radius would have continued its form all the way to the centre of the galaxy.

In section \ref{Mdefobservations} we show that {\it estimates} of 'observed' deficit in real galaxies, made with respect to an inner extrapolation of the SB profile outside the break radius, not only have large variation with respect to predictions from N-body simulations but also vary widely between different researchers for the same galaxy. We also compare signatures of mass-ejection by the binary SMBH with the observed profiles of real massive ellipticals and find that current predictions from N-body simulations are not able to account for some of the largest 'cores' in Virgo galaxies. Prior to such discussion, in section \ref{phases} we briefly describe the various phases of evolution of the binary SMBH and in section \ref{Mdefsimulations} we review the predictions from simulations about the amount of mass ejected, spatial extent and resulting slope of density profiles, in the context of 'dry' dissipationless (gas-free) mergers, believed to be the formation mechanism for the massive 'core' ellipticals. The discussion is thus directed towards systems with the most massive $\gtrsim$ $10^{8}$ $M_{\sun}$ SMBHs, as is the case for the 'core' ellipticals in Virgo. 
\subsection{Phases of Evolution}\label{phases}
 The significant phases in the evolution of the binary SMBH are:

{\it Phase 1}: Subsequent to a galaxy-galaxy merger, a SMBH binary is said to form when the SMBH separation reduces to a distance $r_{inf}$ - the radius of influence of the more massive SMBH.

{\it Phase 2}: Their separation shrinks as the binary loses energy due to dynamical friction and by scattering of stars through the gravitational sling-shot mechanism. A 'hard-binary' is eventually defined to form when the separation reduces to a size,
\begin{align}\label{ahard}
   a_{hard} = \frac{G \mu}{4~\sigma_{*}^{2}} = \frac{q}{4~(1+q)^{2}}\frac{G M_{12}}{\sigma_{*}^2}
 \end{align}  
 where, $\mu=$$M_{1}M_{2}/M_{12}$ is the reduced mass of the binary with $q=$$M_{2}/M_{1}$$\leq$1 and $M_{12}$$=$$M_{1}+M_{2}$ is the combined total mass of the binary. $a_{hard}$ is thus a fraction of $r_{inf}$ (equation \eqref{rhPeebles})

 At this stage, due to depletion of stars, the decrease in binary separation can stall unless additional mechanisms can continue removing energy from the binary to drive it towards Phase 3. \cite{BMSB06}, \cite{HL07}, \cite{BPBMS09}, \cite{KJM11} show mechanisms through which stalling may easily be avoided. Also, see \cite{DSD11} for a review and references therein.

{\it Phase 3}: The binary then continues to harden until its semi-major axis reduces to a size $a_{gw}$ where energy loss due to emission of gravitational waves begins to dominate energy losses due to scattering of stars, driving the binary towards coalesence.

{\it Phase 4}: During the last few stable orbits before coalesence, the centre of mass recoils with a recoil or kick velocity that sensitively depends on the binary mass-ratio and spins, relative spin allignement and orientation of spin-axis with the angular momentum vector. If the kick velocity $V_{kick}$ $>$ $V_{esc}$, the escape velocity, the coalesced SMBH can be completely removed from the galaxy. For lower kick velocities, the SMBH performs damped oscillatory motion in the central regions and eventually settles down to a Brownian motion about the centre.
\subsection{Results of Simulations}\label{Mdefsimulations}
\subsubsection{Mergers in hierarchical models of galaxy formation}

\cite{VHM03} (V+03a) and \cite{VMH03} (V+03b) have shown that, in galaxy mergers:

i) SMBHs of mass ratio $q$$=$1 occur at high redshift $z$$\sim$20, while for 0$\lesssim$$z$$\lesssim$14, $q$$<$0.2 with typical values of $q$$\sim$0.1.

ii) Assuming that the progenitors start with $\rho \propto r^{-2}$ profiles and mergers (accompanied by coalesence of the binary) not only erase the cusps completely to form a constant density core but also preserve such cores, they estimate that at the end of a succession of mergers the resulting mass deficit, $M_{def}$, is given by $M_{def}$$=$8.2$\pm$3.8) $M_{BH}$ (equation 15 of V+03b). Note that most 'core' galaxies have a shallow cusp rather than a constant density core. The above estimate should therefore be considered a limiting value. At the same time, mass deficit due to the scouring effect of SMBHs in Phase 4 is not accounted for in this study.

iii) Even when the entire $r^{-2}$ cusp is erased and a constant density core forms, the resulting core size is only $\sim$60 pc for a $10^{13}$$M_{\sun}$ halo (fig.1 and 2 of V+03b) -- a mass resembling the most massive 'core' ellipticals in Virgo. For example, assuming $M_{halo}/L_{V}$$=$50 the halo mass of M87, the giant 'core' galaxy in Virgo with $L_{VT}$$=$1.53$\times$$10^{11}$$M_{V\sun}$, is $\sim$7.65$\times$$10^{12}$$M_{\sun}$. The break radius for this galaxy is however $\sim$600 pc$>>$60 pc core formed by the merging binaries; a point we will return to in the next section. V+03b also find that such core sizes are typically $\sim$$3GM_{BH}/\sigma^2_{*}$$=$$3r_{inf}$. 
\subsubsection{N-body estimates of $M_{def}$ before coalesence}
Using \cite{Dehnen93} models -- a power-law of the form $\rho(r)$ $\propto$ $r^{-\gamma}$ at small $r$ -- to describe the intial density profiles, \cite{Merritt06} conducted 39 N-body simulations to study the evolution upto the hard-binary stage (Phase 2), for the case of 'dry' mergers in non-rotating spherically symmetric systems, including nine remergers. He showed that, after each merger:

i) $M_{def}$$=$0.7$q^{0.2}$$M_{12}$. For initial density profiles as Dehnen $\gamma$$=$0.5 model and for typical values of $q$$=$(0.1,0.25) (V+03a), $M_{def}$$=$(0.33,0.46)$M_{12}$ (table 1, Merritt06).

ii) there is a lowering of the density profile in a region about the size of the radius of influence. This can be seen by comparing $r^{'}_{h}$ from his table 1 with the radius at which reduction in density occurs after the {\it first} merger in his figs. 6a,6b and 6c.

\cite{MMS07}, MMS07, study the evolution in Phase 3. For intial Dehnen density profiles with $\gamma$$=$0.5, and for $q$$=$0.1, and $M_{12}$$=$$10^{8}$$M_{\sun}$, they obtain $M_{def}$$=$0.9$M_{12}$. When added to the mass deficit upto Phase 2, this yields $M_{def}$$\sim$1.2$M_{12}$. Further, as can be seen from their figure 21, the net deficit at the end of Phase 3 is also within a region $\sim$ $r_{h}$ - the region of influence defined as the radius containing a stellar mass equal to twice the total black hole mass. Note that $r_{h}$ is typically much less than the observed break radius ('core') of massive ellipticals.
\subsubsection{N-body estimates of $M_{def}$ after coalesence}\label{Nbodyaftercoal}
Assuming that the binary coalesces and recoils with a kick velocity (Phase 4), \cite{BMQ04} and \cite{M+04} investigate the resulting effects on the density structure. They find that for $V_{kick}$$>$(0.25-035)$V_{esc}$, cores of size $r_h$ can form as a result of the kicks; with {\it larger} deficits for $V_{kick}$$<$$V_{esc}$.

\cite{GM08}(GM08) explore this phase in detail using a Core-Sersic(CS) profile (in 2D) and its deprojection \citep{TG05} for the density profiles. They found that:

i) For $0.3$ $V_{esc}$ $\lesssim$ $V_{kick}$ $<$ $V_{esc}$, the coalesced SMBH performs damped oscillations about the galactic centre and loses energy during each passage through the {\it pre-existing} 'core'; the later {\it assumed} to be the break radius of the CS profile.  

ii) the flattest profiles with CS $\gamma$$\lesssim$$0.05$, leading to the largest deficit, occurs for $V_{kick}$ $\gtrsim$ $0.8$ $V_{esc}$ (their fig.12). It is important to note that such flat profiles are produced given an initial profile with a fairly low $\gamma$$=$$0.55$. i.e., it is not clear whether such flat profiles can be produced for larger initial $\gamma$, even with near $V_{esc}$ kicks.

iii) the resulting $M_{def}/M_{12}$ $\sim$ 5.08 $(V_{kick}/V_{esc})^{1.75}$, in Phase 4, and depends weakly on $M_{12}/M_{gal}$; where $M_{gal}$ is the mass of the galaxy. Consequently for $V_{kick}$ $=$ (0.1,0.4,0.8)$V_{esc}$,  $M_{def}$ $=$ (0.1,1.0,3.44)$M_{12}$. Further, $V_{kick}/V_{esc}$ decreases with galaxy luminosity. Hence the lower values of $M_{def}$ are more likely for the massive shallow cusp ('core') galaxies; for example, GM08 show that $V_{kick}$$=$0.4$V_{esc}$$=$$550$$km$ $s^{-1}$ is a likely kick-velocity for the most massive core galaxy in Virgo, NGC4472 (M49). 

Self-consistently combining $M_{def}$ from all phases, for mergers of $\sim$$10^8$$M_{\sun}$ SMBHs with typical $q$$=$0.1 and $V_{kick}/V_{esc}$$=$0.4, the total cumulative $M_{def}$ is 2.2$M_{BH}$ per merger. We shall show in the next section that {\it estimates} of $M_{def}$ of $10$,$20$ and sometimes $40$$M_{BH}$, from high-resolution SB profiles, cannnot be explained by the above results of N-body simulations.

iv) and, the radius of the {\it pre-existing} 'core' (here, break radius) expands by $15$ per cent for $V_{kick}$$=$0.2$V_{esc}$ to about $70$ per cent for $V_{kick}$$=$0.9$V_{esc}$ and is {\it lesser} for larger $M_{BH}$(refer to their table 1 and 3). Note that in the simulations upto Phase 3, the radial extent of the core ($\sim$$r_{h}$) formed by the SMBHs is {\it less than} the break radius of massive 'core' ellipticals. Since the break radius has been assumed to correspond to the core formed by the binary through Phase 3, the fractional expansion in size of the break radius is more relevant than the absolute eventual size of the break radius.
\subsection{Estimates of observed mass deficit}\label{Mdefobservations}
In this section we investigate whether the signatures of mass deficit in N-body simulations are in congruence with the {\it estimates} of mass deficit from SB profiles of galaxies and highlight limitations of existing models of galaxy structure - a single sersic, a core-sersic and a Nuker profile - in estimating such mass deficits as well as in confirming predictions from simulations.

In real galaxies, estimates of the amount of mass deficit and the extent of break radius vary widely for the same galaxy. This can be seen by comparing the mass deficits computed in \cite{Graham04} (their table 1), with data from Rest et al.(2001).

For instance, for NGC4168, Graham04 estimated a projected break radius, $R_b$, of 0.72 arcsec (108 pc) and a $M_{def}$$=$1.2$M_{BH}$ using a CS profile, while estimates with a Nuker profile yield $R_b$$=$2.02 arcsec (303 pc) and a $M_{def}$$=$23.5$M_{BH}$; for the later, the prescription in \cite{MM01} assuming that the unscoured intrinsic (3D) profile within the break radius was $\rho(r)$$\propto$$r^{-2}$, has been used. However, \cite{MMRvB02} (MMRvB02) estimate an $R_{b}$$=$1250 pc (their table 1) using the same data of Rest et al., and obtain $M_{def}$$=$48$M_{BH}$. Not only do these estimates differ by a factor of $40$ for the same galaxy, even the most extreme estimates of $M_{def}$ from N-body simulations are not able to account for deficits as large as (20-50)$M_{BH}$.

For NGC2986, MMRvB02 obtain a lesser estimate of $M_{def}$$=$5.24$M_{BH}$ from a {\it larger} break radius of 400 pc compared to the (Nuker,CS) profile estimates (Graham04) of $M_{def}=$(26.7,7.02)$M_{BH}$ over a {\it smaller} $R_{b}=$(174,94) pc. 

Assuming $r^{-2}$ initial profiles, MMRvB02 show that on average $M_{def}$$=$$10$ $M_{BH}$ for the case of ellipticals. This estimate is similar to the predictions of V+03a ($\sim$ 8 $M_{BH}$) and the average value of 10 $M_{BH}$ in KFCB09 for Virgo ellipticals. However they do not agree with the (2.4$\pm$1.8)$M_{BH}$ estimates of \cite{F06} (F+06) for the Virgo ellipticals, the (2.1$\pm$1.1)$M_{BH}$ estimate of Graham04, the 2 $M_{BH}$ estimate of \cite{HL07} (HL07), including Virgo ellipticals, and the (2.29$\pm$0.67) $M_{BH}$ and (1.24$\pm$0.3) $M_{BH}$ estimates of \cite{HBSN08}.

For the giant 'core' elliptical in Virgo, M87(NGC4486), HL07 estimate 2.5 $M_{BH}$, MMRvB02 estimate 8.7 $M_{BH}$ and \cite{KB09} (KB09) estimate 14.1 $M_{BH}$, all scaled to $M_{BH}=$3.5$\times$$10^{9}$$M_{\sun}$. Despite such large deficits, the SB profile of M87 shows a distinct rising trend; apart from the fact that it has an unsually large break radius. NGC4649 also has a fairly large break radius. While KB09 estimate 5$M_{BH}$, HL07 estimate only 1 $M_{BH}$, both for $M_{BH}=$2$\times$$10^9$$M_{\sun}$. Another Virgo galaxy with an apparent 'core' is NGC4382. KB09 estimate a deficit of 13 $M_{BH}$, however \cite{Gultekin11} find that this galaxy has an unusually small black hole mass, consistent with no black hole, yet obtain a mass deficit of 45.6 $M_{BH}$.

Estimates of $M_{def}/M_{BH}$ do depend on a number of factors like uncertainties in estimates of $M_{BH}$, the assumed form of $M/L$ etc. However, the large differences are generally due to discrepancy in -- estimation of $R_{b}$; estimation of the inner power-law index; and the huge uncertainty in our assumptions about what {\it might have been} the shape of the density profile prior to the action of the SMBHs. For instance, KFCB09 defines the central region of light deficit or 'extra-light' based on the region over which a Sersic fits their estimate of the non-central region. F06 on the other hand fits a Core-Sersic model which defines the central region as the region inside $R_b$ of the core-sersic model. Consequently the two methods yield widely varying estimates of what the 'central region' of deficit is. This leads to a large discrepancy in the estimated $M_{def}$. Also see fig.2 of MMRvB02.

 \cite{HH10}(HH10) suggest a new non-parametric method to obtain mass deficit and find that $M_{def}$ varies with radius such that at $\sim$ $100$ pc, $M_{def}/M_{BH}$ '{\it assymptotes to a maximum of 0.5-2}' and for the largest galaxies at core-radius of $\sim$ kpc, $M_{def}$$=$(2-4)$M_{BH}$. While, this is similar to the estimates of F+06 and Graham04, it is at odds with the KFCB09 estimates where the authors reasoned that $M_{def}$ $\sim$ $10$ $M_{BH}$ accounts for the net mass deficit through Phase 4 (results from MMS07 and GM08) and in agreement with V+03a and MMRvB02.

As discussed in section \ref{Nbodyaftercoal}, $M_{def}$$=$(10-20)$M_{BH}$ are hard to explain using the results of N-body simulations. However, even if such were to be true, the resulting region over which KFCB09 and MMRvB02 estimates such deficit is much larger than the region of influence of the SMBH over which simulations predict mass deficits; a point also noted in MMRvB02. This is also true for estimates of $R_b$ using a Core-Sersic or Nuker model; especially for the largest 'cores' as in NGC4486 and NGC4649. The shallow cusps ('core') are thus not entirely due to mass ejection by the SMBHs.

\subsection{Deficit or excess?}
Disagreements in the literature are limited not only to the amount, but also the sign of the deficit. In MMRvB02, NGC4478 is shown to have a mass deficit of 15.85 $M_{BH}$. However, its SB profile shows an apparent 'cusp' (see Fig.\ref{NGC4478}, of this paper), due to which F+06, Cote et al.(2007) (C+07), KFCB09 infer this galaxy to have a light (and mass) excess. Similarly in MMRvB02, NGC4473 has the highest $M_{def}$$=$22.39$M_{BH}$ amongst the Virgo ellipticals. F+06 and C+07 also conclude that this galaxy has a mass deficit. But KFCB09,  argue that this galaxy has 'extra-light' and not a deficit. Further, in this paper and in Hopkins et al. (2009 a,b), the authors argue that all galaxies can be modelled with extra light above an inner extrapolation of the outer components.

 Thus, whether the SB profile of a galaxy 'exhibits' an excess or deficit also depends on ones methodology.

\subsection{Discussion}

 Given the large disagreements in estimates of $M_{def}$; the region affected by the binary SMBHs; and which profile best describes the light of ellipticals; it can not be said that robust estimates of mass deficits, in real galaxies, have been made. Consequently, validating predictions of N-body simulations and especially the stage(s) of SMBH evolution responsible for the observed deficit turn out to be largely uncertain.

 Further, extrapolating the profile from the break radius, $R_b$, presents a number of conceptual difficulties:

i) When the SB profiles of the massive 'core' galaxies are modelled with the Core-Sersic (F+06) or single Sersic (KFCB09) profiles, the non-central regions, beyond $R_b$, generally have Sersic index $m$ $>$ $5$; sometimes as large as $9$ and $11$. Larger the $m$, steeper the density as $R\to0$ and consequently larger the extrapolated density. Since estimates of $M_{def}$ are correlated with an estimate of $R_b$ and in most cases $R_b$ is much larger than the region of influence of the SMBHs (example: NGC4486, NGC4649), extrapolating profiles with large Sersic indices will invariably imply larger, but incorrect, mass deficits (also see HH10).

ii) The SMBHs are believed to affect only the central regions, around the radius of influence, and not the global profile of the galaxy. While it is believed that the mass deficit is due to erosion of the steep cusps, it remains to be explained what forms the large $m$$\sim$9 Sersic profiles, in the outer regions of only the 'core' galaxies. If SMBHs have not influenced these regions, some other processes have shaped the formation and evolution of the 'core' galaxies. Their shallow cusps could well be influenced by these processes as well and not entirelly due to the core-scouring effect.

Infact, F+06, KFCB09 and \cite{L07} observe that shallow- and steep- cusp galaxies have different characteristics, not just profile shapes, and are likely to have had different formation pathways. Their central regions could well have fundamentally different initial shapes as well, than being a mere extrapolation of the profile outside $R_b$. Further, baryonic effects are usually more dominant (over dark matter) in the central regions of all galaxies - core and cuspy alike - leading to relatively more concentrated central, than outer, regions (or components). This is precisely what our modelling indicates (section \ref{centcomp} and Fig.\ref{nMVT}). If this is true, it will not be meaningful to extrapolate the outer component inwards to estimate the 'mass deficit' and neither would comparing possibly fundamentally different profiles of 'cuspy' and 'core' galaxies, to get non-parametric estimates of mass deficits, as in HH10. 
 
iii) If the massive ellipticals have formed through multiple mergers, it is likely that their outer profile has also been built up as a cumulative effect. Consequently their Einasto or Sersic index ought to have evolved with their merger histories. Recall that the large indices, for the massive galaxies, are due to the gradual fall-off of their extended stellar light well outside the central regions. We are inclined to believe that this index was different in the past and has evolved over the merger history of these galaxies. The large spatial extent of these galaxies could also be indicative of a larger number of mergers, compared to the much smaller steep cusp galaxies; a point noted by KFCB09 as well. If this is true, and mergers alter the shape (Sersic index) of the outer profile, then extrapolating the current observed outer Sersic profile (index) to estimate the initial density profile will give us an incorrect estimate of the mass deficit.

In this paper, we have shown that all galaxies, shallow and steep cusps, can be modelled with very high precision over a large dynamic range, as a superposition of DW-components and that they all have a 'light excess' in their central regions with respect to an inner extrapolation of their outer components. As shown in section\ref{comparison}, not only do the DW-models fit the SB profiles better than the Core-Sersic and Nuker profiles, at least for three galaxies the central component correlates with a real physical system and is not a mere mathematical construction of the fitting process.  

Mass ejection and consequently some deficit due to core-scouring binary SMBHs could well have occurred and these are likely to have shaped the central components of our DW models as well. For instance, in this work, we find that the central component of all 'core' galaxies have an Einasto index $n$$\sim$1. It will be instructive to see if such components can be robustly isolated in SPH$+$N-body simulations and how well can they be fit with $n$$\sim$1 Einasto profiles. Similarly, as noted in section \ref{intcomp}, it is possible that the intermediate DW-components within the central kiloparsec or so, could have formed as a result of evolution of the binary SMBHs.

Recent advances in N-body simulations with large particle numbers is a promising development that can shed much light on the dynamical role played by the SMBHs. Unlike simulations, however, observations of real galaxies do not have the advantage of knowing what the unscoured profile was. Hence, estimates of such effects from observations should be made with caution and with respect to a robust model for galaxy structure. 

From the above discussion, we conclude that mass ejection due to SMBHs is unlikely to be the sole cause of the shallow-cusps in massive ellipticals. The role of processes shaping the global structure of galaxies should also be accounted for.
  \section{The outer $n$ of ellipticals}\label{BaryonsDM}
  In this section, we present two empirical speculations about the structure and formation of elliptical galaxies.
  
\begin{figure}
    \begin{minipage}{1.0\columnwidth}
      \includegraphics[width=84mm]{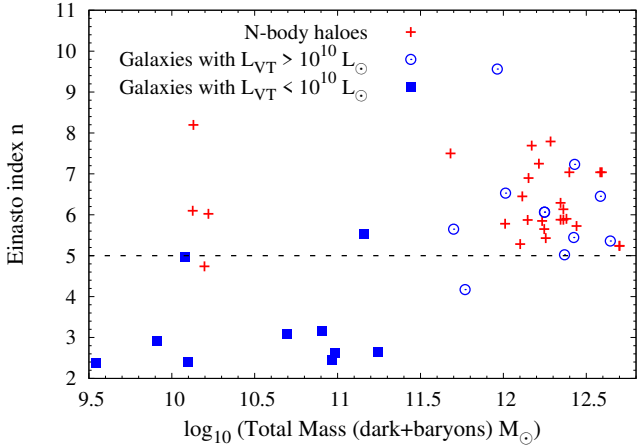}
      \caption{Einasto index $n$ of $\Lambda$CDM N-body haloes and that of the outer component of the Virgo galaxies in this paper. The outer component of the most luminous (massive) galaxies have an $n$ similar to that of the N-body haloes. This is not true for most of the lower luminosity ellipticals. The solid box at $n$$=$5.53 is NGC4434 and the open circle at $n$$=$9.56 is NGC4621 whose outer-$n$ may be uncertain (section~\ref{exceptions}). The other solid box at $n$$=$4.98 is VCC1440, a low luminosity galaxy bordering the dwarf-elliptical population. Not included are NGC4406 and NGC4382, whose outer components are severely perturbed. Refer to text (section~\ref{BaryonsDM}) for discussion. A colour version of this figure is available in the online edition.}
      \label{Ndarkbaryons}
    \end{minipage}
  \end{figure}

  (1)~~  Our multi-component models reveal an interesting property of the most luminous, $\gtrsim$$10^{10}$$L_{V\sun}$ galaxies in our sample. The Einasto shape parameter $n$ of their outer component is very similar to the $n$ of pure dark matter haloes; 5 $\lesssim$ $n$ $\lesssim$ 8, as shown in Fig.~\ref{Ndarkbaryons}.  We remind the reader that because dark matter haloes are well fit with Einasto profiles (N+04, M+06), a direct comparison with the $n$ of our galaxies is possible.\protect \footnote{Papers describing fits to the haloes in N-body simulations use the reciprocal of our shape parameter, $\alpha$$=$$1/n$, and typical $\alpha$ values hover around 0.17 which corresponds to $n$$=$5.88.}
  
  The plus symbols in Fig.~\ref{Ndarkbaryons} represent 26 galaxy-size dark matter haloes compiled from the high resolution  $\Lambda$CDM N-body simulations of \cite{Diem04}, \cite{Nav04}, \cite{M06}, \cite{Pr06}, \cite{SM09} and \cite{Nav10}. We assume that $M_{200}$ of these haloes is $80$ per cent of $M_{gal}$, the combined mass of the dark matter and baryons. To translate the stellar mass of our Virgo galaxies to $M_{gal}$ we have assumed (i) a stellar mass-light ratio $M_{\star}/L_{VT}$$=$5 for all galaxies, and (ii) the stellar mass comprises $20$ per cent of $M_{gal}$. Observe that the masses of the galaxies hosted by these dark matter haloes, $\sim$$10^{10}$-$10^{12}$$M_{\sun}$, are comparable to those of the low and high luminosity ellipticals in our sample, $-24$$<M_{VT}$$<$$-15$.
  
  If we interpret the dark matter simulation results to mean that Einasto shape parameter 5 $\lesssim$ $n$ $\lesssim$ 8 characterizes collisionless relaxation, then we can tentatively conclude that the distribution of stars in the outer regions of massive ellipticals, which have $n_{outer}$$\sim$$n_{N-body}$, was also shaped by collisionless processes. This conclusion is broadly consistent with the prevailing notion that massive ellipticals form through dry, dissipationless mergers.

  (2)~~   \cite{FSW05} and \cite{FSWB07} have shown using a combination of strong lensing and stellar population synthesis models that the outer regions $\gtrsim$$R_{E}$ of massive galaxies are more dark matter dominated than those of smaller galaxies. \cite{Cap06} arrive at a similar conclusion that fast-rotating galaxies, which are generally low-luminosity with steep cusps, have relatively lower dark matter content than the slow-rotating galaxies which are usually massive with shallow cusps. Also, \cite{Auger10} observe that the mean dark matter fraction within $R_{E}/2$ increases with galaxy size and mass. These observations are consistent with Planetary Nebula observations (\cite{Douglas07}, Napolitano et al. (2007,2009,2011), \cite{Tort09}) that the outer regions of small and intermediate mass ellipticals have varying degree of low dark matter content, while dark matter is quite dominant in the outer regions of massive ellipticals. These observations indicate that all massive ellipticals are dark matter dominated.
\pagebreak

  Since all the massive ellipticals in our sample have outer 5 $\lesssim$ $n$ $\lesssim$ 8, this could indicate that such galaxies are likely to be dark matter dominated. There might even be a relation between the outer $n$ of the galaxy's SB profile and its dark matter content. 
  \section{Summary and Discussion}\label{conclusion}  
  Critical to our understanding of galaxies, their formation and evolution, is our ability to accurately quantify the galaxies' structural properties and their variation with mass. The Sersic profile is the single most commonly used function for describing galaxy light distribution. The works of Caon, Capaccioli, Einasto, Ferrarese, Graham, Kormendy, Lauer and their collaborators have revolutionized our understanding of galaxy structure. These authors have shown us that a single Sersic profile does not fit the surface brightness distribution of ellipticals consistent with measurement errors over a radial range larger than $2$-$3$ decades. To extend the radial range of the fit, especially in to the central regions, the core-Sersic ($6$ parameters) and the Nuker ($5$ parameters for the central $10$-$20$ arcsec) profiles have been introduced, with a further addition of a King model ($3$ more parameters) for the nuclear region. 
  
  But even with these flexible and multi-parameter models, the fit residuals often exceed the measurement errors in some radial ranges, in spite of the overall $rms$ of residuals being low due to the large regions over which the Sersic profile fits well. Double Sersic models (Hopkins et al. 2009 a,b) provide an improvement, but still the residuals remain larger than measurement errors. 
  
  In addition to fit residuals exceeding measurement errors, the Sersic profile has another drawback. The galaxy structure, dynamics and evolution exist in 3D, so it is more meaningful to describe galaxies using 3D functions. The Sersic profile is an intrinsically 2D distribution whose deprojections preassumes an infinite 3D extent. If, for example, a galaxy has a truncation radius in 3D, a Sersic profile will not be able to model that. The parametric forms for deprojecting a Sersic profile (and power-laws, for that matter) are not well behaved near the centre. Further, these analytic functions are either reasonable but not extremely accurate (PS97, LGM99), or are accurate but extremely complicated (BG11).   
  
  To overcome the above limitations, we propose that the 3D luminosity density may be described with a multi-component Einasto profile whose parameters can be estimated by modelling the surface brightness using a multi-component DW-function. This is similar to the observations of Einasto and collaborators with a small sample of spiral (and the giant elliptical M87), but extends the idea to a much larger and diverse sample of shallow and steep cusp ellipticals and also allows for a direct parametric description of the 2D surface brightness profiles.
  
  \medskip
  
  We model the surface brightness (SB) profiles of $23$ ellipticals in and around the Virgo cluster with a multi-component DW-function and summarize our observations as follows:
  
  \medskip
  
  1.~~~ Multi-component DW-function fits the SB profiles of ellipticals with residuals consistently comparable to measurement errors over large dynamic ranges $\sim 10^5$ in radii, and $\sim 10^6$ in SB, with a median sample $rms$ of $0.032~\mgasc$.  Nine galaxies are well described with a $2$-component (central and outer) DW model, while for fourteen galaxies a third component was required, and confirmed through an F-test. The third component acts as an intermediate component between the central and outer components, except in NGC4621 and NGC4434 where it appears to be embedded within the outer component.

  \medskip
  
  2.~~~ All steep-cusp and shallow-cusp galaxies reveal a central component that is in excess to an inward extrapolation of the outer component. Its shape parameter $n$ is usually less than that of the outer component, which implies, as expected, that the central component is more concentrated than the outer component. Exceptions are NGC4459, NGC4387 and NGC4467 (section \ref{outcomp}).

  \medskip
  
  3.~~~ The central component of all massive shallow-cusp galaxies have $n \lesssim 1$, while those of steep-cusp galaxies generally have $n>2$, although there are some cases with $1\lesssim n<2$ (section \ref{outcomp}). This indicates that the central components of shallow-cusp galaxies
  
  a)~ are more concentrated than that of the steep-cusp galaxies even though the later are more denser in their central regions; and 
  
  b)~ could signal the presence of disk-like systems; however, this must be verified spectroscopically.

  \medskip
  
  4.~~~  The central component of the shallow-cusp galaxies is far more luminous (and massive) and spatially more extended (large $r_{-2}$ or $r_{3E}$) than that of the steep-cusp galaxies. Further, within a factor of two, all galaxies appear to host a similar fraction of total light in their central+intermediate components, with a weak indication that massive galaxies may be hosting a larger fraction. The last point is inconclusive due to the small number of massive ellipticals in our sample.

  \medskip
  
  5.~~~  In most of the shallow-cusp galaxies the outer component makes a comparable contribution to the density in the central regions with respect to that of the central component, while in the steep-cusp galaxies the central component is dominant.

  From the modelling point of view (section \ref{outcomp}), the shallow-cusp feature usually seen in massive galaxies is due to a combination of a larger $r_{-2}$, a low $n \lesssim 1$ implying a more concentrated component, and the non-negligible contribution to the density in the central regions from the $n\gtrsim 5$ outer component.
  
  \medskip

  6.~~~ Galaxy formation models indicate that in massive ellipticals, the formation, evolution and subsequent coalesence of binary SMBHs can remove nearly (2-3)$M_{BH}$ of stellar material from the central regions, in a typical merger (refer to iii, section \ref{Nbodyaftercoal}), leading to mass deficits. Our observation that the central component in shallow-cusp galaxies are already massive, can be used to constrain galaxy formation models by accounting for the amount of mass that must have assembled in these galaxies prior to mass ejection by the binary SMBH's.

  It is possible that the current shapes of the central as well as the intermediate DW-components, within the central kiloparsec or so, have been influenced by the evolution of the binary SMBHs; and consequently some deficit could well have occured. However, the amount and sign of such deficits depend on estimating a break radius $R_b$, and assuming that the functional form of the density profile beyond $R_b$ can be extrapolated all the way to the centre. This may not be meaningful and has led to large disagreements in the literature -- both for a given galaxy as well as for averages over a sample. We note that core-scouring by SMBHs is unlikely to be the sole mechanism for producing some of the largest 'cores' (shallow-cusps), and other processes in galaxy formation and evolution are likely to have played their role as well in forming this feature. In such a case extrapolating the density profile inwards from $R_b$, to {\it estimate} the mass deficit, will yield misleading results.

\medskip

  7.~~~ For $14$ of the $23$ galaxies we could describe the intrinsic 3D luminosity density distribution fairly uniquely with a multi-component Einasto model (section~\ref{lumdensity}). Since these galaxies span a wide range of luminosity $-24$$<$$M_{VT}$$<$$-15$ and come from both the steep- and shallow-cusp families, it is likely that other galaxies also have an intrinsic structure that can be modelled with a multi-component Einasto profile.

  \medskip
  
  8.~~~ Pure dark matter haloes are known to be well fit with the Einasto function. We have shown here that the same fitting function that describes the intrinsic density distribution of $\Lambda$CDM N-body haloes can also be used to model the intrinsic baryonic density of ellipticals, but for baryons a multi-component Einasto model is required. There thus appears to be an universality in the functional form of the density profile of baryons (stars) and dark matter.

  \medskip

  9.~~~ The Einasto shape parameter $n$ of dark matter haloes in N-body simulations is very similar to the $n$ of the outer component of our massive ellipticals; both are 5 $\lesssim$ $n$ $\lesssim$ 8. Further, PNe and strong lensing observations indicate that massive ellipticals are more dark matter dominated than less massive ellipticals. This indicates that our result -- the outer component of the surface brightness profiles of massive galaxies has 5 $\lesssim$ $n$ $\lesssim$ 8 -- could imply i) a common feature of collisionless systems; and ii) that galaxies with such $n$ for their outer (major) component are dark matter dominated.

  \medskip

 10. ~~~ In section \ref{massdeficit}, we have shown that the shapes of central and global profiles of shallow- and steep- cusp galaxies differ markedly and could be a result of differing formation pathways. In section \ref{outcomp} we have shown that the shallow-cusps can be typically described as systems with - i) central DW-components of {\it low} $n$ ($\lesssim 1$) and large scale radius $r_{-2}$ or $r_{3E}$, and ii) outer DW-component of {\it large} $n$ ($\gtrsim 5$) and central density comparable to that of the central DW-component. And in section \ref{BaryonsDM} we have shown that large outer $n$ systems are likely to be dark matter dominated.

   Hence, if dark matter has played a role in the formation and evolution of massive shallow-cusp galaxies, it will be instructive to explore its role, if any, in shaping the central regions and consequently in forming the shallow cusps.

  Finally we note that the galaxies modelled in this paper are all in and around the Virgo Cluster. Galaxy structure, formation and evolution is known to depend on the environment, and even though we suspect that our conclusion about the multi-component Einasto structure should be more widely applicable than Virgo, only more data can confirm that.

  \section*{Acknowledgments} 
  The authors would like to thank David Merritt, who provided valuable feedback and Michael Boylan-Kolchin for fruitful discussions. We would also like to thank the referee for providing constructive comments and suggestions that helped us improve the paper.

  \label{lastpage}
  \end{document}